\documentclass[%
reprint,
numerical,
prd,
superscriptaddress,
twocolumn,
nofootinbib,
amsmath,
amssymb,
]{revtex4-2}

\usepackage[utf8]{inputenc}
\usepackage{graphicx} 
\usepackage{dcolumn}
\usepackage{color}
\usepackage{booktabs}
\usepackage{acronym}
\usepackage{ifthen}
\usepackage{blindtext}
\usepackage[normalem]{ulem}
\usepackage[colorlinks=true, linkcolor=blue, urlcolor=blue, citecolor=blue]{hyperref}
\usepackage{etoolbox}
\usepackage{fancyhdr}
\usepackage{xspace}
\usepackage{textcomp}
\usepackage{multirow}
\usepackage[caption=false]{subfig}
\usepackage{lineno}
\usepackage{tabularx}
\usepackage{aas_macros}
\usepackage{orcidlink}
\usepackage{appendix}

\def\gwh{gravitational-wave\xspace}
\def\gws{gravitational waves\xspace}

\def\pbhs{primordial black holes\xspace}
\def\pbh{primordial black hole\xspace}
\newcommand{\ftilde}{\tilde{f}}

\newcommand{\fpbh}{f_\text{PBH}}
\newcommand{\fsup}{f_\text{sup}}
\newcommand{\mone}{m_1}
\newcommand{\mtwo}{m_2}
\newcommand{\fmone}{f(m_1)}
\newcommand{\fmtwo}{f(m_2)}

\def\pspat{p_{\rm spatial}}
\newacro{S/N}{signal-to-noise ratio}
\newacro{PN}{post-Newtonian}

\newacro{O3}{the third observing run}

\newacro{O2}{the second observing run}

\newacro{MSP}{millisecond pulsar}
\def\MSP{\ac{MSP}\xspace}
\def\CW{CW\xspace}

\newacro{FFT}{fast Fourier transform}
\newacro{CNN}{convolutional neural network}

\newcommand{\pf}{\textrm{P}_{\textrm{f}}}

\newcommand{\Tcoh}{T_{\textrm{coh}}}

\newcommand{\pfa}{p_{\mathrm{fa}}}
\newcommand{\Tfft}{T_{\mathrm{FFT}}}

\newcommand{\Nwalk}{N_{\mathrm{w}}}
\newcommand{\Ntemp}{N_{\mathrm{t}}}
\newcommand{\Ntot}{N_{\mathrm{tot}}}

\newcommand{\fstat}{{\em $\mathcal{F}$-statistic}}

\newcommand{\soap}{{\em SOAP}}

\newtoggle{endauthorlist}
\toggletrue{endauthorlist}

\newtoggle{fullauthorlist}
\toggletrue{fullauthorlist}

\def\beq{\begin{equation}}
\def\eeq{\end{equation}}
\def\bea{\begin{eqnarray}}
\def\eea{\end{eqnarray}}

\def\Tcoh{T_{\textrm{\mbox{\tiny{coh}}}}}

\def\max{\textrm{\mbox{\tiny{max}}}}

\def\ec{\textrm{~,}}
\def\ed{\textrm{~.}}
\def\fdot{\dot f_{\rm GW}}

\newif\ifshowfigs
\showfigstrue

\newcommand{\F}{\mathcal{F}}
\newcommand{\SNR}{\textrm{SNR}}
\newcommand{\Fstat}{$\mathcal{F}$-statistic}
\newcommand{\nseg}{$N_{\textrm{seg}}$}
\newcommand\T{\rule{0pt}{2.6ex}}       
\newcommand\B{\rule[-1.2ex]{0pt}{0pt}} 

\def\pf{{\footnotesize\textsc{PowerFlux}}\xspace}
\def\fh{{\footnotesize\textsc{FrequencyHough}}\xspace}
\def\sh{{\footnotesize\textsc{SkyHough}}\xspace}
\def\so{{\footnotesize\textsc{SOAP}}\xspace}
\def\eh{{\footnotesize\textsc{Einstein@Home}}\xspace}
\def\pyfstat{{\footnotesize\textsc{PyFstat}}\xspace}

\def\fgw{f_{\rm GW}}
\def\frot{f_{\rm rot}}
\def\Izz{I_{\rm zz}}

\def\sci#1#2{#1\times10^{#2}}

\def\vs{{\it vs.}}

\def\loweststrain{$\sci{9.7}{-26}$}
\def\loweststrainfreq{290}
\def\ratiolow{1.6}

\def\RAJ{\textrm{RA}_{\textrm J2000}}
\def\DECJ{\textrm{DEC}_{\textrm J2000}}

\begin{document}

\title{All-sky Search for Continuous Gravitational Waves from Isolated Neutron Stars in the First Part of the Fourth LIGO-Virgo-KAGRA Observing Run}

\iftoggle{endauthorlist}{
  %
  %
  \let\mymaketitle\maketitle
  \let\myauthor\author
  \let\myaffiliation\affiliation
  \author{The LIGO Scientific Collaboration}
  \author{The Virgo Collaboration}
  \author{The KAGRA Collaboration}
  \noaffiliation
}
{
\iftoggle{fullauthorlist}{
  \author{A.~G.~Abac\,\orcidlink{0000-0003-4786-2698}}
\affiliation{Max Planck Institute for Gravitational Physics (Albert Einstein Institute), D-14476 Potsdam, Germany}
\author{I.~Abouelfettouh}
\affiliation{LIGO Hanford Observatory, Richland, WA 99352, USA}
\author{F.~Acernese}
\affiliation{Dipartimento di Farmacia, Universit\`a di Salerno, I-84084 Fisciano, Salerno, Italy}
\affiliation{INFN, Sezione di Napoli, I-80126 Napoli, Italy}
\author{K.~Ackley\,\orcidlink{0000-0002-8648-0767}}
\affiliation{University of Warwick, Coventry CV4 7AL, United Kingdom}
\author{A.~Adam}
\affiliation{OzGrav, University of Western Australia, Crawley, Western Australia 6009, Australia}
\author{C.~Adamcewicz\,\orcidlink{0000-0001-5525-6255}}
\affiliation{OzGrav, School of Physics \& Astronomy, Monash University, Clayton 3800, Victoria, Australia}
\author{S.~Adhicary\,\orcidlink{0009-0004-2101-5428}}
\affiliation{The Pennsylvania State University, University Park, PA 16802, USA}
\author{D.~Adhikari}
\affiliation{Max Planck Institute for Gravitational Physics (Albert Einstein Institute), D-30167 Hannover, Germany}
\affiliation{Leibniz Universit\"{a}t Hannover, D-30167 Hannover, Germany}
\author{N.~Adhikari\,\orcidlink{0000-0002-4559-8427}}
\affiliation{University of Wisconsin-Milwaukee, Milwaukee, WI 53201, USA}
\author{R.~X.~Adhikari\,\orcidlink{0000-0002-5731-5076}}
\affiliation{LIGO Laboratory, California Institute of Technology, Pasadena, CA 91125, USA}
% ...

}
{
  \author{The LIGO Scientific Collaboration}
  \affiliation{LSC}
  \author{The Virgo Collaboration}
  \affiliation{Virgo}
  \author{The KAGRA Collaboration}
  \affiliation{KAGRA}
}
}
\date[\relax]{compiled \today}

\begin{abstract}
  We present results from an all-sky search for continuous gravitational waves, using three
  different methods applied to the first eight months of LIGO data from the fourth
  LIGO-Virgo-KAGRA Collaboration's observing run.
  We aim at signals potentially emitted by
  rotating, non-axisymmetric isolated neutron star
  in the Milky Way. The analysis spans a
  frequency range from 20 Hz to 2000 Hz and accommodates
  frequency derivative magnitudes up to $10^{-8}$ Hz/s.
  No statistically significant periodic gravitational wave signals were detected.
  We establish 95\%\ confidence-level (CL) frequentist upper limits on
  the dimensionless strain amplitudes. The most stringent population-averaged strain upper limits
  reach \loweststrain\ near \loweststrainfreq\ Hz, matching
  the best previous constraints from 250 to $\sim$1700 Hz while extending coverage to a much broader
  spin-down range. At higher frequencies,
  the new limits improve upon previous results by factors of approximately $\sim$\ratiolow.
  These constraints are applied to three astrophysical scenarios:
  1) the distribution of galactic
  neutron stars as a function of spin frequency and ellipticity;
  2) the contribution of millisecond pulsars
  to the ``GeV excess'' near the galactic center; and
  3) the possible dark matter fraction composed of nearby inspiraling
  primordial binary black holes with asteroid-scale masses.

\end{abstract}

\maketitle

\section{Introduction}
\label{sec:introduction}

We report new results of all-sky searches for continuous-wave (CW), nearly monochromatic gravitational waves
using the data from the two LIGO detectors~\cite{bib:aligodetector1} during the first eight months (O4a) of
the fourth LIGO-Virgo-KAGRA observing run.
Potential sources include conventional but slightly non-axisymmetric, fast-spinning galactic neutron stars~\cite{bib:GandG,bib:Laskyreview,bib:OwenReview} and
more exotic sources: very nearby and light primordial binary black hole systems~\cite{bib:MillerEtalPBH}.
All-sky searches for continuous gravitational waves from isolated neutron stars
have been carried out in Advanced LIGO and Virgo data previously~\cite{bib:cwallskyO1paper1,bib:cwallskyO1paper2,bib:cwallskyO1EatH,bib:cwallskyO2,bib:PalombaEtalAxion,bib:cwallskyO2EatH,bib:cwallskyFalconO2MidFreq,bib:cwallskyFalconO2HighFreq,bib:cwallskyFalconO2LowFreq,bib:WetteEtalDeep,bib:cwallskyO3a,bib:cwallskyO3FourPipelines,bib:cwallskyO3EatH,bib:TripatheeRilesO3,bib:cwallskyO3EatHbucket,bib:cwallskyO3EatHHighFreq}.
The all-sky results presented here are the most sensitive to date in strain amplitude for the full frequency and frequency derivative
parameter space covered, although less sensitive than some previous O3 searches in some subsets of that parameter space.
All-sky searches for isolated gravitational CW sources belong to a suite of searches for
quasi-monochromatic gravitational wave emitters from the galaxy and beyond (recent reviews:~\cite{bib:TenorioKeitelSintesreview,bib:PiccinniReview,bib:RilesReview,bib:WetteReview}).

Three different semi-coherent search programs (``pipelines'')  are used here:
1) \pf~\cite{bib:cwallskyS4,bib:cwallskyearlyS5,bib:cwallskyS5,bib:cwallskyS6,bib:cwallskyO1paper1,bib:cwallskyO1paper2}
with loose-coherence follow-up~\cite{bib:loosecoherence,bib:LooseCoherenceWellModeledSignals} (templated);
2) \fh~\cite{bib:cwallskyO1paper2,bib:cwallskyO2,bib:cwallskyO3FourPipelines,bib:freqhough2}
(templated); and
3) \so~\cite{bib:ViterbiGlasgow,bib:SOAP2} (hidden Markov model; non-templated).
Collectively, these searches span a signal frequency band from 20 Hz to as high as 2000 Hz and allow
frequency time derivative magnitudes as high as $10^{-8}$ Hz/s.
Finding no credible signals from any search, we set upper limits on signal strain amplitudes and present
astrophysical population inferences.

This article is organized as follows:
Section~\ref{sec:data} describes the data set used, including steps taken to mitigate extremely loud and relatively frequent instrumental glitches seen in the O4 LIGO data, a phenomenon also seen in third observing run (O3).
Section~\ref{sec:methodology} describes the methodologies used by the \pf, \fh\ and \so\ search pipelines,
while Section~\ref{sec:interpretation} describes the methodologies used to interpret the results for different astrophysical populations.
Section~\ref{sec:search_results} presents the results of the searches, 
and Section~\ref{sec:interpretation_results} interprets those results astrophysically.
Section~\ref{sec:conclusions} concludes with a summary of the results
and prospects for future searches. The appendices provide more details on the search methodologies and validations.

\section{Data sets used}
\label{sec:data}

We report here results from searches of early LIGO data from
the fourth observing run O4~\cite{bib:CapoteEtal,bib:GanapathyEtal,bib:JiaEtal}.
Intrinsic detector noise has been reduced with respect to O3 levels, especially at frequencies
higher than $\sim$100 Hz for the LIGO Hanford (H1) and LIGO Livingston (L1) interferometers.
The O4 run began on May 24, 2023 and completed on November 18, 2025. This article
presents results based on analysis of the first eight months of O4, known as the O4a period,
which ended on January 16, 2024, at the start of a 2-month commissioning break.

The Virgo interferometer~\cite{bib:avirgodetector} did not operate during the O4a period, while the KAGRA
interferometer~\cite{bib:KAGRA} observed for only one month and with a much lower sensitivity than the LIGO interferometers.
Hence, results from analyzing only the H1 and L1 data are presented here.
We use data from the \texttt{GDS-CALIB\_STRAIN\_CLEAN} frame channel,
corresponding to the standard online calibration \citep{viets2018reconstructing, Sun:2020wke, Sun:2021qcg,Wade:2025tgt}
with some noise subtraction applied \citep{T2100058,1911.09083,O4gwosc}.

Prior to searching for CW signals, the quality of the O4a data was assessed~\cite{bib:SoniEtal}
and steps taken to mitigate the effects of instrumental artifacts, including vetoes of data segments
with severe instrumental disturbances~\cite{bib:O4segments}. As in previous Advanced LIGO observing
runs~\cite{bib:linesO1O2}, instrumental ``lines'' (sharp peaks in fine-resolution run-averaged
H1 and L1 spectra) are marked, and where possible, their instrumental or environmental sources
identified~\cite{bib:O4aLinesList}.
The resulting database of artifacts proved helpful in eliminating spurious signal candidates
emerging from the search.

Another type of artifact observed in the O4a data for both H1 and L1 were relatively frequent and loud
``glitches'' (short, high-amplitude instrumental transients) with most of their spectral power
lying below $\sim$500 Hz. As was true in the O3 run, these transients can be loud enough and frequent
enough to distort the noise floor appreciably in spectra taken over even long durations.
To mitigate the effects of these glitches on O4 CW searches at frequencies,
a glitch-gating algorithm~\cite{bib:SelfGatingO4} was applied by the \pf\ and \so\ pipelines that is similar in approach, but somewhat different in design,
from the approach~\cite{bib:SelfGating2} used in an O3 \eh\ all-sky search~\cite{bib:cwallskyO3EatH}.
The algorithm iteratively excises short intervals of time around detected glitches, verifying that
the resulting change in spectral noise is an improvement, starting with the loudest glitches and
considering smaller glitches in turn, until the preserved live time falls below 99\%, or no further
loud glitches are detected.
The excision is carried out via inverse Planck windowing in which the data stream is tapered to
zero over a half-second interval preceding the glitch and tapering back to unexcised data
over another half second following the glitch. The tapered intervals are included when computing
the deadtime losses capped at 1\%. Glitches are defined by large excursions from nominal values of absolute power or of whitened power in the
25-50 Hz and 70-110 Hz bands. ``Hardware injections'' (see Section~\ref{sec:hwinjections}) are recovered with
higher signal-to-noise ratio (SNR) in gated data than in the
original, ungated data, as was true for the O3 data~\cite{bib:SelfGating1}. The \fh\ method, on the other hand, uses pre-processed
input data where short time-domain disturbances have been subtracted following the procedure described in \cite{bib:SFTdatabase}, see Sec.
\ref{sec:frequencyhough_methodology} for more details.

All three search methods described in Section~\ref{sec:methodology} use ``short'' discrete Fourier transforms
of the strain time series, but with different choices of coherence time. For the computationally demanding
semi-coherent searches (\pf\ and \fh), there is a tradeoff between sensitivity and cost that disfavors
longer coherence times at higher frequencies. The \pf\ method uses three distinct choices of coherence time:
7200 s for 20-475 Hz, 3600s for 475-1475 Hz and 1800s for 1475-2000 Hz when computing
``short'' Fourier transforms (SFTs);
the 7200 s SFTs are created from gated data. The \fh\ method uses a band-sampled-data (BSD~\cite{bib:BSD})
approach that enables many more discrete choices of SFT coherence time, ranging from 16384 s at 20 Hz
to 2048 s at 1024 Hz, with discrete steps at each 1 Hz boundary.
The \so\ method uses spectral averages over 1-day intervals based on 1800s SFTs created from gated data.

\section{Search Methodology}
\label{sec:methodology}

The three methods applied in this analysis have been used previously and described in detail elsewhere
(see sections~\ref{sec:powerflux_methodology}-\ref{sec:soap_methodology} below for references).
In the following subsections, we summarize briefly their different approaches after describing the signal
model used explicitly in the \pf\ and \fh\ semi-coherent searches. We also present validation of the
search programs via recovery of ``hardware injections.''

\subsection{Signal model}
\label{sec:model}

The signal templates used in the \pf\ and \fh\ searches
assume a classical model of a spinning neutron star with a time-varying quadrupole moment that produces
circularly polarized gravitational radiation along the rotation axis, linearly polarized radiation
in the directions perpendicular to the rotation axis and elliptical polarization for the general case.
The star's orientation, which determines the polarization, is parametrized by the inclination angle $\iota$ of its
spin axis relative to the detector line-of-sight and by the angle $\psi$ of the axis projection on the plane of the sky.
The linear polarization case ($\iota=\pi/2$) is the most unfavorable
because the gravitational wave flux impinging on the detectors
is smallest for an intrinsic strain amplitude $h_0$,
possessing eight times less incident strain power than
for circularly polarized waves ($\iota = 0,\>\pi$).

The strain signal model $h(t)$ for a periodic source is assumed to be the following function of time $t$:
\begin{eqnarray}
  h(t) & = & h_0\bigl(F_+(t, \alpha_0, \delta_0, \psi)\frac{1+\cos^2(\iota)}{2}\cos(\Phi(t))\vphantom{\frac{1+\cos^2(\iota)}{2}}\nonumber\\
  & & \quad\>+F_\times(t, \alpha_0, \delta_0, \psi)\cos(\iota)\sin(\Phi(t))\bigr)\ec
\end{eqnarray}

\noindent where $h_0$ is the intrinsic strain amplitude, $\Phi(t)$ is the signal phase, $F_+$ and $F_\times$ characterize the detector responses to signals with ``$+$'' and ``$\times$''
quadrupolar polarizations \cite{bib:cwallskyS4}, and the sky location is described by right ascension $\alpha_0$ and
declination $\delta_0$.

In a rotating triaxial ellipsoid model for a
star at distance $r$ spinning at frequency $\frot$ about its (approximate) symmetry axis ($z$),
the amplitude $h_0$ can be expressed as
\begin{eqnarray}
  \label{eqn:hexpected}
h_0 & = & {4\,\pi^2G\epsilon\Izz\fgw^2\over c^4r} \nonumber \\
    & = & [1.1\times10^{-26}]\Bigl[{\epsilon\over10^{-6}}\Bigr]\Bigl[{\Izz\over I_0}\Bigr]\Bigl[{\fgw\over{\rm 100\>Hz}}\Bigr]^2
\Bigl[{1\>{\rm kpc}\over r}\Bigr],
\end{eqnarray}
where $I_0=10^{38}$ kg$\cdot$m$^2$ (10$^{45}$ g$\cdot$cm$^2$) is a nominal neutron star moment of inertia $\Izz$ about $z$,
and the gravitational radiation is emitted at frequency $\fgw=2\,\frot$.
The equatorial ellipticity $\epsilon$ is a convenient, dimensionless measure of stellar non-axisymmetry:
\begin{equation}
\label{eqn:ellipticity}
\epsilon \quad \equiv \quad {I_{xx}-I_{yy}\over I_{zz}}.
\end{equation}

The phase evolution of the signal is given in the reference frame of the Solar System barycenter (SSB) by the second-order approximation:
\begin{equation}
\label{eqn:phase_evolution}
\Phi(t)=2\pi\left(f_\textrm{GW}\cdot (t-t_0)+\fdot\cdot (t-t_0)^2/2\right)+\phi\ec
\end{equation}
where $f_\textrm{GW}$ is the SSB source frequency, $\fdot$ is the first frequency derivative (which, when negative,
is termed the spin-down), $t$ is the SSB time, and the initial phase $\phi$ is
computed relative to reference time $t_0$. When expressed as a function of the local time of ground-based detectors,
Eq.~\ref{eqn:phase_evolution} acquires sky-position-dependent Doppler shift terms.

All known isolated pulsars spin down more slowly
than the maximum value of $|\dot{f}|_\max$ used here, and as seen in the results
section, the equatorial ellipticity required for higher $|\dot{f}|$ is improbably high
for a source losing rotational energy primarily via gravitational radiation at low
frequencies. More plausible is a source with spin-down dominated by electromagnetic
radiation energy loss, but for which detectable gravitational radiation is also emitted.

\subsection{The \pf\ search}
\label{sec:powerflux_methodology}
The \pf\ search starts with a semi-coherent
stage~\cite{bib:cwallskyS4,bib:cwallskyO1paper1,bib:cwallskyO1paper2,bib:cwallskyS5} and uses loose coherence to follow up
outliers~\cite{bib:loosecoherence,bib:LooseCoherenceWellModeledSignals}. In brief (see Appendix~\ref{sec:powerflux_appendix} for details), strain power is summed over many SFTs after
correcting for Doppler modulations, for a large bank of templates based on sky location, frequency, frequency
derivative and stellar orientation. The maximum strain powers detected over the entire sky and for all
frequency derivatives in each narrow frequency sub-band (see Table~\ref{tab:powerflux_tuning_info}
in Appendix~\ref{sec:powerflux_appendix} for the frequency-dependent widths) define strict frequentist upper limits.
The search is carried out over the frequency band 20-2000 Hz and the frequency derivative range
$[-10^{-8}$--$10^{-9}]$ with three discrete choices of SFT coherence time (see Appendix~\ref{sec:powerflux_appendix}).
Figure~\ref{fig:parameterspace} depicts the parameter space coverage for the \pf\ search and for the \fh\ search described
in Sec.~\ref{sec:frequencyhough_methodology}.

\begin{figure}[htbp]
  \begin{center}
    \ifshowfigs
    \includegraphics[width=3.25in]{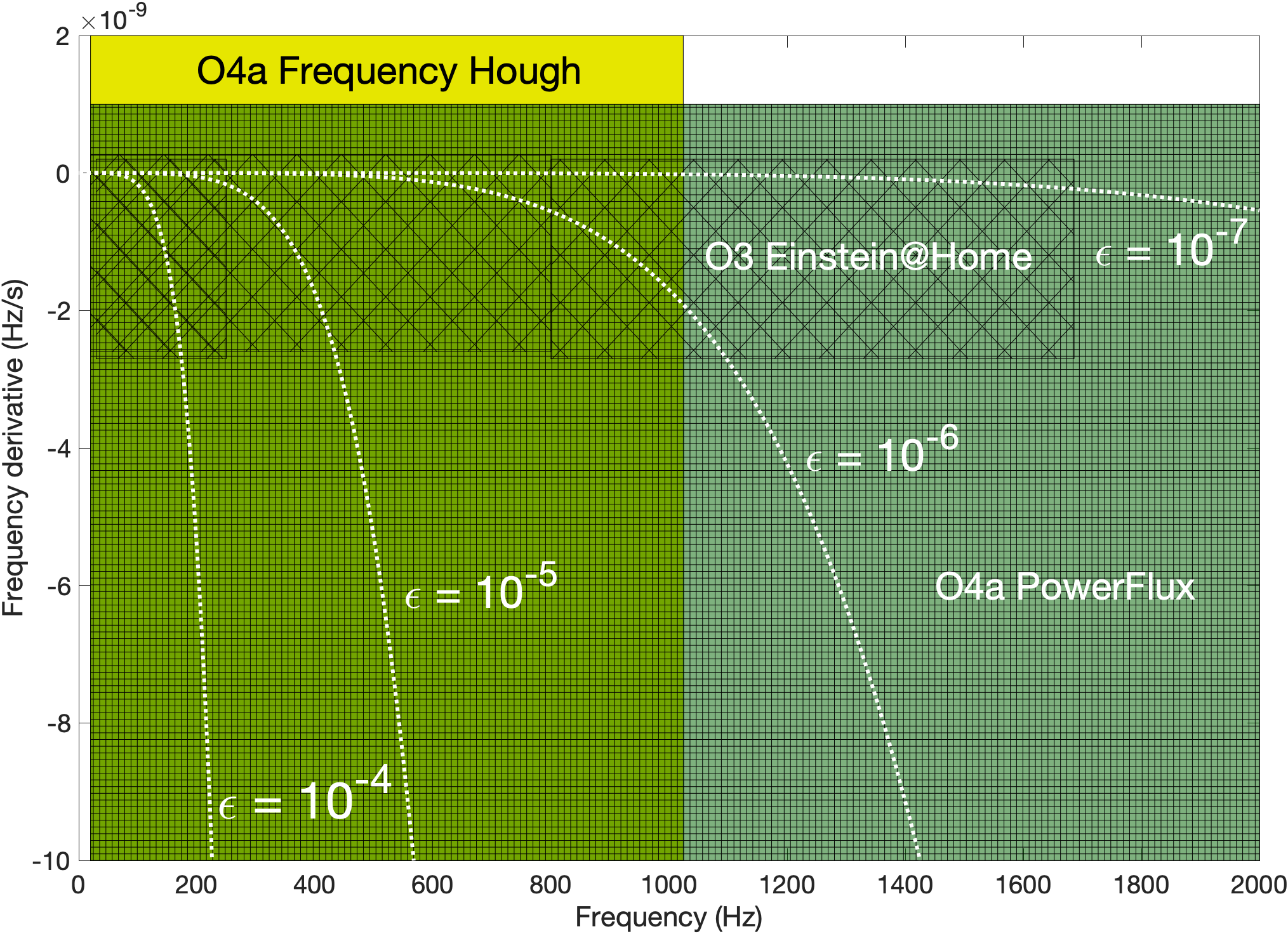}
    \fi
    \caption[Comparison of parameter space coverage]{Comparison of frequency and spin-down ranges for the templated searches presented here,
      along with those of previous \eh\ searches~\cite{bib:cwallskyO3EatH,bib:cwallskyO3EatHbucket,bib:cwallskyO3EatHHighFreq} of the O3 data
      discussed in section~\ref{sec:constraints}.
      The dark-green-shaded rectangle with fine gridding bars shows the 20--2000 Hz and
      $-10^{-8}$--$10^{-9}$ Hz/s range for the O4a \pf\ search. The slightly taller but narrower yellow rectangle shows the
      region searched in the O4a \fh\ analysis. The smaller rectangles with diagonal hatching show the three \eh\ search ranges for the
      O3 data~\cite{bib:cwallskyO3EatH,bib:cwallskyO3EatHbucket,bib:cwallskyO3EatHHighFreq}, which have small variations in $\fdot$ ranges searched.
      The white curves show corresponding nominal ellipticities implied by a given frequency and spin-down for the gravitar model
      in which all rotational energy loss can be attributed to gravitational wave emission due to a mass quadrupole. Searching for
      high ellipticity values at a given frequency requires a high $\fdot$ range.
      (color online)}
    \label{fig:parameterspace}
  \end{center}
\end{figure}

The \pf\ pipeline\footnote{The \pf\ infrastructure used here is nearly identical to that used for
the O3a analysis~\cite{bib:cwallskyO3a}, but has been upgraded to use AVX floating point operations in
frequently executed code.}
has a hierarchical structure
that permits systematic follow-up of loud outliers from the
initial stage. The later stages improve intrinsic strain sensitivity by
increasing effective coherence time while dramatically
reducing the parameter space volume over which the follow-up
is pursued. The pipeline uses loose coherence~\cite{bib:loosecoherence}
with stages of improving refinement via steadily increasing effective
coherence times. Any outliers that survive
all stages of the search pipeline are examined manually for
contamination from known instrumental artifacts and for evidence of
contamination from a previously unknown single-interferometer artifact.
Those for which no artifacts are found are subjected to further
follow-up described below.

In the pipeline's initial stage, the main \pf\
algorithm \cite{bib:cwallskyS4,bib:cwallskyO1paper1,bib:cwallskyO1paper2,bib:cwallskyearlyS5,bib:cwallskyS5,bib:cwallskyS6}
establishes upper limits and produces lists of outliers.
The program sets strict frequentist upper limits
on detected strain power in circular and linear polarizations that
apply everywhere on the sky except for small regions near the ecliptic
poles, where signals with small Doppler modulations can be masked by
stationary instrumental spectral lines.
The procedure defining these excluded regions is described in \cite{bib:cwallskyS5} and applies to less than $0.2$\% of the sky over the entire run,
where the precise shapes of the regions near the poles depend on assumed signal frequency and spin-down.
Initial outliers are defined by a
joint H1-L1 signal-to-noise ratio (SNR) greater than a threshold of 7, with consistency among
corresponding H1, L1 and joint H1-L1
outliers (criteria described in section~\ref{sec:powerflux_outlier_followup}). These outliers are then followed up with
a loose-coherence detection pipeline \cite{bib:loosecoherence,bib:LooseCoherenceWellModeledSignals,bib:cwallskyS5}, which is
used to reject or confirm the outliers.

\subsubsection{Upper limits determination}
\label{sec:powerflux_upperlimit_determination}

The 95\% confidence-level (CL) upper limits presented in section~\ref{sec:powerflux_results}
are reported in terms of the worst-case value of $h_0$ (linear polarization) and for
the most sensitive case of circular polarization.

These upper limits,
produced in stage 0, are based
on the overall noise level and largest outlier in strain found for every template in each narrow sub-band in
the first stage of the pipeline. Sub-bands are analyzed by separate instances of \pf~\cite{bib:cwallskyS5}

To allow robust analysis of the entire spectrum, including regions with severe spectral artifacts, a
{\em Universal statistic} algorithm~\cite{bib:universalstatistic,bib:cwallskyO1paper1} is used for establishing upper limits.
The algorithm is derived from the Markov inequality and shares its independence from the underlying noise distribution.
It produces upper limits less than $5$\% above optimal in case of Gaussian noise. In non-Gaussian bands,
it can report values larger than what would be obtained if the true underlying
distribution were known, but the upper limits are always
at least 95\% valid. Appendix~\ref{sec:powerflux_upperlimits_app} gives details
on the validation of the \pf\ upper limits derived in this analysis.

\subsubsection{Outlier follow-up}
\label{sec:powerflux_outlier_followup}

A follow-up search for detection is carried out for high-SNR outliers found in stage 0.
The outliers are subject to an initial coincidence test.  For each
outlier with $\SNR>7$ in the combined H1 and L1 data, we require there
to be outliers in the individual detector data of the same small sky patch,
approximately square with side length $\sim$30 mrad $\times$ (100 Hz / frequency) that have $\SNR>5$ and
match the parameters of the combined-detector outlier within 2.5~mHz in frequency and
$\sci{3}{-10}$\,Hz/s in spin-down.  The combined-detector SNR is additionally required to be above both
single-detector SNRs, in order to suppress single-detector instrumental artifacts, except for unusually
loud outliers (H1, L1 and combined SNRs all greater than 20).

The identified outliers using combined data are then passed to the follow-up stage using a loose-coherence
algorithm~\cite{bib:loosecoherence} with progressively
improved determination of frequency, spin-down, and sky location.

As the initial stage 0 sums only powers, it does not use the relative phase between interferometers,
which results in some degeneracy among sky position, frequency, and spin-down. The first loose-coherence
follow-up stage (1) demands greater temporal
coherence within each interferometer, which should
boost the SNR of viable outliers, but combines H1 and L1 power sums incoherently,
The subsequent stage (2) uses combined H1 and L1 data coherently, providing tighter bounds on outlier location.
Details concerning parameters used in outlier follow-up may be found in Table~\ref{tab:powerflux_followup_parameters} in Appendix~\ref{sec:powerflux_config_app}.
Validation of the \pf\ loose-coherence outlier follow-up is described in Appendix~\ref{sec:powerflux_outlier_app}.

As in previous \pf\ analyses with loose-coherence follow-up~\cite{bib:cwallskyO1paper1,bib:cwallskyO1paper2},
only a mild influence from parameter mismatch is expected, as the parameters are chosen to accommodate the worst
few percent of injections. The follow-up procedure establishes wide margins for outlier follow-up.
For example, when transitioning from the semi-coherent stage 0 to the loose-coherence stage 1 below 475 Hz, the effective
coherence length increases by a factor of 4. The average true signal SNR should then increase by more than $40$\%.
But the threshold used in follow-up is only 15--20\%, depending on frequency, which accommodates unfavorable noise conditions,
template mismatch, and detector artifacts.

As in the O3a \pf\ search, we apply a combination of manual inspection and systematic deep follow-up
to any outliers that survive the second stage of follow-up.
After a simple clustering in the parameter space of ($f$, $\dot f$, $\alpha$, $\delta$), the loudest outlier in each cluster is
examined manually via a ``strain histogram''~\cite{bib:TripatheeRilesO3} and frequency trajectory graph that helps assess the degree of
contamination in the putative outlier signal by known or obvious instrumental contamination.
Surviving outliers not exhibiting clear contamination are subjected to a final
follow-up method using
the Python-based \pyfstat~\cite{bib:pyFstat,bib:PyFstatZenodo} software
infrastructure to combine a
MCMC approach~\cite{bib:AshtonPrix,bib:TenorioEtal} with semi-coherent summing of the well known \Fstat\ detection
statistic~\cite{bib:JKS}. In this approach, the parameter space near to those values from a stage-2 survivor is sampled
randomly according to a certain probability density function determined by the \Fstat\ likelihood function.
We follow the same implementation~\cite{bib:TenorioEtal} applied to the O3a \pf\ outliers~\cite{bib:cwallskyO3a}.

Briefly, the O4a observation time is divided into \nseg\ segments, for each of which the \Fstat\ is computed
over a coherence time approximately equal to the observation time divided by \nseg. For each point sampled
in parameter space, the sum of the \Fstat\ values is computed to form a total detection statistic.
This procedure is repeated, decreasing the number of segments (increasing the coherence time),
using the resulting MCMC-maximized  \Fstat\ sum as the seed for the next stage, with a
consequent reduction in parameter space volume searched.
For this O4a analysis we choose six successive stages of follow-up with decreasing values of \nseg\ = 500 (coherence
time of 0.47 day), 250, 150, 60, 5 and 1. A random sample of 600 off-source sky locations having the same declination as the putative
signal direction, but separated by more than 90 deg from that direction, is used to determine a non-signal expectation for
the background distributions in the same frequency band~\cite{bib:TenorioEtal}.
A Bayes factor $B_{\rm SN}$ is computed from the change in
\Fstat\ values for a nominal signal compared to the  empirical background  distribution in the last stage (\nseg=5 to \nseg=1).

\subsection{The \fh\ search}
\label{sec:frequencyhough_methodology}
The \fh\ pipeline is a semi-coherent procedure in which interesting outliers are selected in the signal parameter space, and then are followed-up in order to confirm or reject them. This method has been used in several past all-sky searches of Virgo and LIGO data~\cite{bib:cwallskyfreqhoughVSR2VSR4,bib:cwallskyO1paper2,bib:cwallskyO2,bib:PalombaEtalAxion,bib:cwallskyO3FourPipelines}. A detailed description of the methodology can be found in~\cite{bib:freqhough2}, but some changes have been made to the procedure, in order to improve the sensitivity of the search and also to optimize its computational cost. We thus describe in the following the main analysis steps, leaving the details to Appendix~\ref{app:fh_details}. A relevant novelty with respect to the past is that the \fh\ pipeline now uses the BSD framework~\cite{bib:BSD}. This change allows us to evaluate Fast Fourier Transforms (FFTs) with duration $\Tfft$ optimized for each 1 Hz band (the previous implementation used four different bands, each of fixed $\Tfft$ duration). This duration fixes the coherent time of the search, hence a non optimal choice of this parameter would reduce the final sensitivity. 

From the collection of FFTs computed over the run duration, the so-called ``peakmap'' is built. For each sky position\footnote{Over a suitable grid, for which the bin size depends on the frequency and sky location.} the time-frequency peaks of the peakmap are properly shifted, to compensate for the Doppler effect due to the detector motion~\cite{bib:freqhough2}. The shifted peaks are then fed to the \fh\ algorithm~\cite{bib:freqhough2}, which transforms each peak to the frequency/spin-down plane of the source. A new implementation of the {\fh} algorithm has been used for the first time in this search,  see Appendix~\ref{app:newfh} for more details. The output of a \fh\, transform is a 2-D histogram in the frequency/spin-down plane of the source. Outliers, that are statistically significant points in this plane, are selected using a criterion described in Appendix~\ref{app:fh_details}.
As in past analyses~\cite{bib:cwallskyO1paper2,bib:cwallskyO2}, coincidences among outliers of the two detectors are required, using a distance metric $d_{\rm FH}$ built in the four-dimensional
parameter space of sky position $(\lambda,~\beta)$ (in ecliptic coordinates), frequency $f$ and spin-down $\dot{f}$.
Coincident outliers are ranked according to the value of a specific statistic and the most significant are selected and subject to the follow-up.

\subsubsection{Follow-up}\label{sec:fh_fu}
The \fh\ follow-up runs on each outlier of each coincident pair. It is based on the construction of a new peakmap, over $\pm 3$ {\it coarse} bins around the frequency of the outlier, with a tripled $T_{\rm FFT}$ with respect to the all-sky stage.
This new peakmap is built after heterodyning the data for the Doppler and spin-down (or spin-up) of the outlier. 
A refined sky grid is then constructed around the outlier position, spanning $\pm3$ coarse bins, to account for uncertainties in the outlier's parameters. For each point on this refined grid, we correct the peakmap by shifting the frequency peaks to remove the residual Doppler. 
Each corrected peakmap is the  input for the \fh\ transform, which explores the relevant ranges in frequency and spin-down ($\pm±3$ coarse bins in both dimensions). Among all the resulting FrequencyHough histograms, the most significant peak associated with a {\it refined} set of parameters is selected and passed on for further post-processing steps to reduce the number of false outliers as described in \cite{bib:cwallskyO3FourPipelines} and in Appendix~\ref{app:fh_fu_details}. Only those outliers that pass all these tests proceed to the subsequent follow-up stages, where the coherence time is further increased.
These additional stages are carried out with \pyfstat{}~\cite{bib:AshtonPrix,bib:pyFstat,Mirasola:2024lcq} in a similar way to what is described in Sec.~\ref{sec:powerflux_methodology}.

\subsubsection{Upper limits}\label{sec:fhulmethod}
``Population averaged'' upper limits are computed for every 1~Hz sub-band in the range of the search (i.e., 10--1024~Hz). First, for each detector we use the following analytical relation (where an error in the numerical coefficient has been corrected w.r.t. the original formula presented in~\cite{bib:freqhough2}, see Appendix~\ref{app:sens}):
\begin{equation}
h_{\mathrm{UL},95\%}\approx
\frac{4.71}{N^{1/4}}\sqrt{\frac{S_n(f)}{T_{\rm FFT}}}\sqrt{\rm CR_{\rm max}+1.6449},
\label{eq:hul}
\end{equation} 
where $N=\mathrm{round}\left(\frac{T_\mathrm{obs}}{T_\mathrm{FFT}}\right)$, $S_n(f)$ is the detector average noise power spectrum, computed through a weighted mean over time segments of duration $T_{\rm FFT}$ (in order to take into account noise non-stationarity), and $\rm CR_{\rm max}$ is the maximum outlier Critical Ratio\footnote{The Critical Ratio of an outlier is defined as $CR=\frac{n_\mathrm{HM}-\mu_\mathrm{HM}}{\sigma_\mathrm{HM}}$, where $n_\mathrm{HM}$ is the outlier number count in the map where it is selected and $\mu_\mathrm{HM},~\sigma_\mathrm{HM}$ are, respectively, the average and the standard deviation of the number count computed over the same map.} (CR) for each 1~Hz band. The coefficient 4.71 has been obtained taking the parameter $\theta_\mathrm{thr}=2.5$ (the values of $p_0,~p_1$ directly depend on it), while the term 1.6449 comes from taking the confidence level parameter $\Gamma=0.95$. For each 1 Hz band, the final upper limit is the worse among those computed separately for Hanford and Livingston. 
Such upper limits implicitly assume an average over the source population parameters. 

As verified through a detailed comparison based on LIGO and Virgo O2 and O3 data, this procedure produces conservative upper limits with respect to those obtained through the injection of simulated signals, which is computationally more intensive~\cite{bib:O2_bosonclouds_FH}. 
This conservativeness has been verified also through software injections for the first part of the fourth observing run for a set of 40 1~Hz frequency bands, showing that the upper limits estimation given by Eq.~\eqref{eq:hul} overestimates limits obtained by performing software injections across the whole search frequency band, see Appendix \ref{app:ul_fh} for more details.

\subsection{The \so\ search}
\label{sec:soap_methodology}
{\soap} \cite{bib:ViterbiGlasgow} is a fast, model-agnostic search for long duration signals based on the Viterbi algorithm \cite{Viterbi1967}. 
It is intended as both a rapid initial search for isolated NSs, quickly providing candidates for other search methods to 
investigate further, as well as a method to identify long duration signals which may not follow a conventional model of CW frequency evolution.
In its simplest form {\soap} analyzes a spectrogram to find the continuous time-frequency track which gives the highest sum of fast Fourier transform power.
If there is a detectable signal present within the data, then this track is the most likely to correspond to that signal.
The search pipeline consists of three main stages, the initial {\soap} search \cite{bib:ViterbiGlasgow}, the post processing step using convolutional neural 
networks \cite{Bayley2020} and a parameter estimation stage \cite{bib:SOAP2}. 

\subsubsection{Data preparation \label{ssect:SOAPm:data}}

The calibrated detector data are used to create a set of H1 and L1 SFTs  with a coherence time of 1800 s. 
The power spectra of these SFTs are then summed over one day, {\it i.e.}, over up to 48 SFTs. 
Assuming that the signal remains within a single bin over the day, this averages out the antenna pattern modulation and increases the SNR in a given frequency bin.
As the frequency of a CW signal increases, the magnitude of the daily Doppler modulation also increases, therefore the assumption that a signal remains in a single frequency bin within one day no longer holds. 
Hence, the analysis is split into 4 separate bands (40-500 Hz, 500-1000 Hz, 1000-1500 Hz, 1500-2000 Hz) where for each band the Doppler modulations are accounted for by taking the sum of the powers in adjacent frequency bins.
For the bands starting at 40, 500, 1000 and 1500 Hz, the sum is taken over every one (no change), two, three and four adjacent bins respectively such that the resulting time-frequency plane has one, two, three or four times the original bin width.
The broad bands are  further split into `sub-bands' of widths 0.1, 0.2, 0.3 and 0.4 Hz wide respective to the four band sizes above. 
These choices ensure that the maximum yearly Doppler shift $\Delta f_{\rm{orb}}^{(\rm{max})}$ is half the sub-band width, where the maximum is given by
\begin{equation}
    \Delta f_{\rm{orb}}^{(\rm{max})} = f\frac{v_{\rm{orb}}^{\rm{(max)}}}{c} \lesssim 10^{-4} f,
\end{equation}
where $v_{\rm{orb}}^{\rm{(max)}}$ is the maximum orbital velocity of the earth relative to the source, $c$ is the speed of light and $f$ is the initial pulsar frequency.
The sub-bands overlap by half of the sub-band width such that any signal should be fully contained within a sub-band.

\subsubsection{Search pipeline \label{ssect:SOAPm:SOAP}}

{\soap} searches through each of the summed and narrow-banded spectrograms described in Sec.~\ref{ssect:SOAPm:data}, returning four main outputs for each sub-band: 
the Viterbi track, the Viterbi statistic, the Viterbi map and a \ac{CNN} statistic. 
The Viterbi statistic from the core pipeline and the \ac{CNN} statistc from the \ac{CNN} followup are used to select candidates for further analysis, as described in Sec.~\ref{ssect:SOAPm:cand}. 
For more details on the search algorithm see Sec.~\ref{ssect:SOAPm}.

\subsubsection{Candidate selection\label{ssect:SOAPm:cand}}

At this stage there is a set of Viterbi statistics and \ac{CNN} statistics for each sub-band that is analysed, from which a set of candidate signals is selected for followup.
Before doing this, any sub-bands that contain known instrumental artifacts are removed from the analysis.
The sub-bands corresponding to the top 1\% of the Viterbi statistics from each of the four analysis bands are then combined with the sub-bands corresponding to the top 1\% of CNN statistics, leaving us with a maximum of 2\% of the sub-bands as candidates.
It is at this point where we begin to reject candidates by manually removing sub-bands that contain clear instrumental artifacts and cross the detection threshold for either the Viterbi or CNN statistic.
There are a number of features we use to reject candidates including: strong detector artifacts that only appear in a single detector's spectrogram, broad ($> 1/5$ sub-band width) long duration signals, individual time-frequency bins that contribute large amounts to the final statistic and very high-power signals in both detectors. 
Any remaining candidates are then passed on for parameter estimation.

\subsection{Hardware injections}
\label{sec:hwinjections}

During the O4 run 18 hardware injections~\cite{bib:hwinjectionpaper,bib:TauWhelan,bib:HWInjectionsMonitoring} were used to simulate
particular CW signals, as part of detector response validation~\cite{bib:HWInjectionsMonitoring}, including long-term
phase fidelity. The injections were imposed via radiation pressure from auxiliary lasers~\cite{bib:hwinjectionpaper}.
For reference, Table~\ref{tab:hwinjections} lists the key source parameters, including injected strain amplitude $h_0$, for the
14 injections relevant to this analysis (labeled Inj0-Inj12 and Inj14), namely those that simulate isolated
neutron stars with nominal frequencies between 20 and 2000 Hz.

The last three columns of Table~\ref{tab:hwinjections} state whether or not each injection in one of the three
pipelines was detected, including survival of outlier follow-up. Entries marked with ``-'' indicate injection parameters
outside the nominal search range for that pipeline. In addition, for the \pf\ pipeline, which can set valid upper
limits even in the presence of a loud signal, there is a column labeled  ``UL sig bin''
giving an injection's obtained \pf\ 95\%\ upper limit in the signal's frequency bin. Ideally,
the ``UL sig bin'' value should exceed the true injected amplitude $h_0$ for at least 95\%\ of the injections.
In this case, that statement holds for all 14 injections. The column labeled ``UL ctrl bins'' shows the average
95\%\ ULs for the six nearest neighboring frequency subbands as a rough guide to the expected value in
the absence of an injection (or true signal).

\begin{table*}[htbp]
\begin{center}
{
\setlength{\tabcolsep}{5pt}
\begin{tabularx}{\textwidth}{cD{.}{.}{6}rD{.}{.}{5}D{.}{.}{5}rrrccc}
\hline\hline
Inj & \multicolumn{1}{c}{Frequency} & \multicolumn{1}{c}{Spindown} & \multicolumn{1}{c}{$\RAJ$} & \multicolumn{1}{c}{$\DECJ$} & \multicolumn{1}{c}{$h_0$} & \multicolumn{1}{c}{P.F. UL} & \multicolumn{1}{c}{P.F. UL} & \multicolumn{3}{c}{Detected?}\\ 
 & \multicolumn{1}{c}{Hz} & \multicolumn{1}{c}{nHz/s} & \multicolumn{1}{c}{degrees} & \multicolumn{1}{c}{degrees} & \multicolumn{1}{c}{true} & \multicolumn{1}{c}{sig bin}  & \multicolumn{1}{c}{ctrl bins} & P.F. & F.H. & SOAP \\
\hline
0 & 265.574500 & $\sci{-4.15}{-3}$ & 71.55193 & -56.21749 & $\sci{4.01}{-26}$ & $\sci{1.2}{-25}$ & $\sci{1.3}{-25}$ & N & N & N \\ 
1 & 848.894997 & $\sci{-3.00}{-1}$ & 37.39385 & -29.45246 & $\sci{2.79}{-25}$ & $\sci{3.0}{-25}$ & $\sci{1.9}{-25}$ & Y & Y & N \\ 
2 & 575.163487 & $\sci{-1.37}{-4}$ & 215.25617 & 3.44399 & $\sci{3.20}{-26}$ & $\sci{1.6}{-25}$ & $\sci{1.6}{-25}$ & N & N & N \\ 
3 & 108.857159 & $\sci{-1.46}{-8}$ & 178.37257 & -33.43660 & $\sci{1.30}{-25}$ & $\sci{1.7}{-25}$ & $\sci{1.5}{-25}$ & Y & Y & N \\ 
4 & 1387.220675 & $\sci{-2.54}{+1}$ & 279.98768 & -12.46662 & $\sci{4.90}{-25}$ & $\sci{2.7}{-25}$ & $\sci{2.8}{-25}$ & Y\footnote{True spin-down value outside of nominal search range. Injection is recovered when search range is extended for this band.} & - & - \\ 
5 & 52.808324 & $\sci{-4.03}{-9}$ & 302.62664 & -83.83914 & $\sci{3.99}{-25}$ & $\sci{5.6}{-25}$ & $\sci{2.7}{-25}$ & Y & Y & Y \\ 
6 & 144.494850 & $\sci{-6.73}{0\mathrm{\;\;\;}}$ & 358.75095 & -65.42262 & $\sci{3.20}{-25}$ & $\sci{3.3}{-25}$ & $\sci{1.5}{-25}$ & Y & N & Y \\ 
7 & 1220.276598 & $\sci{-1.12}{0\mathrm{\;\;\;}}$ & 223.42562 & -20.45063 & $\sci{7.16}{-26}$ & $\sci{2.5}{-25}$ & $\sci{2.5}{-25}$ & N & - & N \\ 
8 & 188.879028 & $\sci{-8.65}{0\mathrm{\;\;\;}}$ & 351.38958 & -33.41852 & $\sci{8.28}{-26}$ & $\sci{1.3}{-25}$ & $\sci{1.3}{-25}$ & N & N & N \\ 
9 & 763.847316 & $\sci{-1.45}{-8}$ & 198.88558 & 75.68959 & $\sci{6.87}{-26}$ & $\sci{1.8}{-25}$ & $\sci{1.8}{-25}$ & N & N & N \\ 
10 & 26.320768 & $\sci{-8.50}{-2}$ & 221.55565 & 42.87730 & $\sci{6.26}{-25}$ & $\sci{1.6}{-24}$ & $\sci{1.1}{-24}$ & Y & Y & Y \\ 
11 & 31.424632 & $\sci{-5.07}{-4}$ & 285.09733 & -58.27209 & $\sci{3.17}{-25}$ & $\sci{5.0}{-25}$ & $\sci{5.8}{-25}$ & Y & N & N \\ 
12 & 36.922850 & $\sci{-6.25}{0\mathrm{\;\;\;}}$ & 331.85267 & -16.97288 & $\sci{2.63}{-25}$ & $\sci{3.7}{-25}$ & $\sci{3.6}{-25}$ & N & N & N \\ 
14 & 1991.092151 & $\sci{-1.00}{-3}$ & 300.80284 & -14.32394 & $\sci{7.61}{-25}$ & $\sci{8.6}{-25}$ & $\sci{6.1}{-25}$ & Y & - & N \\ 
\hline\hline
\end{tabularx}
}

\caption[Parameters of hardware injections]{Key parameters of the hardware-injected simulated isolated-source continuous wave
  signals during the O4 data run (20-2000 Hz).
  The frequencies listed here correspond to the start of the run (GPS $1238166018$), but
  upper limits obtained by the search pipelines are indexed to frequency bands with
  a reference time at the approximate midpoint of the
  O4a epoch ($\sim$4 months later).
  Also shown are \pf\ strain amplitude upper limits in the nominal signal bins and averaged over the six nearest
  control bins. The last three columns state whether or not an injection is detected by the three pipelines, including
  outlier follow-up. The relatively loud injection 6 at $\sim$144.5 Hz is detected by \fh\ as an initial-stage outlier,
  but is correctly vetoed in follow-up because of the overlap of the frequency band it covers with identified instrumental lines. The injection 11 is missed 
  by \fh\ because the outliers it produces do not meet the strict distance criterion used to flag an injected signal as detected, see Appendix \ref{app:ul_fh}.}
\label{tab:hwinjections}
\end{center}
\end{table*}

\section{Astrophysical interpretation -- Methodology}
\label{sec:interpretation}

Results of all-sky CW searches can be used to constrain certain astrophysical scenarios.
In the following subsections, we discuss how constraints can be applied to models of 1) the population of
galactic neutron stars; 2) the millisecond pulsar explanation for the GeV excess; and
3) the potential contributions to dark matter from asteroid-scale primordial black holes.

\subsection{Implications for the neutron star population}
\label{sec:nspop_methodology}
To interpret the astrophysical significance of strain amplitude upper limits, we translate them into constraints on the Galactic abundance of rapidly spinning, highly deformed neutron stars. For a neutron star with ellipticity $\epsilon$, emitting at GW frequency $\fgw$, the horizon distance $r_{H}(\epsilon,\fgw)$ of the search is computed using Eq.~\ref{eqn:hexpected}:

\begin{equation}
    r_{H}(\epsilon,\fgw) =\frac{4\pi^2 G I_{zz}}{c^4}\frac{\fgw^2\epsilon}{ h_{\mathrm{UL}}^{95\%}(\fgw)},
    \label{horizon}
\end{equation}

Any neutron star emitting at frequency $\fgw$ with ellipticity $\geq \, \epsilon$ located within the horizon distance $r_{H}(\epsilon,\fgw)$ is a detectable source. Assuming a spatial distribution of neutron stars allows us to find the mean fraction of the total Galactic population with ellipticity $\geq \, \epsilon$ that lie within the horizon distance $r_H$ (i.e., they are detectable) at frequency $\fgw$, considering a nominal moment of inertia of $I_{zz} = I_0$.  We denote this fraction by $\alpha (\epsilon, \fgw) $, defined as:

\begin{equation}
 \alpha (\epsilon,\fgw)=\int_0^{r_H}\int _0^{\pi}\int_0^{2\pi}\pspat(r,\theta,\phi)\,r^2\, \sin\theta \, dr \, d\theta \, d\phi,
 \label{mean_fraction}
\end{equation}

\noindent where $\pspat(r,\theta,\phi)$ represents the spatial distribution in spherical-polar coordinates centered at Earth. We consider four such distributions: the progenitor model in \cite{Paczynski}, which assumes a distribution for young neutron stars that follow the same spatial distribution as their progenitor stars and three specific models in~\cite{Reed}, where the neutron star population is time-evolved in the galactic potential, accounting for natal kick velocities. The likelihood of a total of $N$ galactic neutron stars with ellipticity $ \geq \, \epsilon$ and frequency $\fgw$, given $n$ detections within the distance reach can be approximated as the following Poisson distribution:
\begin{align}
    p\left( n \middle| N, \alpha(\epsilon, \fgw) \right) &= \frac{(\alpha N)^n \, e^{- \alpha N}}{n!}
\end{align}

As no detections are reported, we assert $n=0$ and use the Bayes' theorem to construct the posterior distribution of $N$ as:

\begin{equation}
      p(N|n=0,\alpha)  \propto \exp(-\alpha N)p(N)
      \label{Ntotalprobability}
\end{equation}
where the prior $p(N)$ is considered to be uniform. By construction, the posterior on $N$ is agnostic to any assumed spin or ellipticity distribution of the Galactic neutron star population. Instead, it provides constraints in the ($\fgw$, $\epsilon$) space denoting the maximum number of sources exceeding ellipticity $\epsilon$ at emission frequency $\fgw$. This prescription follows from \cite{Prabhu_2024}.

\subsection{Implications for the GeV excess}
\label{sec:gev_methodology}
In a similar vein, using the lack of detection of \CW signals in this search, we can place constraints on the millisecond pulsar (MSP) hypothesis for the ``GeV excess'' of gamma radiation detected from near the Galactic Center. The enigmatic excess was first observed over a decade ago by the Fermi-LAT satellite \cite{Fermi-LAT:2013svs,Fermi-LAT:2015sau}, which has led to two hypotheses being put forward to explain its origins: (1) annihilating Weakly Interacting Massive Particle (WIMP) dark matter \cite{Goodenough:2009gk,Hooper:2010mq,Hooper:2011ti,Gordon:2013vta,Daylan:2014rsa,Calore:2014nla,Abazajian:2014fta} and (2) an unresolvable, electromagnetically faint population of MSPs \cite{Abazajian:2010zy,Calore:2014xka,Yuan:2014rca,Petrovic:2014xra,Ye:2022yxt}. If there is a concentration of electromagnetically dim MSPs in the Galactic Center that are invisible to Fermi-LAT, these pulsars could, in principle, still be detected via their \gwh emission \cite{Calore:2018sbp,Miller:2023qph}. While two recent works consider a stochastic \gwh background arising from the Galactic Center  \cite{Calore:2018sbp,Bartel:2024jjj}, we consider a more powerful way of constraining the GeV excess in this paper through individually resolvable MSPs \cite{Miller:2023qph}.  We note that Ref.  \cite{Lei:2025jsu} recently reached a pessimistic conclusion regarding the possibility of seeing MSPs from the galactic center with current-generation detectors; however, we note that this is heavily dependent on the unknown MSP ellipticity distribution function. 

To determine whether our searches could be sensitive to the GeV excess, we first calculate the probability of detecting \gws from MSPs, $P_{\rm GW}$ by assuming frequency and ellipticity distributions $P(\log \fgw),P(\log \epsilon)$ for the unknown MSP population:

\begin{eqnarray}
  P_{\rm GW} & = & \int_{\log f_{\rm min}}^{\log f_{\rm max}} d(\log \fgw) P(\log \fgw) \nonumber \\
  & & \times \int_{\log \epsilon_{\rm UL}(f)}^{\log\epsilon_{\rm max}} d(\log \epsilon) P(\log \epsilon)
\end{eqnarray}
where $f_{\rm min}$ and $f_{\rm max}$ refer to the minimum and maximum frequencies searched (we define a MSP to have a rotational frequency of at least 60 Hz \cite{manchester2004observational}), $\epsilon_{\rm UL}(f)$ is the minimum detectable equatorial deformation in the \CW searches at the center of the galaxy (8 kpc), and $\epsilon_{\rm max}$ is the maximum ellipticity to which a search is sensitive (fixed by max($\fdot$)). In this analysis, we assume a rotational frequency distribution for the unknown MSP that follows that of the known pulsars in the ATNF catalog \cite{bib:ATNFdb}, although there is significant selection bias towards higher frequencies of detected pulsars \cite{Lorimer:2008se,Liu:2023qnf}. In practice, however, the frequency distribution turns out to be an $\mathcal{O}(1)$ factor that is minor with respect to the ellipticity.
We also assume a log10 exponential distribution $P(\log\epsilon) \propto \exp(-\Lambda\log\epsilon )$ for the ellipticities between $10^{-9}$ -- motivated by the potentially minimum allowable ellipticity of a pulsar \cite{bib:WoanEtalMSP} -- and $10^{-5}$ -- motivated by the canonical maximum strain that a neutron star could support \citep{morales2022}, with a power-law of $\Lambda=2.1$. The choice of $\Lambda$ is arbitrary, which reflects our ignorance of the ellipticity distribution of millisecond pulsars, but we study its impact on the constraints later in this section. For $\Lambda=2.1$, we simulate 100 distributions of ellipticities and take the average of the predicted $P_{\rm GW}$, which results in $P_{\rm GW}=5.88\times 10^{-5}$. The averaging allows us to take different realizations of the ellipticity distributions, ensuring robustness against random variations in the distributions.

The other ingredient necessary to put a constraint on the MSP hypothesis for the GeV excess is an estimation of the number of MSPs in the Galactic Center, which is called the gamma ray ``luminosity function''. In this work, we use a standard prescription for the luminosity function~\cite{Hooper:2016rap}:

\begin{align}
    \frac{dN}{dL} \propto P(L)= \frac{\log_{10} e}{\sigma_L \sqrt{2\pi} L}\exp{\left(-\frac{{\log_{10}^2 (L/L_0) }}{2\sigma_L^2}\right)},
    \label{eqn:log-normal}
\end{align}
where $L$ is the luminosity, and $L_0$ and $\sigma_L$ are two free parameters. The choice of  $L_0$ and $\sigma_L$ fixes the number of MSPs in the Galactic Center $N_{\rm MSP}$ \cite{Dinsmore:2021nip}:

\begin{align}
     L_\text{GCE} = N_\text{MSP}\int_{L_\text{min}}^\infty L P(L) dL.
     \label{eqn:lgce}
\end{align}
where $L_\text{GCE}\approx10^{37}$ erg/s (see \cite{Miller:2023qph} for more details) and  $L_{\rm min}$ is the minimum detectable luminosity by Fermi-LAT. Our work assumes that MSPs are emitting both gravitationally and electromagnetically.

In the absence of a detection, if $N_{\rm MSP}P_{\rm GW}> 1$, we would have been able to detect a \gwh signal from a \MSP in the Galactic Center, and can thus rule out that choice of $L_0$ and $\sigma_L$.

\subsection{Implications for light primordial black holes}
\label{sec:pbh_methodology}
 The \CW searches presented here could also be sensitive to GWs emitted by quasi-infinite inspirals of asteroid-mass ultra-compact objects that could be \pbhs \cite{Hawking:1971ei,Chapline:1975ojl,Bird:2016dcv,Clesse:2016vqa,Sasaki:2016jop,Carr:2019kxo}. \CW searches offer a promising avenue to probe the currently-unconstrained asteroid-mass regime for \pbhs \cite{Miller:2020kmv,Miller:2024rca,Miller:2024khl}. In particular, the \fh and \pf methods require that the \gwh signal follows a linear frequency evolution in order to detect it.  Thus, binary systems with chirp masses\footnote{The chirp mass of a binary black hole system with stellar masses $m_1$ and $m_2$ is defined to be $\mathcal{M} \equiv (m_1*m_2)^{3/5}/(m_1+m_2)^{1/5}$.} $\mathcal{M}\lesssim \mathcal{O}(10^{-5})M_\odot$  would emit \CW signals that would mimic those that arise from spinning up, non-axisymmetric rotating neutron stars \cite{Miller:2021knj,Andres-Carcasona:2023zny,Andres-Carcasona:2024jvz,Alestas:2024ubs}.
The upper limits derived in Fig.~\ref{fig:combined_upper_limits} can therefore be mapped to a constraint on the maximum distance away that we could have detected an inspiraling binary via the relation:

\begin{align}
h_0&=\frac{4}{d}\left(\frac{G \mathcal{M}}{c^2}\right)^{5/3}\left(\frac{\pi f_{\rm gw}}{c}\right)^{2/3}~ \label{eqn:hamp}\\
&\simeq1.61\times 10^{-26}\! \left[\frac{d}{10 \text{ pc}}\right]^{-1}\! \left[\frac{\mathcal{M}}{10^{-6}M_\odot}\right]^{5/3}\! \left[\frac{f_{\rm gw}}{50 \text{ Hz}}\right]^{2/3}, \nonumber
\end{align}
We can then compute a rate density $\mathcal{R}$ associated with the distance reach $d$, as described in \cite{Miller:2021knj}, as a function of chirp mass. This rate density is independent of the existence and formation mechanisms of \pbhs, and can be interpreted as a formation rate density, as opposed to a merger rate density that is frequently constrained in other \gwh searches \cite{LIGOScientific:2025slb}.

After obtaining a rate density, we must make the assumption that these rates would arise from the formation of \pbhs. It is important to note that such constraints on \pbhs is highly model-dependent. Using the rate density prescriptions for early two-body binaries, the dominant channel of \pbh formation relevant for this work, we can write~\cite{Clesse:2015wea,Clesse:2016vqa,Raidal:2018bbj,Hutsi:2020sol}:

 \begin{eqnarray}
   \mathcal{R} &=& 5.28 \times 10^{-7}\, \mathrm{kpc}^{-3} \mathrm{yr}^{-1} \nonumber \\
   & & \times \left(\frac{m_1}{M_\odot}\right)^{-32/37} \left(\frac{m_2}{m_1}\right)^{-34/37}\ftilde^{53/37}~,
\label{eqn:rate_asymm}
\end{eqnarray}
where $m_2<m_1$

\begin{equation}
\ftilde(\mone,\mtwo) \equiv \fpbh \left[\fsup\fmone\fmtwo\right]^{37/53}
\label{eqn:ftilde}
\end{equation}
is a parameter that relates to the fraction $\fpbh$ of dark matter that \pbhs could compose and the mass distribution functions $f(m)$ for $m_1$ and $m_2$. $f_{\rm sup}$ is the suppression factor that accounts for {various mechanisms affecting the binary orbital evolution throughout cosmic time~\cite{Raidal:2018bbj},  which could change their merger time or destroy them. For no suppression, $f_{\rm sup}=1$. We note that  $f(m_1)$, $f(m_2)$ and $f_{\rm sup}$ depend on the specific \pbh formation model chosen; thus we decide to simply constrain $\ftilde$, which parameterizes our ignorance of these parameters, but allows interested readers to set constraints on $\fpbh$ for their chosen model.

\section{Search results}
\label{sec:search_results}

None of the three searches detected a credible CW signal. In the following subsections, we provide more
details on these negative results.

\subsection{The \pf\ results}
\label{sec:powerflux_results}



\newcounter{BandAcountzero} \newcounter{BandBcountzero}\newcounter{BandCcountzero} \newcounter{BandDcountzero}\newcounter{BandEcountzero}
\newcounter{BandFcountzero} \newcounter{BandGcountzero}\newcounter{BandHcountzero} \newcounter{BandIcountzero}\newcounter{BandJcountzero}
\newcounter{BandKcountzero} \newcounter{BandLcountzero}\newcounter{BandMcountzero} \newcounter{BandNcountzero}\newcounter{BandOcountzero}
\newcounter{BandPcountzero} \newcounter{BandQcountzero}\newcounter{BandRcountzero} \newcounter{BandScountzero}\newcounter{BandTcountzero}
\newcounter{VlowBandcountzero} \newcounter{LowBandcountzero} \newcounter{MediumBandcountzero} \newcounter{HighBandcountzero}

\newcounter{BandAcountone} \newcounter{BandBcountone}\newcounter{BandCcountone} \newcounter{BandDcountone}\newcounter{BandEcountone}
\newcounter{BandFcountone} \newcounter{BandGcountone}\newcounter{BandHcountone} \newcounter{BandIcountone}\newcounter{BandJcountone}
\newcounter{BandKcountone} \newcounter{BandLcountone}\newcounter{BandMcountone} \newcounter{BandNcountone}\newcounter{BandOcountone}
\newcounter{BandPcountone} \newcounter{BandQcountone}\newcounter{BandRcountone} \newcounter{BandScountone}\newcounter{BandTcountone}
\newcounter{VlowBandcountone} \newcounter{LowBandcountone} \newcounter{MediumBandcountone} \newcounter{HighBandcountone}

\newcounter{BandAcounttwo} \newcounter{BandBcounttwo}\newcounter{BandCcounttwo} \newcounter{BandDcounttwo}\newcounter{BandEcounttwo}
\newcounter{BandFcounttwo} \newcounter{BandGcounttwo}\newcounter{BandHcounttwo} \newcounter{BandIcounttwo}\newcounter{BandJcounttwo}
\newcounter{BandKcounttwo} \newcounter{BandLcounttwo}\newcounter{BandMcounttwo} \newcounter{BandNcounttwo}\newcounter{BandOcounttwo}
\newcounter{BandPcounttwo} \newcounter{BandQcounttwo}\newcounter{BandRcounttwo} \newcounter{BandScounttwo}\newcounter{BandTcounttwo}
\newcounter{VlowBandcounttwo} \newcounter{LowBandcounttwo} \newcounter{MediumBandcounttwo} \newcounter{HighBandcounttwo}

\newcounter{BandAcountcluster} \newcounter{BandBcountcluster}\newcounter{BandCcountcluster} \newcounter{BandDcountcluster}\newcounter{BandEcountcluster}
\newcounter{BandFcountcluster} \newcounter{BandGcountcluster}\newcounter{BandHcountcluster} \newcounter{BandIcountcluster}\newcounter{BandJcountcluster}
\newcounter{BandKcountcluster} \newcounter{BandLcountcluster}\newcounter{BandMcountcluster} \newcounter{BandNcountcluster}\newcounter{BandOcountcluster}
\newcounter{BandPcountcluster} \newcounter{BandQcountcluster}\newcounter{BandRcountcluster} \newcounter{BandScountcluster}\newcounter{BandTcountcluster}
\newcounter{VlowBandcountcluster} \newcounter{LowBandcountcluster} \newcounter{MediumBandcountcluster} \newcounter{HighBandcountcluster}

\newcounter{VlowBandcountHWinj} \newcounter{LowBandcountHWinj} \newcounter{MediumBandcountHWinj} \newcounter{HighBandcountHWinj}
\newcounter{VlowBandcountStrainhist} \newcounter{LowBandcountStrainhist} \newcounter{MediumBandcountStrainhist} \newcounter{HighBandcountStrainhist}
\newcounter{VlowBandcountPyFStat} \newcounter{LowBandcountPyFStat} \newcounter{MediumBandcountPyFStat} \newcounter{HighBandcountPyFStat}


\setcounter{BandAcountzero}{4916}   
\setcounter{BandBcountzero}{1501}   
\setcounter{BandCcountzero}{1047}   
\setcounter{BandDcountzero}{2916}   
\setcounter{BandEcountzero}{1856}   

\setcounter{BandFcountzero}{1860}   
\setcounter{BandGcountzero}{1961}   
\setcounter{BandHcountzero}{6817}   
\setcounter{BandIcountzero}{3876}   
\setcounter{BandJcountzero}{4961}   
\setcounter{BandKcountzero}{1917}   
\setcounter{BandLcountzero}{1961}   
\setcounter{BandMcountzero}{6581}   
\setcounter{BandNcountzero}{1916}   
\setcounter{BandOcountzero}{1557}   

\setcounter{BandPcountzero}{617}   
\setcounter{BandQcountzero}{4964}   
\setcounter{BandRcountzero}{1966}   
\setcounter{BandScountzero}{195}   
\setcounter{BandTcountzero}{9176}   


\setcounter{BandAcountone}{4267}   

\setcounter{BandBcountone}{1916}   
\setcounter{BandCcountone}{2961}   
\setcounter{BandDcountone}{4357}   
\setcounter{BandEcountone}{3861}   

\setcounter{BandFcountone}{1985}   
\setcounter{BandGcountone}{1051}   
\setcounter{BandHcountone}{5920}   
\setcounter{BandIcountone}{2917}   
\setcounter{BandJcountone}{5916}   
\setcounter{BandKcountone}{1847}   
\setcounter{BandLcountone}{1582}   
\setcounter{BandMcountone}{7731}   
\setcounter{BandNcountone}{2961}   
\setcounter{BandOcountone}{556}   

\setcounter{BandPcountone}{196}   
\setcounter{BandQcountone}{1972}   
\setcounter{BandRcountone}{1368}   
\setcounter{BandScountone}{256}   
\setcounter{BandTcountone}{3061}   


\setcounter{BandAcounttwo}{5816}   

\setcounter{BandBcounttwo}{1301}   
\setcounter{BandCcounttwo}{810}   
\setcounter{BandDcounttwo}{280}   
\setcounter{BandEcounttwo}{603}   

\setcounter{BandFcounttwo}{192}   
\setcounter{BandGcounttwo}{326}   
\setcounter{BandHcounttwo}{184}   
\setcounter{BandIcounttwo}{172}   
\setcounter{BandJcounttwo}{591}   
\setcounter{BandKcounttwo}{185}   
\setcounter{BandLcounttwo}{107}   
\setcounter{BandMcounttwo}{472}   
\setcounter{BandNcounttwo}{936}   
\setcounter{BandOcounttwo}{185}   

\setcounter{BandPcounttwo}{20}   
\setcounter{BandQcounttwo}{16}   
\setcounter{BandRcounttwo}{193}   
\setcounter{BandScounttwo}{54}   
\setcounter{BandTcounttwo}{3716}   


\setcounter{BandAcountcluster}{666}   

\setcounter{BandBcountcluster}{161}   
\setcounter{BandCcountcluster}{8}   
\setcounter{BandDcountcluster}{0}   
\setcounter{BandEcountcluster}{0}   

\setcounter{BandFcountcluster}{343}   
\setcounter{BandGcountcluster}{326}   
\setcounter{BandHcountcluster}{0}   
\setcounter{BandIcountcluster}{2524}   
\setcounter{BandJcountcluster}{56}   
\setcounter{BandKcountcluster}{3}   
\setcounter{BandLcountcluster}{0}   
\setcounter{BandMcountcluster}{0}   
\setcounter{BandNcountcluster}{0}   
\setcounter{BandOcountcluster}{0}   

\setcounter{BandPcountcluster}{0}   
\setcounter{BandQcountcluster}{33}   
\setcounter{BandRcountcluster}{0}   
\setcounter{BandScountcluster}{5}   
\setcounter{BandTcountcluster}{153}   


\setcounter{VlowBandcountzero}{\value{BandAcountzero}}

\setcounter{LowBandcountzero}{\value{BandBcountzero}}
\addtocounter{LowBandcountzero}{\value{BandCcountzero}} \addtocounter{LowBandcountzero}{\value{BandDcountzero}}
\addtocounter{LowBandcountzero}{\value{BandEcountzero}}

\setcounter{MediumBandcountzero}{\value{BandFcountzero}}
\addtocounter{MediumBandcountzero}{\value{BandGcountzero}} \addtocounter{MediumBandcountzero}{\value{BandHcountzero}}
\addtocounter{MediumBandcountzero}{\value{BandIcountzero}} \addtocounter{MediumBandcountzero}{\value{BandJcountzero}}
\addtocounter{MediumBandcountzero}{\value{BandKcountzero}} \addtocounter{MediumBandcountzero}{\value{BandLcountzero}}
\addtocounter{MediumBandcountzero}{\value{BandMcountzero}} \addtocounter{MediumBandcountzero}{\value{BandNcountzero}}
\addtocounter{MediumBandcountzero}{\value{BandOcountzero}}

\setcounter{HighBandcountzero}{\value{BandPcountzero}}
\addtocounter{HighBandcountzero}{\value{BandQcountzero}} \addtocounter{HighBandcountzero}{\value{BandRcountzero}}
\addtocounter{HighBandcountzero}{\value{BandScountzero}} \addtocounter{HighBandcountzero}{\value{BandTcountzero}}


\setcounter{VlowBandcountone}{\value{BandAcountone}}

\setcounter{LowBandcountone}{\value{BandBcountone}}
\addtocounter{LowBandcountone}{\value{BandCcountone}} \addtocounter{LowBandcountone}{\value{BandDcountone}}
\addtocounter{LowBandcountone}{\value{BandEcountone}}

\setcounter{MediumBandcountone}{\value{BandFcountone}}
\addtocounter{MediumBandcountone}{\value{BandGcountone}} \addtocounter{MediumBandcountone}{\value{BandHcountone}}
\addtocounter{MediumBandcountone}{\value{BandIcountone}} \addtocounter{MediumBandcountone}{\value{BandJcountone}}
\addtocounter{MediumBandcountone}{\value{BandKcountone}} \addtocounter{MediumBandcountone}{\value{BandLcountone}}
\addtocounter{MediumBandcountone}{\value{BandMcountone}} \addtocounter{MediumBandcountone}{\value{BandNcountone}}
\addtocounter{MediumBandcountone}{\value{BandOcountone}}

\setcounter{HighBandcountone}{\value{BandPcountone}}
\addtocounter{HighBandcountone}{\value{BandQcountone}} \addtocounter{HighBandcountone}{\value{BandRcountone}}
\addtocounter{HighBandcountone}{\value{BandScountone}} \addtocounter{HighBandcountone}{\value{BandTcountone}}


\setcounter{VlowBandcounttwo}{\value{BandAcounttwo}}

\setcounter{LowBandcounttwo}{\value{BandBcounttwo}}
\addtocounter{LowBandcounttwo}{\value{BandCcounttwo}} \addtocounter{LowBandcounttwo}{\value{BandDcounttwo}}
\addtocounter{LowBandcounttwo}{\value{BandEcounttwo}}

\setcounter{MediumBandcounttwo}{\value{BandFcounttwo}}
\addtocounter{MediumBandcounttwo}{\value{BandGcounttwo}} \addtocounter{MediumBandcounttwo}{\value{BandHcounttwo}}
\addtocounter{MediumBandcounttwo}{\value{BandIcounttwo}} \addtocounter{MediumBandcounttwo}{\value{BandJcounttwo}}
\addtocounter{MediumBandcounttwo}{\value{BandKcounttwo}} \addtocounter{MediumBandcounttwo}{\value{BandLcounttwo}}
\addtocounter{MediumBandcounttwo}{\value{BandMcounttwo}} \addtocounter{MediumBandcounttwo}{\value{BandNcounttwo}}
\addtocounter{MediumBandcounttwo}{\value{BandOcounttwo}}

\setcounter{HighBandcounttwo}{\value{BandPcounttwo}}
\addtocounter{HighBandcounttwo}{\value{BandQcounttwo}} \addtocounter{HighBandcounttwo}{\value{BandRcounttwo}}
\addtocounter{HighBandcounttwo}{\value{BandScounttwo}} \addtocounter{HighBandcounttwo}{\value{BandTcounttwo}}


\setcounter{VlowBandcountcluster}{\value{BandAcountcluster}}

\setcounter{LowBandcountcluster}{\value{BandBcountcluster}}
\addtocounter{LowBandcountcluster}{\value{BandCcountcluster}} \addtocounter{LowBandcountcluster}{\value{BandDcountcluster}}
\addtocounter{LowBandcountcluster}{\value{BandEcountcluster}}

\setcounter{MediumBandcountcluster}{\value{BandFcountcluster}}
\addtocounter{MediumBandcountcluster}{\value{BandGcountcluster}} \addtocounter{MediumBandcountcluster}{\value{BandHcountcluster}}
\addtocounter{MediumBandcountcluster}{\value{BandIcountcluster}} \addtocounter{MediumBandcountcluster}{\value{BandJcountcluster}}
\addtocounter{MediumBandcountcluster}{\value{BandKcountcluster}} \addtocounter{MediumBandcountcluster}{\value{BandLcountcluster}}
\addtocounter{MediumBandcountcluster}{\value{BandMcountcluster}} \addtocounter{MediumBandcountcluster}{\value{BandNcountcluster}}
\addtocounter{MediumBandcountcluster}{\value{BandOcountcluster}}

\setcounter{HighBandcountcluster}{\value{BandPcountcluster}}
\addtocounter{HighBandcountcluster}{\value{BandQcountcluster}} \addtocounter{HighBandcountcluster}{\value{BandRcountcluster}}
\addtocounter{HighBandcountcluster}{\value{BandScountcluster}} \addtocounter{HighBandcountcluster}{\value{BandTcountcluster}}


\setcounter{VlowBandcountHWinj}{452}
\setcounter{LowBandcountHWinj}{20}
\setcounter{MediumBandcountHWinj}{2513}
\setcounter{HighBandcountHWinj}{29}

\setcounter{VlowBandcountStrainhist}{145}
\setcounter{LowBandcountStrainhist}{142}
\setcounter{MediumBandcountStrainhist}{706}
\setcounter{HighBandcountStrainhist}{149}

\setcounter{VlowBandcountPyFStat}{69}
\setcounter{LowBandcountPyFStat}{7}
\setcounter{MediumBandcountPyFStat}{33}
\setcounter{HighBandcountPyFStat}{13}


\def\VlowBandcountzero{{\color{black}\arabic{VlowBandcountzero}}}
\def\LowBandcountzero{{\color{black}\arabic{LowBandcountzero}}}
\def\MediumBandcountzero{{\color{black}\arabic{MediumBandcountzero}}}
\def\HighBandcountzero{{\color{black}\arabic{HighBandcountzero}}}

\def\VlowBandcountone{{\color{black}\arabic{VlowBandcountone}}}
\def\LowBandcountone{{\color{black}\arabic{LowBandcountone}}}
\def\MediumBandcountone{{\color{black}\arabic{MediumBandcountone}}}
\def\HighBandcountone{{\color{black}\arabic{HighBandcountone}}}

\def\VlowBandcounttwo{{\color{black}\arabic{VlowBandcounttwo}}}
\def\LowBandcounttwo{{\color{black}\arabic{LowBandcounttwo}}}
\def\MediumBandcounttwo{{\color{black}\arabic{MediumBandcounttwo}}}
\def\HighBandcounttwo{{\color{black}\arabic{HighBandcounttwo}}}

\def\VlowBandcountcluster{{\color{black}\arabic{VlowBandcountcluster}}}
\def\LowBandcountcluster{{\color{black}\arabic{LowBandcountcluster}}}
\def\MediumBandcountcluster{{\color{black}\arabic{MediumBandcountcluster}}}
\def\HighBandcountcluster{{\color{black}\arabic{HighBandcountcluster}}}

\def\VlowBandcountHWinj{{\color{black}\arabic{VlowBandcountHWinj}}}
\def\LowBandcountHWinj{{\color{black}\arabic{LowBandcountHWinj}}}
\def\MediumBandcountHWinj{{\color{black}\arabic{MediumBandcountHWinj}}}
\def\HighBandcountHWinj{{\color{black}\arabic{HighBandcountHWinj}}}

\def\VlowBandcountStrainhist{{\color{black}\arabic{VlowBandcountStrainhist}}}
\def\LowBandcountStrainhist{{\color{black}\arabic{LowBandcountStrainhist}}}
\def\MediumBandcountStrainhist{{\color{black}\arabic{MediumBandcountStrainhist}}}
\def\HighBandcountStrainhist{{\color{black}\arabic{HighBandcountStrainhist}}}

\def\VlowBandcountPyFStat{{\color{black}\arabic{VlowBandcountPyFStat}}}
\def\LowBandcountPyFStat{{\color{black}\arabic{LowBandcountPyFStat}}}
\def\MediumBandcountPyFStat{{\color{black}\arabic{MediumBandcountPyFStat}}}
\def\HighBandcountPyFStat{{\color{black}\arabic{HighBandcountPyFStat}}}

Carrying out the \pf\ stage-0 analysis described above leads to a set of all-sky upper limits
(95\%\ CL) on strain amplitude for worst-case  (unfavorable stellar spin orientation),
linear polarization and for best-case, circular polarization. That analysis also leads to an
initial set of outliers for follow-up with later analysis stages.
Whether or not a particular narrow frequency band contains an outlier,
the upper limits obtained remain valid.
As shown in Table~\ref{tab:hwinjections}, hardware injections were reliably recovered when
their injected strain was at least 0.6 times the recorded upper limit.
Figure~\ref{fig:powerflux_upper_limits} in Appendix~\ref{sec:powerflux_appendix} presents strict \pf\ upper limits
on signals that are circularly polarized or linearly polarized, along with \pf\ upper limits from
searches in the O3 data set~\cite{bib:cwallskyO3a,bib:TripatheeRilesO3}.
The circular polarization limits
are then used to estimate population-averaged upper limits, as described in Appendix~\ref{sec:powerflux_appendix}.

Outliers seen in the stage-0 search are followed up with loose-coherence stages of
increasing effective coherence time, as described in section~\ref{sec:powerflux_outlier_followup},
with the SNR expected to increase for true signals. Table~\ref{tab:powerflux_outlier_counts}
shows the counts of outliers seen at each stage for major sub-bands.
In the presence of Gaussian noise and no signal, one expects the outlier counts to
decrease monotonically with increasing stage as SNR increases are required for each advancement.
For particular sub-bands, however, one can see count increases from at least two contributions,
both associated with the finer sampling of parameter space in successive stages.
1) Hardware injections (section~\ref{sec:hwinjections}) from
a simulated signal naturally satisfy the SNR increase requirements demanded of a signal;
and 2) stationary line artifacts of finite bandwidth can be compatible with
signal templates having limited Doppler modulation or having partial cancellation
between seasonal modulation and assumed frequency derivative
for a certain region of the sky~\cite{bib:cwallskyS4}.

In the extreme, the number of outliers produced by a particular instrumental artifact
can be so large as to make systematic follow-up impracticable and pointless.
Artifacts include loud mechanical resonances, such as higher harmonics of violin modes, and especially
loud hardware injections. The widest bands excluded lie in the regions of test-mass violin modes and their higher harmonics. Details concerning these vetoed bands can be found in
Appendix~\ref{sec:powerflux_vetoedbands_app}.

\def\mc#1{\multicolumn{1}{c}{#1}}
\begin{table*}[htb]
\begin{center}
  \begin{tabular}{lrrrr}\hline\hline
    \T\B                                       &    \mc{20-60  Hz}        &  \mc{60-475 Hz}         & \mc{475-1475 Hz}           & \mc{1475-2000 Hz} \\
    \hline
    \T\B Stage 0                                     &  \VlowBandcountzero &  \LowBandcountzero  & \MediumBandcountzero   & \HighBandcountzero \\
    \T\B Stage 1                                     &  \VlowBandcountone  &  \LowBandcountone   & \MediumBandcountone    & \HighBandcountone  \\
    \T\B Stage 2                                     &  \VlowBandcounttwo  &  \LowBandcounttwo   & \MediumBandcounttwo    & \HighBandcounttwo   \\
    \T\B Outlier clusters                             &  \VlowBandcountcluster  &  \LowBandcountcluster   & \MediumBandcountcluster    & \HighBandcountcluster   \\
    \hline
    \T\B HW injections                               &  \VlowBandcountHWinj  &  \LowBandcountHWinj   & \MediumBandcountHWinj    & \HighBandcountHWinj   \\
    \T\B Visible artifacts                           &  \VlowBandcountStrainhist  &  \LowBandcountStrainhist   & \MediumBandcountStrainhist    & \HighBandcountStrainhist   \\
    \T\B \pyfstat\ follow-up                           &  \VlowBandcountPyFStat  &  \LowBandcountPyFStat   & \MediumBandcountPyFStat    & \HighBandcountPyFStat   \\
    \hline
    \T\B \pyfstat\ survivors                           &  0 & 0 & 0 & 0 \\
    \hline\hline
  \end{tabular}
  \caption{Counts of \pf\ outliers surviving different stages of the hierarchical
    follow-up for four frequency bands.
    Survivors from stage-2 follow-up are broken down into the categories of hardware injections,
    visible artifacts identified via strain histograms and candidates for \pyfstat\ follow-up.
    Additional follow-up of these weak candidates, which appear in a visibly disturbed band of L1 data,
    was carried out using the MCMC \pyfstat\ follow-up procedure described in section~\ref{sec:powerflux_outlier_followup}.
    The \pyfstat\ follow-up confirmed these 122 outliers to be uninteresting.}
\label{tab:powerflux_outlier_counts}
\end{center}
\end{table*}

Nearly all outliers that survived all stages of the loose-coherence
follow-up correspond to hardware injections (see Table~\ref{tab:hwinjections})
or lie in highly disturbed bands, for which contamination of the putative signal by an instrumental spectral line is
apparent. To identify these contaminations, we construct ``strain histograms''~\cite{bib:cwallskyO3a} in which the summed power over the
observation period from a simulation of the nominal
signal candidate is superposed on a background estimate of the noise estimated via interpolation between neighboring
frequency bands.
Except for signal templates with high-magnitude spin-downs, the histograms typically display at least one ``horn''
(narrow peak) from
an epoch during the 8-month O4a period when the orbitally modulated frequency is relatively stationary. We discard
outliers for which the signal template's shape aligns with a spectral artifact known to be instrumental or appearing loudly
in one detector but not the other. Before visual inspection, outliers are clustered in frequency, spin-down and sky location.
For reference, Table~\ref{tab:histogramvetoes} lists the parameters of
the single loudest outlier in each cluster discarded after inspecting strain histograms.

\begin{table}[htb]
  \begin{center}
    \def\mc#1{\multicolumn{1}{c}{#1}}
\begin{tabular}{rrrr}\hline\hline
\mc{$f$} & \mc{$df/dt$} & \mc{R.A.} & \mc{Dec.} \\ 
\mc{(Hz)} & \mc{(nHz/s)} & \mc{(radians)} & \mc{(radians)} \\ 
\hline
  31.4965 & -0.525 & 1.790 &  0.051 \\
  62.4994 & -0.060 & 4.614 &  1.125 \\
  77.7271 &  0.150 & 3.584 &  1.150 \\
  78.2360 & -0.900 & 4.483 &  0.005 \\
  83.3327 & -0.140 & 4.555 &  1.094 \\
 497.3866 &  0.283 & 4.029 &  1.277 \\
 497.7838 & -6.900 & 2.920 & -1.511 \\
 499.9568 &  1.150 & 1.548 &  1.169 \\
 517.7407 & -8.433 & 4.773 &  0.029 \\
 523.1429 & -1.900 & 1.376 & -1.345 \\
 530.6302 & -3.333 & 3.513 &  0.839 \\
 599.9255 & -9.667 & 1.826 &  0.845 \\
 944.9449 & -0.150 & 4.513 &  1.133 \\
1600.1460 & -0.985 & 6.283 &  0.008 \\
1930.2163 & -9.640 & 3.849 & -1.456 \\
1980.7634 & -5.880 & 1.702 & -1.049 \\
\hline\hline
\end{tabular}

    \caption{Parameters of the \pf\ outliers surviving stage-2 follow-up, but discarded after visual inspection of a strain histogram confirmed instrumental
      contamination. Each outlier listed is the loudest after clustering in frequency, spin-down and sky location.}
\label{tab:histogramvetoes}
\end{center}
\end{table}

Any outliers surviving manual inspection are subjected
to the MCMC \pyfstat\ follow-up procedure described in section~\ref{sec:powerflux_outlier_followup}.
For signal strengths near the upper limit sensitivity, we would expect a resulting final Bayes factor $B_{\rm SN}$
to lie well above 50 (confirmed for detected hardware injections).
To be conservative, however, we subject any outliers surviving the criterion $B_{\rm SN} > 20$
(indicated in Table~\ref{tab:powerflux_followups}) to further 
follow-up using additional data from the O4 run. We carry out a \pyfstat\ follow-up using both O4a and O4b data on
survivors including separate H1-only and L1-only \pyfstat\ follow-ups. No non-injection outliers
survive this scrutiny, failing either to increase in significance with more data or revealing
complete dominance of the combined SNR by one detector's data.
We verified that hardware injections in the vicinity of instrumental artifacts survived
all of the criteria used in outlier follow-up. 

We conclude that there is no significant evidence for a continuous wave signal from the \pf\ search.

\begin{table*}[htb]
  \begin{center}
    \def\mc#1{\multicolumn{1}{c}{#1}}
\def\mrp#1{\multicolumn{1}{r}{#1\>}}
\def\mrq#1{\multicolumn{1}{r}{#1\>\>\>\>}}
\def\mlp#1{\multicolumn{1}{l}{\>#1}}
\begin{tabular}{rrrr@{\hskip0.3in}rrrr@{\hskip0.5in}rrrr}\hline\hline
\mrq{$f$} & \mc{$df/dt$} & \mc{R.A.} & \mlp{Dec.} & \mrq{$f$} & \mc{$df/dt$} & \mc{R.A.} & \mlp{Dec.} & \mrq{$f$} & \mc{$df/dt$} & \mc{R.A.} & \mlp{Dec.} \\
\mrp{(Hz)} & \mc{(nHz/s)} & \mc{(rad)} & \mlp{(rad)} & \mrp{(Hz)} & \mc{(nHz/s)} & \mc{(rad)} & \mlp{(rad)} & \mrp{(Hz)} & \mc{(nHz/s)} & \mc{(rad)} & \mlp{(rad)} \\
\hline
21.4265 &  -1.390 & 5.628 & -0.979
& 
*22.2226 &  -0.015 & 4.595 & 1.171
& 
(inj)\>\>\>\> 26.3189 &  -0.315 & 3.231 & 1.020
\\
*26.4996 &  0.040 & 5.281 & 1.247
& 
  27.2539 & -6.615 & 3.992 &  0.504 & 
  27.8156 & -8.725 & 4.041 &  0.248 \\
  28.3764 & -4.725 & 1.809 & -0.715 & 
  28.4245 & -5.250 & 4.763 & -0.687 & 
(inj)\>\>\>\> 31.4251 &  -0.160 & 4.583 & -0.742
\\
32.5014 &  -0.525 & 3.500 & -0.392
& 
  33.3337 & -0.175 & 2.323 & -1.319 & 
  33.5023 & -0.315 & 3.864 & -0.025 \\
  34.5024 & 0.315 & 4.743 &  0.571 & 
36.5018 & 0.535 & 5.406 & 0.366
& 
  *37.8785 & -0.415 & 1.551 & -0.424 \\
  37.9467 & -5.015 & 6.025 &  0.858 & 
  38.5034 & 0.565 & 5.048 &  0.117 & 
  39.5028 & 0.425 & 4.902 &  0.459 \\
  40.8167 & -6.915 & 0.702 & -0.218 & 
  40.8792 & -8.035 & 0.275 & -0.122 & 
*41.5014 & 0.515 & 5.435 & 0.585
\\
  *42.8571 & -0.015 & 1.533 & -1.142 & 
  44.5346 & -0.025 & 3.172 & -0.995 & 
  45.2384 & -0.015 & 0.747 & -0.156 \\
  49.9938 & -1.115 & 1.957 &  0.085 & 
  51.5020 & 0.935 & 5.926 & -0.340 & 
(inj)\>\>\>\> 52.8083 & 0.000 & 5.287 & -1.465
\\
  53.6688 & -0.035 & 3.684 &  0.916 & 
  54.1702 & -0.015 & 0.867 & -1.211 & 
  54.8307 & -2.135 & 2.424 & -0.883 \\
76.5053 & -1.790 & 4.611 & -0.835
& 
94.2399 & -1.240 & 3.506 & -0.950
& 
(inj)\>\> 108.8568 &  -1.100 & 3.111 & -0.494
\\
(inj)\>\>\> 144.4957 &  -6.720 & 6.260 & -1.137
& 
182.3303 &  -5.610 & 0.270 & -0.824
& 
 211.9994 & -6.140 & 1.360 &  0.799 \\
491.5714 &  -1.900 & 6.170 & 1.087
& 
499.8757 & 0.283 & 2.200 & -1.123
& 
506.5055 &  -4.183 & 3.109 & 1.147
\\
515.5096 &  -9.433 & 4.544 & -0.867
& 
517.8990 &  -3.483 & 1.187 & -1.266
& 
536.6181 &  -9.000 & 2.771 & 0.703
\\
 598.9800 & -7.000 & 2.138 &  1.308 & 
 800.0543 & -0.100 & 5.434 & -1.203 & 
 829.1247 &  0.383 & 5.058 &  0.903 \\
 842.1306 & -8.850 & 3.251 &  0.995 & 
(inj)\> 848.8950 &  -0.283 & 0.651 & -0.512
& 
 *961.9981 & -0.433 & 4.723 &  1.182 \\
1027.4462 & -1.233 & 0.008 & -0.001 & 
1610.2238 & -7.385 & 6.278 & -0.001 & 
1861.2849 & -0.315 & 1.738 &  1.056 \\
1979.0183 & -7.420 & 0.140 & -1.209 & 
1991.0059 & 0.250 & 5.278 & -0.376 
& 
(inj) 1991.0922 & 0.000 & 5.250 & -0.251 \\
\hline\hline
\end{tabular}

    \caption{Parameters of the \pf\ outliers surviving stage-2 follow-up and followed up using \pyfstat.
      Each outlier listed is the loudest after clustering in frequency, spin-down and sky location.  Outliers caused by hardware injections are indicated by ``(inj)''. Survivors of requiring the final Bayes factor  $B_{\rm SN} > 20$, which were
      subjected to additional \pyfstat\ follow-up (O4a+O4b data and H-only, L1-only),
      are indicated by ``*''.}
\label{tab:powerflux_followups}
\end{center}
\end{table*}

\subsection{The \fh\ results}
\label{sec:frequencyhough_results}
The \fh\ search covers the frequency range [10, 1024] Hz, a spin-down range between $-$$10^{-8}$ Hz/s to $2\times10^{-9}$ Hz/s and the whole sky. 
The frequency and spin-down resolutions are given by $\delta\!f = 1/T_{\rm FFT}$ and $\delta\!\dot{f} = \delta\!f/T_{\rm obs}$. The sky resolution, on the other hand, is a function of the frequency and of the sky position and is defined in such a way that for two nearby sky cells the maximum frequency variation, due to the Doppler effect, is within one frequency bin, see~\cite{bib:freqhough2} for more details.

\subsubsection{Follow-up\label{sec:fhufollowup}}
Outliers produced by the \fh\ search are followed up with the procedure described in Sec.~\ref{sec:fh_fu}.
Although none of the outliers survived the entire chain of vetoes, we summarize the output of the search.

The O4a \fh\ follow-up focused only on those outliers with $CR\geq5$ (computed in the \fh\ map) in both detectors due to limitations on the available computing power.
We separate the dataset into two subgroups, one consisting of outliers interpreted in terms of the gravitar model (see Appendix.~\ref{app:fh_space}) and the other made up of outliers inconsistent with the gravitar model.
These sets are identified comparing the first order spin-down of each outlier to the value given by Eq.~\eqref{eq:selection_gravitar_cands}, calculated with the coherence time of the first follow-up stage.
The former set, of $\mathcal{O}(6\times10^5)$ elements, is made up of outliers falling within the orange curve in Fig.~\ref{fig:overlap_FH_par_space_gravitar_condition-label}, while the latter, of $\mathcal{O}(3\times10^5)$ elements, of those outside of it.
Both sets are subjected to the same first follow-up stage as described in Sec.~\ref{sec:fh_fu}.

Immediately below we detail the results for the first set, while
at the end of the section, we summarize the results for the second batch of outliers and the differences between the approaches.

After the first follow-up stage, we apply the veto chain described in~\ref{app:fh_fu_details} (see also \cite{bib:cwallskyO3FourPipelines}).
Table~\ref{tab:surv_cands_O4a_first_stage} summarises the number of outliers discarded by each of the vetoes.
\begin{table}[!ht]
    \centering
    \renewcommand*{\arraystretch}{1.1}
    \begin{tabular}{cccccccc}
    \hline\hline
        Starting & V1 & V2 & V3 & V4 & V5\\
        \hline
        583590 & 109358 & 1555 & 1481 & 798 & 453\\
        \% discarded & 81\% & 98\% & 5\% & 46\% & 43\% \\
        \hline\hline
    \end{tabular}
    \caption{\fh\ outliers surviving at the first follow-up stage after each selection cut summarized in Sec.~\ref{app:fh_fu_details}; refer to~\cite{bib:cwallskyO3FourPipelines} for the vetoes' labels. Percentages are normalised to the number of surviving outliers of the previous veto.}
    \label{tab:surv_cands_O4a_first_stage}
\end{table}
The surviving outliers are further analysed with an increased coherence time to check their compatibility with an astrophysical signal.
This second follow-up stage is carried out with \pyfstat{}.
We explicitly consider $\Ddot{f}_0$ only for those outliers for which the expected second-order spin-down should be larger than the bin dimension in the second stage.
The MCMC follow-up is run with priors described in Sec.~\ref{app:fh_fu_details} and initial $\Tcoh^{(0)}=0.5$~days.
Following the discussion in~\cite{Mirasola:2024lcq}, we evaluate the $\pfa$ for each outlier by running $500$ off-sourced MCMCs.
Those outliers with $\pfa<1\%$ are carefully checked to discard outliers related to detector non-stationarities (see Sec.~\ref{app:fh_fu_details}).

Only 18 outliers survive these tests and are further analysed in a third stage where the coherence time is obtained through~\cite{Mirasola:2024lcq}
\begin{equation}
    \Tcoh =10^{3/D} \, \Tcoh^{(0)}\,.
    \label{eq:ladder_approx_O4a}
\end{equation}
where $D = 5$ (4) for a (non)resolved $\ddot{f}_0$.
The resulting coherence-time sequence is less steep than what is suggested in~\cite{Mirasola:2024lcq}, and reflects a more conservative approach where the priority is given to convergence rather than computational savings.

Not only do we compare the loudest template (i.e., template with the highest $\F$) from each followed up outlier with the noise distribution, but we also check the compatibility of the stages' results by means of Eq.~(11) of~\cite{bib:cwallskyO3FourPipelines}.
Out of the 18 outliers analysed in the third stage, 9 survive all the tests and are related to the HIs 1, 3, 5, 10.
Table~\ref{tab:fh_HI-reconstruction} collects the dimensionless distance of the closest outlier to each HI.

\begin{table}[!ht]
    \centering
    \renewcommand*{\arraystretch}{1.4}
    \begin{tabular}{crrcrrr}
        \hline\hline
         Inj & $\Delta f_0/\delta f$ &  $\Delta\dot{f}_0/\delta\dot{f}$ &  $\Delta\ddot{f}_0/\delta\ddot{f}$ &  $\Delta\lambda/\delta\lambda$ &  $\Delta\beta/\delta\beta$ & $d_{\rm FH}$\\
         \hline
         10 & $-\!$1.7$\cdot10^{-2}$ &	3$\cdot10^{-1}$ &	0 &	3$\cdot10^{-4}$ &	2$\cdot10^{-3}$ & 0.3\\
         5 & $-\!$9$\cdot10^{-3}$ &	$-\!$4$\cdot10^{-2}$ &	0 &	$-\!$1.5$\cdot10^{-4}$ &	1.9$\cdot10^{-4}$ &  0.05 \\
         3 & 4$\cdot10^{-2}$ &	$\!-$4$\cdot10^{-1}$ &	0 &	2$\cdot10^{-4}$ &	$-\!$5$\cdot10^{-4}$ &0.4 \\
         1 & $-\!$6$\cdot10^{-3}$ &	1.01$\cdot10^{-1}$ &	0 &	3$\cdot10^{-6}$ &	$-\!$1.8$\cdot10^{-5}$ & 0.1\\
         \hline\hline
    \end{tabular}
    \caption{Hardware injection recovery by the \fh pipeline. For each HI, we report the dimensionless distances of the closest outlier that survived all the follow-up stages. The values are calculated as $(x_{\rm cand} - x_{\rm HI})/\delta x$, with $x$ the CW's parameters and $\delta x$ the resolution calculated with the coherence time of the third stage (see Eq.~\ref{eq:ladder_approx_O4a}).}
    \label{tab:fh_HI-reconstruction}
\end{table}

Even if not shown, the other five HI-related outliers reported a distance from the injection well below the follow-up resolution and show perfect agreement between the two MCMC stages.

We have followed up outliers from the second set (i.e., those not interpreted in terms of the gravitar model) following an almost identical procedure.
Only 125 outliers survive the entire chain of vetoes out of the $\mathcal{O}(3\times10^5)$ starting ones.
We follow them up with \pyfstat{} now without introducing any $\ddot{f}$ contribution at any stage.
After the second stage, only 5 survived the $\pfa= 1\%$ threshold calculated as described above.
After the additional checks (see Sec.~\ref{app:fh_fu_details}), only one outlier has been brought to the third stage.
The result of this last stage is compatible with the noise distribution; therefore, we ruled out this last outlier as most likely not astrophysical.

Since all the non-HI-related outliers have been discarded, we stopped our follow-up at the third stage and set upper limits as described in Sec.~\ref{sec:fhulresults}.

\subsubsection{Upper limits\label{sec:fhulresults}}
Since none of the outliers survived the {\it follow-up} stage, we proceeded to set a 95\% confidence level on $h_{0\rm{,min}}$, according to the Eq.~\ref{eq:hul}. 
This estimation of the upper limits leads to conservative results, validated through the injection of simulated signals in real data. 
The upper limits from software injections were computed for a set of 40 1 Hz frequency bands, that are \{20--45, 50, 70, 95, 150, 200, 250, 325, 400, 432, 500, 555, 600, 750, 909\}~Hz. 
These choices were made in order to heterogeneously investigate all the regions of the real search, with an eye toward the low-frequency region, contaminated heavily by instrumental spectral lines.
We were not able to set upper limits through software injections in three specific low-frequency 1 Hz bands, \{27, 35, 40\}~Hz, because of intense instrumental disturbances.
In all the other cases, the upper limit estimations via software injections were lower than the ones obtained with Eq.~\eqref{eq:hul}.
We show the results in Fig.~\ref{fig:fh_ul}. 
The method used to establish the 95\% upper limit from injections is described in Appendix~\ref{app:ul_fh}, as well as the results from all the software injections and the comparison with the theoretical estimation.

\subsection{The \so\ results}
\label{sec:soap_results}
\soap \, was run on the O4a dataset from 20-2000 Hz where we are sensitive to a broad range of signals from the entire sky. To contain an entire signal within a single sub-band, its spin-down must be within about $[1.3, 2.6, 5.2, 10]\times10^{-9}$ Hz/s up to [500, 1000, 1500, 2000] Hz, respectively, therefore when values are outside this range we lose sensitivity.

We start from a set of 1800s long \acp{FFT} of cleaned time-series data from the two LIGO detectors H1 and L1. 
As described in Sec.~\ref{ssect:SOAPm:data}, the \acp{FFT} are normalised to the running median of width 100 bins before being split into 0.1 (0.2, 0.3, 0.4) Hz wide sub-bands overlapping by half of their width.
For each of the sub-bands, time segments and frequency bins are summed together, where along the time axis, 48 \acp{FFT} (1 day) are summed along the frequency axis, and every 1 (2, 3, 4) frequency bins are summed respective to the analysis band. 
\soap \, is then run on each of these sub-bands, returning the Viterbi statistic, Viterbi map and Viterbi tracks, which can be input to the \ac{CNN} to return a second statistic.
The number of sub-bands searched totals to 20 231 across all four analysis bands.
Candidates are then selected by taking the sub-bands that contribute to the top 2\% of both the remaining Viterbi and \ac{CNN} statistics. 
These candidates can then be investigated further to identify whether a real GW signal is present.
Sub-bands that contain an instrumental line identified by the calibration group but also cross the 2\% threshold are also investigated to check whether it is the instrumental line which causes the high statistic value.
There were 690 sub-bands in this category, with 128 of these being common to both the Viterbi statistic and the \ac{CNN} statistic. After an initial investigation 7 sub-bands remained that were not clearly associated with instrumental artifacts.

\subsubsection{Outliers}

The remaining 7 candidates were then investigated further by analysing the outputs of the Viterbi search, i.e., the Viterbi maps, Viterbi tracks and Viterbi statistics, alongside the \ac{CNN} statistic and the spectrograms from each detector.
Plots of each of these allow the identification of features that are not astrophysical but originate from the instrument or environment. 
The spectrograms from both detectors summed over time and frequency, as described in Sec.~\ref{ssect:SOAPm:data}, along with the optimal Viterbi track, allow us to identify what features within the data contribute towards the final statistic. 
For example, many of the spectrograms contain spectral features that are far above the noise level and appear in only a single detector, but still cross the detection threshold for one of the statistics. 
These sub-bands can be visually inspected, and if found to contain a non-astrophysical artifact that contributes to the statistic, be removed from the analysis.
Of the sub-bands that were investigated further, 4 were removed because of the presence of an instrumental artifact showing up in a list of known instrumental lines.
The remaining sub-bands contain hardware injections (see section~\ref{sec:hwinjections}).

\subsubsection{Hardware injections}

Of the 14 hardware injections listed in Table~\ref{tab:hwinjections}, 13 fall within our search parameter space (as for other search pipelines, injection 4 has a spin-down magnitude outside the nominal SOAP range). Of these 13 injections, three cross our detection threshold without being excluded (injections 5, 6 and 10). These signals appear in multiple sub-bands due to the 50\% overlapping sub-bands, therefore the sub-band containing a larger fraction of the signal is used for any followup. The remaining injections that did not cross the threshold had SNRs that were below or close to our expected sensitivity for isolated neutron stars, therefore not expected to cross our detection threshold. 

\subsubsection{Sensitivity}
\label{sec:soap_sensitivity}
The sensitivity of \soap \, can be tested by running the search on a set of CW signals injected into real O4a data. 
A total of $2.59 \times 10^5$ signals were injected across each of the four frequency bands described in Sec.~\ref{ssect:SOAPm:data}, where the signals have Doppler parameters which are drawn uniformly on the sky, uniformly within the respective frequency range and uniformly in the range [$-10^{-9}, \; -10^{-16}$] Hz s$^{-1}$ for the frequency derivative. 
The other amplitude parameters varied in the same ranges as described in Sec.~\ref{ssect:SOAPm:CNN}.
A false alarm value of 1\% can be set for each of the odd and even data-sets within the four analysis bands by taking the corresponding statistic value at which 1\% of the noise only bands exceed.
Both the Viterbi and \ac{CNN} statistics are calculated separately for each of the odd and even bands.
Each of the bands containing injected signals can then be classified as detected or not depending on if a statistic crossed its respective false alarm value.
The efficiency curves can then be found by computing the fraction of detected signals as a function of $h_0$.
Each of the four analysis bands are further split into bands of width 40 Hz, where for each of these bands a detection efficiency curve is generated. 
The false alarm values for each band are set based on which of the four larger analysis bands that it falls within.
Our false alarm values are then contaminated by the strongest artifacts within each ~500 Hz wide analysis band, meaning that this is a conservative estimate of our sensitivity.
Values of $h_0$ for each frequency band can then be selected where the detection efficiency reaches 95\%, defining our sensitivity shown in Fig.~\ref{fig:combined_upper_limits}.

\subsection{Summary of constraints on signal strain amplitudes, ellipticities and ranges}
\label{sec:constraints}

Figure~\ref{fig:combined_upper_limits} shows the strain amplitude
results obtained from all three search pipelines.
The O4a \pf\ curve is a population-averaged 95\%\ confidence level (CL) upper limit estimate derived from
strict circular-polarization upper limits shown below in
Fig.~\ref{fig:powerflux_upper_limits}. The O4a Frequency Hough curve
shows the conservative (over-estimated) population-averaged upper limits from
Fig.~\ref{fig:fh_ul} below. The O4a \so\ curve shows the 
estimated sensitivities described in section~\ref{sec:soap_results}.

Given differences in technical implementation between \pf\ and \fh, each
one provides a somewhat complementary backup for the other templated search,
reducing the chances of missing a detectable signal because of configuration
tuning or the handling of spectral line contamination. The untemplated \so\
program, while less sensitive than the templated pipelines to signals that
follow the assumed model, can potentially detect signals that deviate enough from
the model to escape detection by \pf\ and \fh.

Also shown in Fig.~\ref{fig:combined_upper_limits} are 90\%\ CL upper limits from 
\eh\ searches of the O3 data, searches which leverage the
computing capacity of the distributed-computing network to carry out
long-coherence-time semi-coherent searches~\cite{bib:cwallskyO3EatH,bib:cwallskyO3EatHbucket,bib:cwallskyO3EatHHighFreq}.
It is difficult to compare
90\%\ CL limits with 95\%\ limits precisely, but a rough estimate
based on typical detection efficiency curves at a fixed strain amplitude,
suggest that the 95\%\ CL limits presented here are $\sim$10\%\ or more
 higher than would be the corresponding 90\%\ CL limits. Adjusting the
estimated \pf\ upper limits downward correspondlingly would make them comparable to the \eh\ limits above 250 Hz
for that region of spindown parameter space shared by the \pf\ and \eh\ searches. 

\begin{figure*}[htbp]
  \begin{center}
    \ifshowfigs
    \includegraphics[width=6.0in]{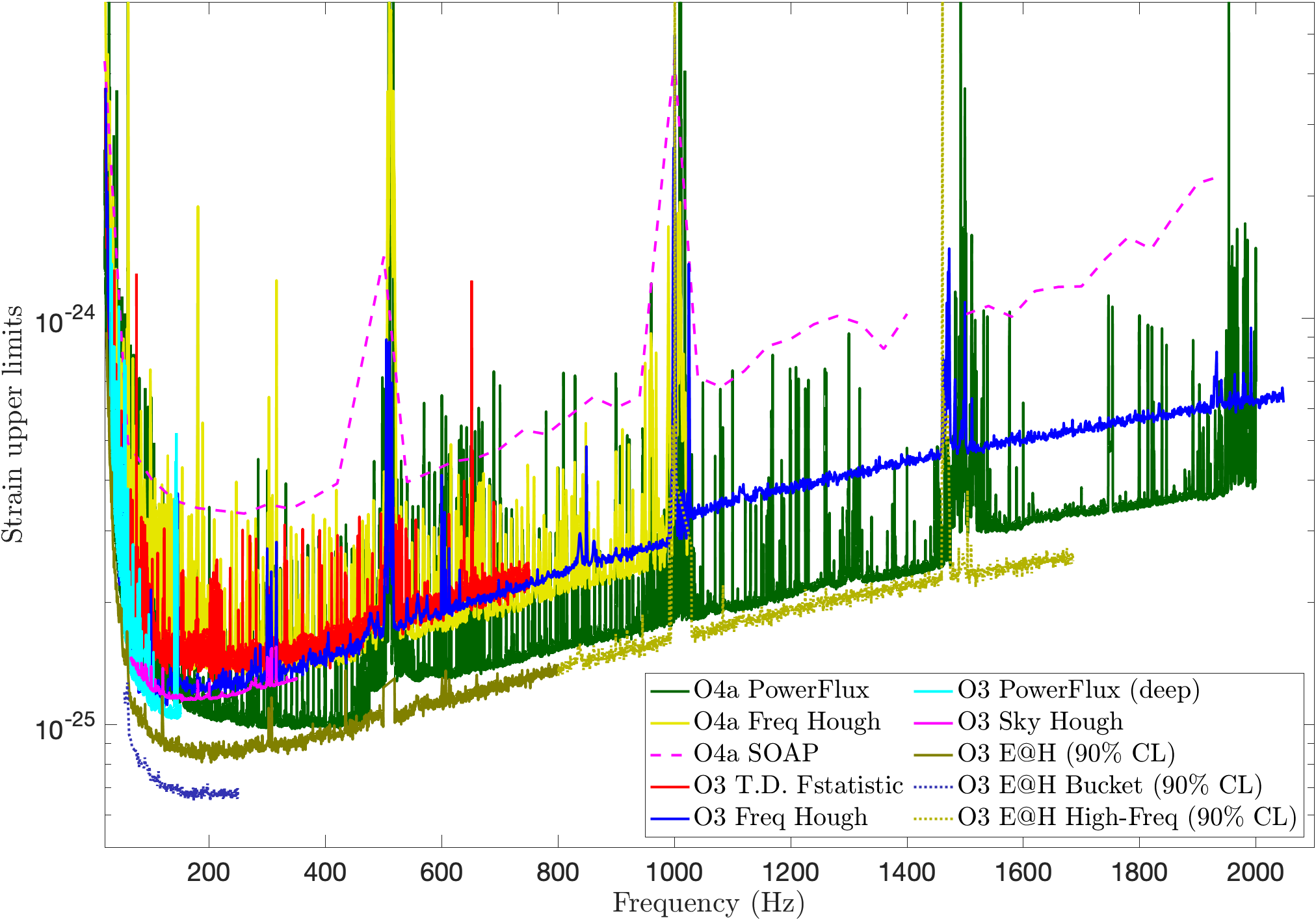}
    \includegraphics[width=6.0in]{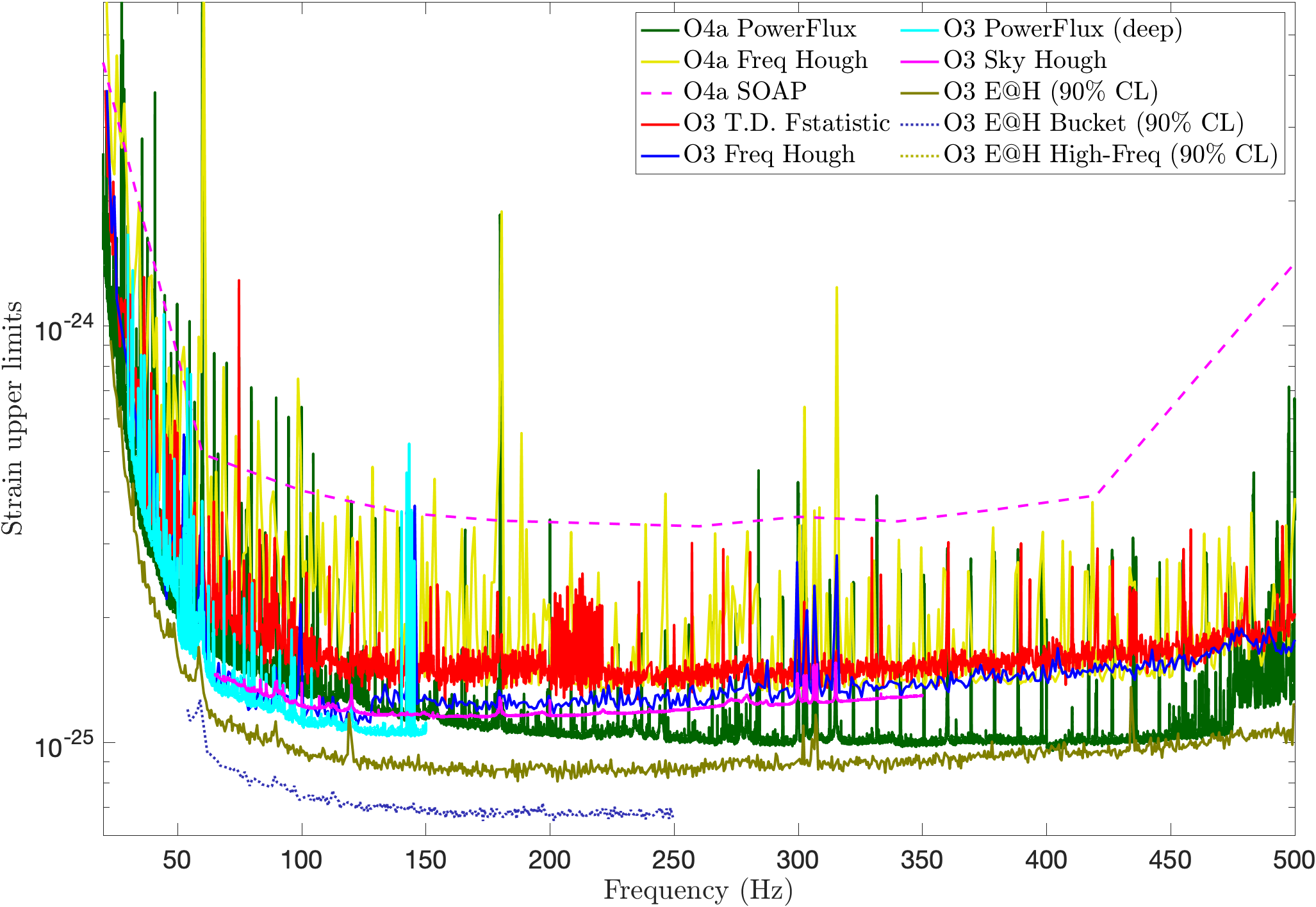}
    \fi
    \caption{{\it Top panel:} Population-averaged 95\%\ confidence level (CL) estimated upper limits on gravitational strain amplitude
      for the three searches presented here: \pf\ (dark green), \fh\ (yellow) and \so\ (magenta dashed).
      The \pf\ limits are estimates based on strict circular-polarization upper limits
      (see section~\ref{sec:powerflux_upperlimits}).
      The \fh\ and \so\ estimated limits are somewhat conservative (see sections~\ref{sec:fhulresults}
      and \ref{sec:soap_sensitivity}).
      Also shown are results from previous searches in the full LIGO O3 data set:
      Time-domain \fstat~\cite{bib:cwallskyO3FourPipelines} (red);
      \fh~\cite{bib:cwallskyO3FourPipelines} (blue);
      \sh~\cite{bib:cwallskyO3FourPipelines} (magenta);
      \pf\ with loose coherence~\cite{bib:TripatheeRilesO3} (cyan);
      and \eh~\cite{bib:cwallskyO3EatH,bib:cwallskyO3EatHbucket,bib:cwallskyO3EatHHighFreq} (dark olive, violet, light olive,
      90\%\ CL, covering $\sim$26\% of the spindown range of the other templated searches).
      {\it Bottom panel:} Magnification in the frequency range 20-500 Hz to show finer detail. (color online)}
    \label{fig:combined_upper_limits}
  \end{center}
\end{figure*}

Figure~\ref{fig:parameterspace} shows, however, that the spin-down (and spin-up) range searched here is nearly four times
greater than in the O3 \eh\ searches. In the ``gravitar'' model in which all stellar rotational energy loss is attributed
to gravitational wave radiation, searching higher spin-down magnitudes allows probing of stars with higher ellipticities,
and more important, allows probing farther reaches of the Milky Way since more distant neutron stars must have higher
ellipticies in order to be detectable and hence must have higher spin-down magnitudes. In addition, searching for higher spin-down magnitudes
allows for a more realistic model,
in which spin-down has contributions other than from GW radiation, most notably from magnetic dipole radiation.
Another consideration in comparing sensitivities is that the long coherence times (60-244 hours) used in the \eh\ searches
impose greater demands on the fidelity of a signal to the model used. The shorter coherence times used by \pf\ and \fh\
permit more deviation from modeling over both short and long time scales. The \so\ pipeline is still more robust than any
of these templated approaches.

Figure~\ref{fig:ellipticity} translates the estimated limits on strain amplitude into ellipticity sensitivities for different
assumed source ranges (kpc). At higher frequencies and ellipticities, however, the maximum allowed spin-down magnitude comes into play,
as shown by the smooth dotted and dashed curves depicting maximum detectable ellipticity \vs\ frequency for
the $|\fdot|$ maxima used here and in the \eh\ searches.
Figure~\ref{fig:range} translates the results into range sensitivity \vs\ frequency for different assumed ellipticities,
taking into account the maximum $\fdot$ magnitude
by truncating curves of fixed ellipticity at the appropriate frequencies. Also shown for reference are ellipticity-independent
maximum ranges for different maximum $\fdot$ magnitudes.

\begin{figure}[htbp]
  \ifshowfigs
  \includegraphics[width=3.25in]{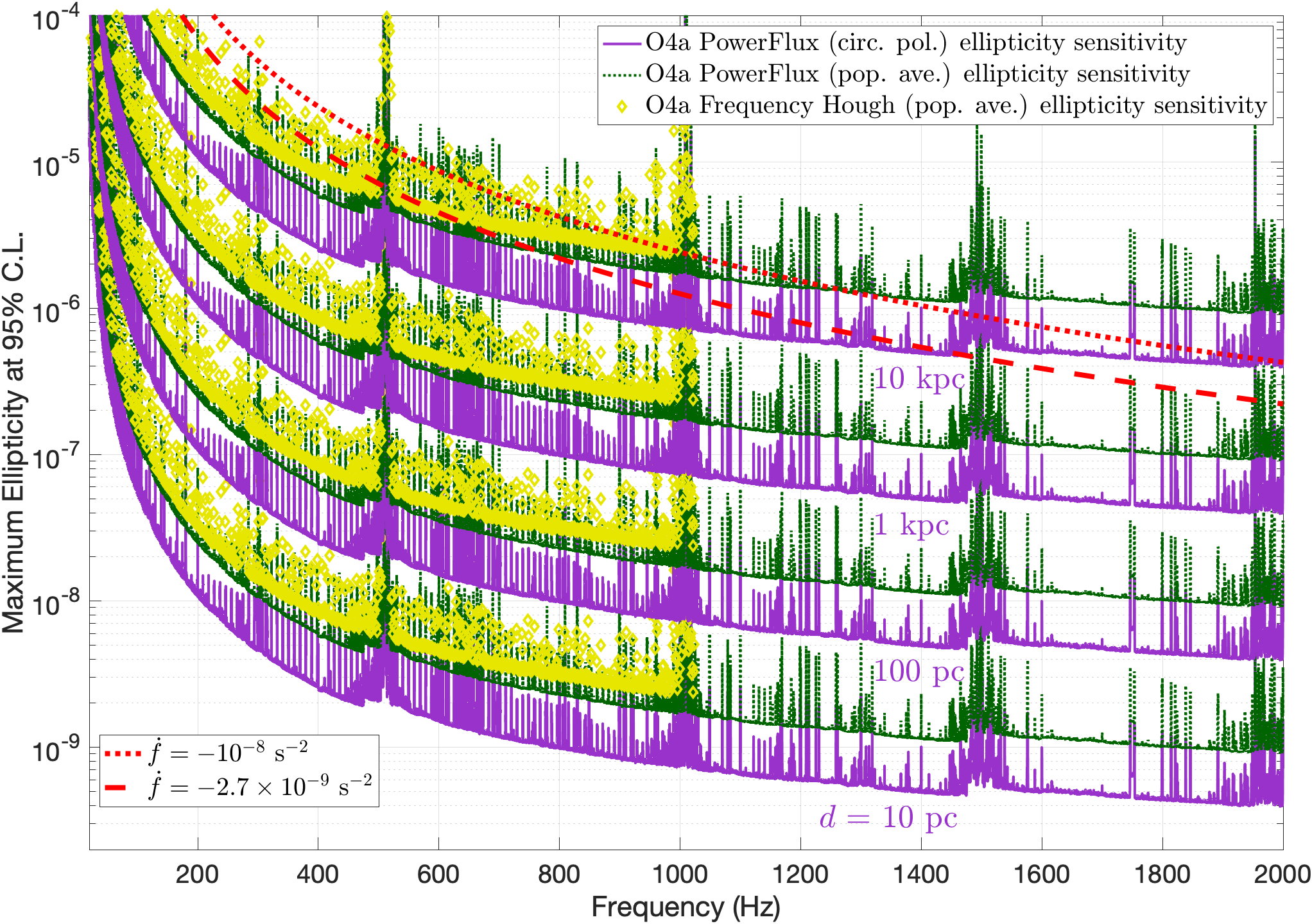}
  \fi
  \caption[Ellipticity sensitivity]{
    Ellipticity sensitivities of the \pf\ and \fh\ searches for neutron stars
    spinning down solely due to gravitational radiation (``gravitars'') under different assumed distances.
    Each family of three curves (orchid, dark green dotted, yellow diamonds) show the sensitivities for
    the O4a \pf\ circular-polarization,
    population-averaged and \fh\ population-averaged sensititivies that apply for four
    different assumed distances of $d$ = 10 pc, 100 pc, 1 kpc and 10 kpc.
    The smooth red dotted curve shows, however, the maximum detectable ellipticity for the maximum allowed
    spindown magnitude of 10$^{-8}$ Hz/s used here, while the lower, dashed red curve shows the corresponding
    maximum detectable ellipticity for the maximum allowed spindown magnitude of 2.7$\times$10$^{-9}$ Hz/s used by
    the O3 Einstein@Home search~\cite{bib:cwallskyO3EatH,bib:cwallskyO3EatHbucket,bib:cwallskyO3EatHHighFreq}.
    Where the search sensitivity curves for optimum stellar orientation (circular-polarization, shown for \pf)
    lie above the $\fdot$ boundary,
    one cannot actually detect the corresponding ellipticity for that range. See Fig.~\ref{fig:range}
    for graphs of range sensitivity for different assumed ellipticities, taking into account the $\fdot$ constraint.
    (color online)}
  \label{fig:ellipticity}
\end{figure}

\begin{figure}[htbp]
  \ifshowfigs
  \includegraphics[width=3.25in]{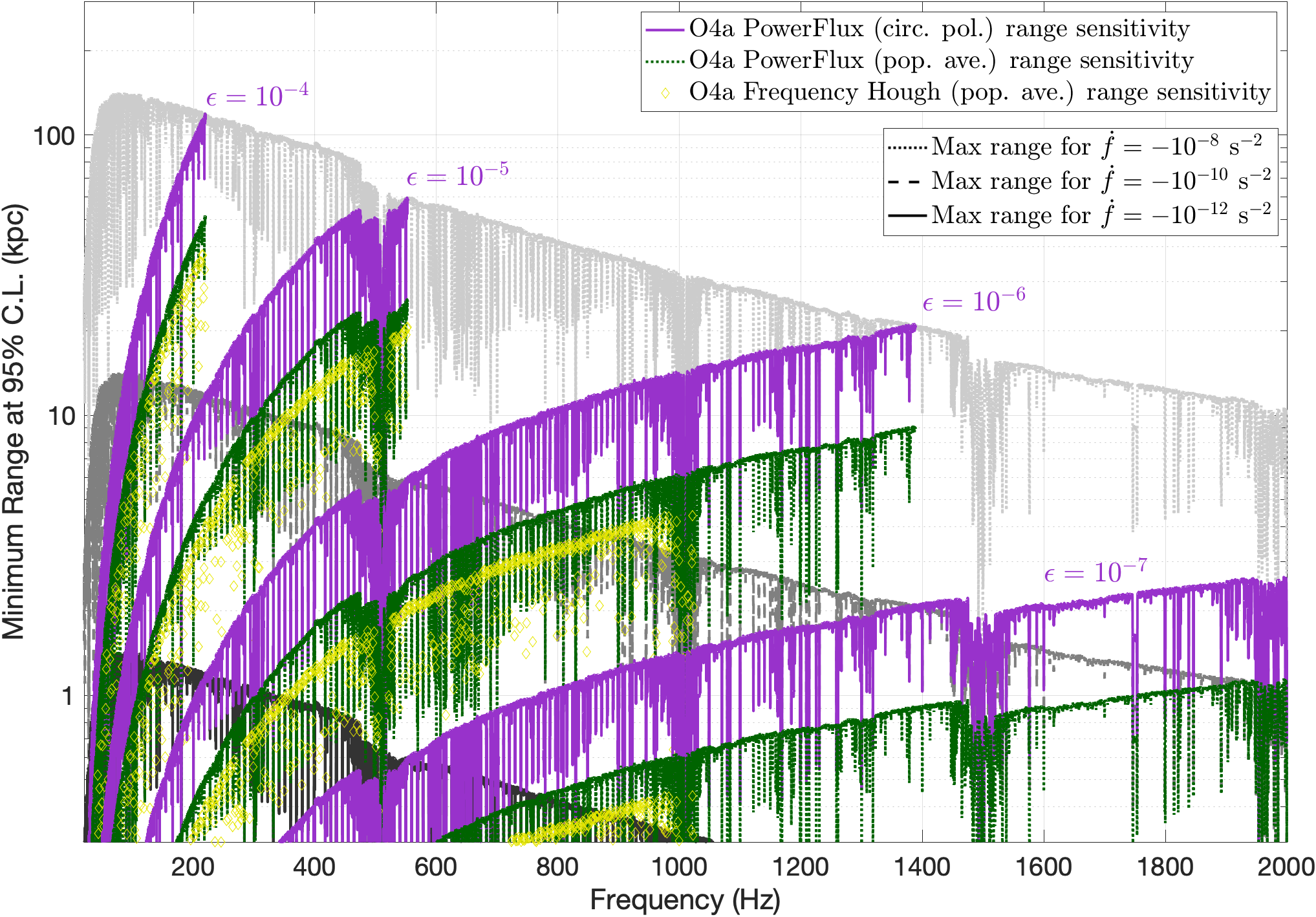}
  \fi
  \caption[Search range]{
    Ranges (kpc) of the \pf\ and \fh\ searches for the gravitar model under different assumptions.
    The four sets of three curves (orchid, dark green dotted, yellow diamonds)
    that generally rise with frequency are the ranges for which the O4a \pf\ circular-polarization,
    population-averaged and \fh\ population-averaged sensititivies apply for four
    different assumed equatorial ellipticities of $\epsilon$ = $10^{-4}$, $10^{-5}$, $10^{-6}$ and $10^{-7}$.
    In each case, the curve stops at the frequency at which the implied spindown magnitude
    at the corresponding range exceeds the maximum range allowed by the maximum spindown magnitude ($10^{-8}$ Hz/s) used here.
    The three gray curves (light dotted, medium dashed, dark solid) that peak below 200 Hz represent the maximum possible ranges
    for which the O4a \pf\ circular-polarization limits apply under the assumption that the
    maximum spindown magnitude is ($10^{-8}$, $10^{-10}$, $10^{-12}$) Hz/s.
    Searching for high spindown magnitudes simultaneously probes higher
    ellipticies and larger regions of the galaxy, reaching beyond the galactic center ($\sim$ 8.5 kpc)
    for $\epsilon = 10^{-6}$ over the full search band. More realistically, a neutron star's spin-down may be dominated
    by electromagnetic radiation, in which case searching for high spindown magnitude achieves a shorter search
    range, but enables finding stars that would not be detectable under the gravitar assumption.
    (color online)}
  \label{fig:range}
\end{figure}

\section{Astrophysical interpretation -- Results}
\label{sec:interpretation_results}

In the following subsections, we discuss the astrophysical implications of these upper limits
in Fig.~\ref{fig:combined_upper_limits}.

\subsection{Implications for the neutron star population}
\label{sec:nspop_results}
Using Eqn.~\ref{Ntotalprobability} for the galactic neutron star population, we report the most conservative $90\%$ quantile on the number constraint from all the considered distributions on a two-dimensional grid of ellipticity and GW frequency in Fig.~[\ref{CO_number_constraints}]. These estimates imply a severe lack of highly deformed, rapidly rotating neutron stars, with ellipticities $\epsilon \gtrsim 10^{-6}\,$ and frequencies $\, \fgw \gtrsim 200$ Hz,  having abundance upper limits at $90\%$ confidence of $N^{90\%} \lesssim 10^6$. Given that the total number of neutron stars in our Galaxy is estimated to be $10^8$--$10^9$ \cite{Rozwadowska2021NewA, Sartore2010A&A}, this strongly disfavors the majority of them to be both rapidly spinning and significantly deformed. Beyond neutron stars, which can support representative ellipticities of about $\epsilon \sim 7.4\times10^{-6}$ (\cite{morales2022}), certain exotic compact objects—like neutron stars with quark‐matter cores (hybrid stars) or pure quark stars—may support exceptionally high ellipticities (\cite{haskell2006mountains}, \cite{haskell2007}, \cite{glampedakis2018}, \cite{nathan2013}). Our results imply that rapidly rotating compact objects with these ellipticities $\epsilon \gtrsim 10^{-5},$ and frequencies $\fgw \gtrsim 200$ Hz have abundance upper limits at $90\%$ confidence of $N^{90\%} \lesssim 10^2$. Such tight constraints on abundance at high ellipticities indicate that exotic compact objects in our galaxy with such theoretically permissible high ellipticities are, at best, exceedingly rare, if they exist at all.

\begin{figure}[htbp]
  \begin{center}
    \ifshowfigs
    \includegraphics[width=\linewidth]{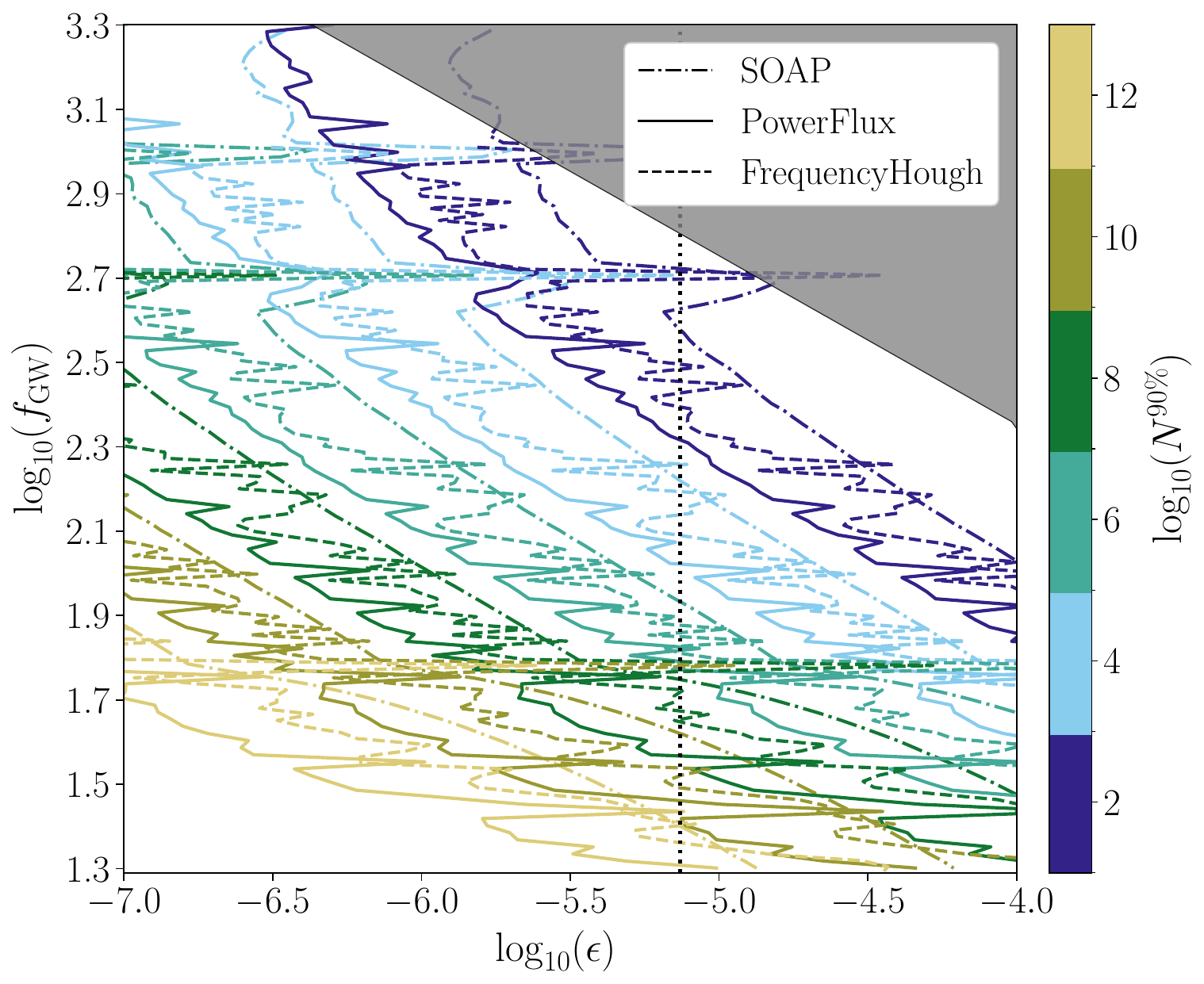}
    \fi
    \caption{
              Contours of constant $\log_{10}(N^{90\%})$ as a function of gravitational-wave frequency $\fgw$ and ellipticity $\epsilon$, for the \pf\ (solid lines), \fh\ (dashed lines), and \so\ (dashed-dotted lines) searches. Here, $N$ denotes the total number of neutron stars with frequency $\fgw$ and ellipticity $\geq \epsilon$. In the absence of detection, we place upper limits on N, and $N^{90\%}$ represents the corresponding 90\% confidence upper limit. The color of each contour indicates the value of $\log_{10}(N^{90\%})$, as shown in the color bar. At each point on the 2D grid, we adopt the most conservative constraint among four spatial distributions viz., the Progenitor \citep{Paczynski} and the three ``Reed'' \citep{Reed} distributions. The filled region in the top right corner is the excluded parameter space for a maximum spin-down magnitude of $10^{-8}$ Hz/s. The vertical dotted line in black demarcates a representative maximum ellipticity the NS crust can support \citep{morales2022}}
    \label{CO_number_constraints}
  \end{center}
\end{figure}

\subsection{Implications for the GeV excess}
\label{sec:gev_results}
Given the absence of any CW signal detections, we follow the methodology described in section~\ref{sec:gev_methodology} to set constraints on the MSP explanation of the GeV excess.

 In Fig.~\ref{fig:GeVexcess}, we show an exclusion plot of regions with  $N_{\rm MSP}P_{\rm GW}> 1$, using upper limits from the \pf search. We note that each search has different sensitivity, covers different frequency ranges, and fixes different maximum $\fdot$, thus different searches would result in different exclusion regions on this plot. We show only \pf results because they are the most constraining.

\begin{figure}[htbp]
  \begin{center}
    \ifshowfigs
    \includegraphics[width=\linewidth]{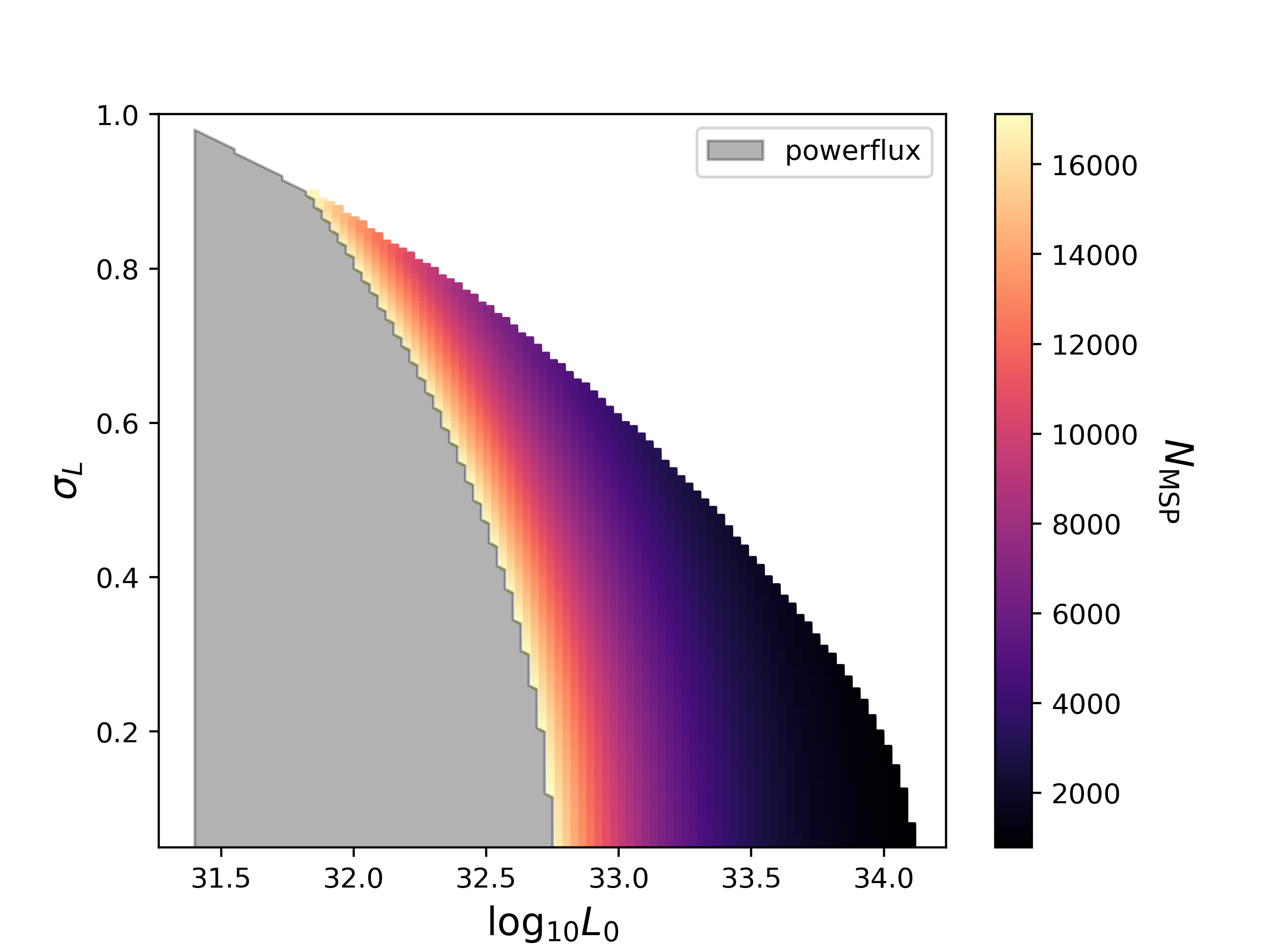}
    \fi
    \caption{Constraints on the GeV excess for \pf.
      The gray region is ruled out by this search for the chosen ellipticity and frequency PDFs. White regions are ruled out by Fermi-LAT. Here, $P_{\rm GW}=1/17000$.
      }
    \label{fig:GeVexcess}
  \end{center}
\end{figure}

\begin{figure}[htbp]
  \begin{center}
    \ifshowfigs
    \includegraphics[width=\linewidth]{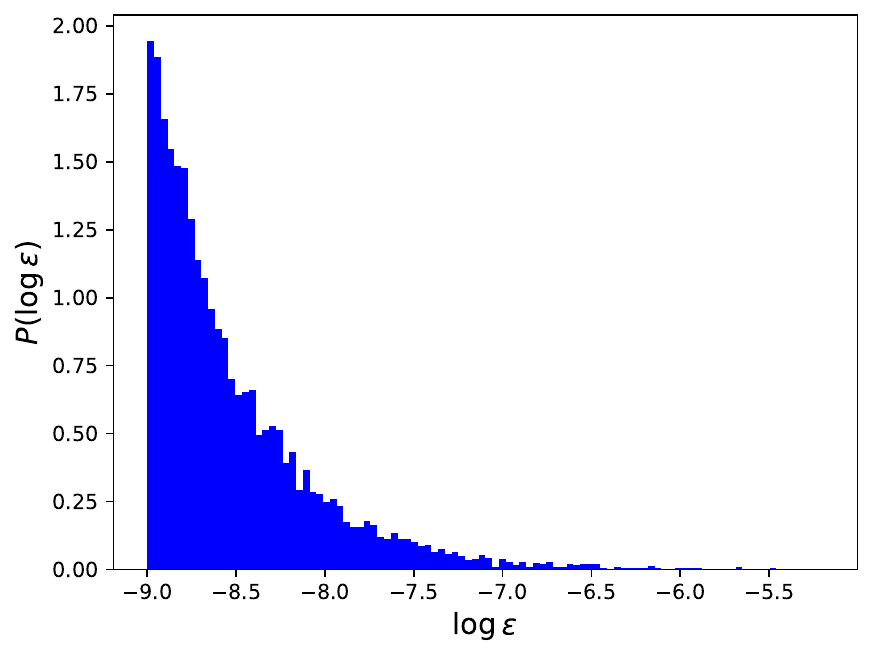}
    \fi
    \caption{One realization of the ellipticity PDF used to compute constraints on the GeV excess. We use 100 realizations of this PDF and average over the number of \gwh signals we could have detected to arrive at the constraints in Fig. \ref{fig:GeVexcess}. The power law of this  distribution is fixed to be $\Lambda=2.1$. }
    \label{fig:ellip_pdf}
  \end{center}
\end{figure}

Note that $P_{\rm GW}$ is strongly sensitive to the value of $\Lambda$. In Fig. \ref{fig:lambdapgw}, we show this dependence, and in particular the smallest possible value of $\Lambda$ that would result in a constraint from \pf in this search. $\Lambda$ controls the fall-off of the exponential distribution on ellipticities: smaller values indicate a slower fall-off, while larger values imply a quicker fall-off. Thus, for larger values of $\Lambda$, the constraints weaken, since there are fewer MSPs with larger ellipticities that would be more likely to be detected than those with smaller ellipticities.

\begin{figure}[htbp]
  \begin{center}
    \ifshowfigs
    \includegraphics[width=\linewidth]{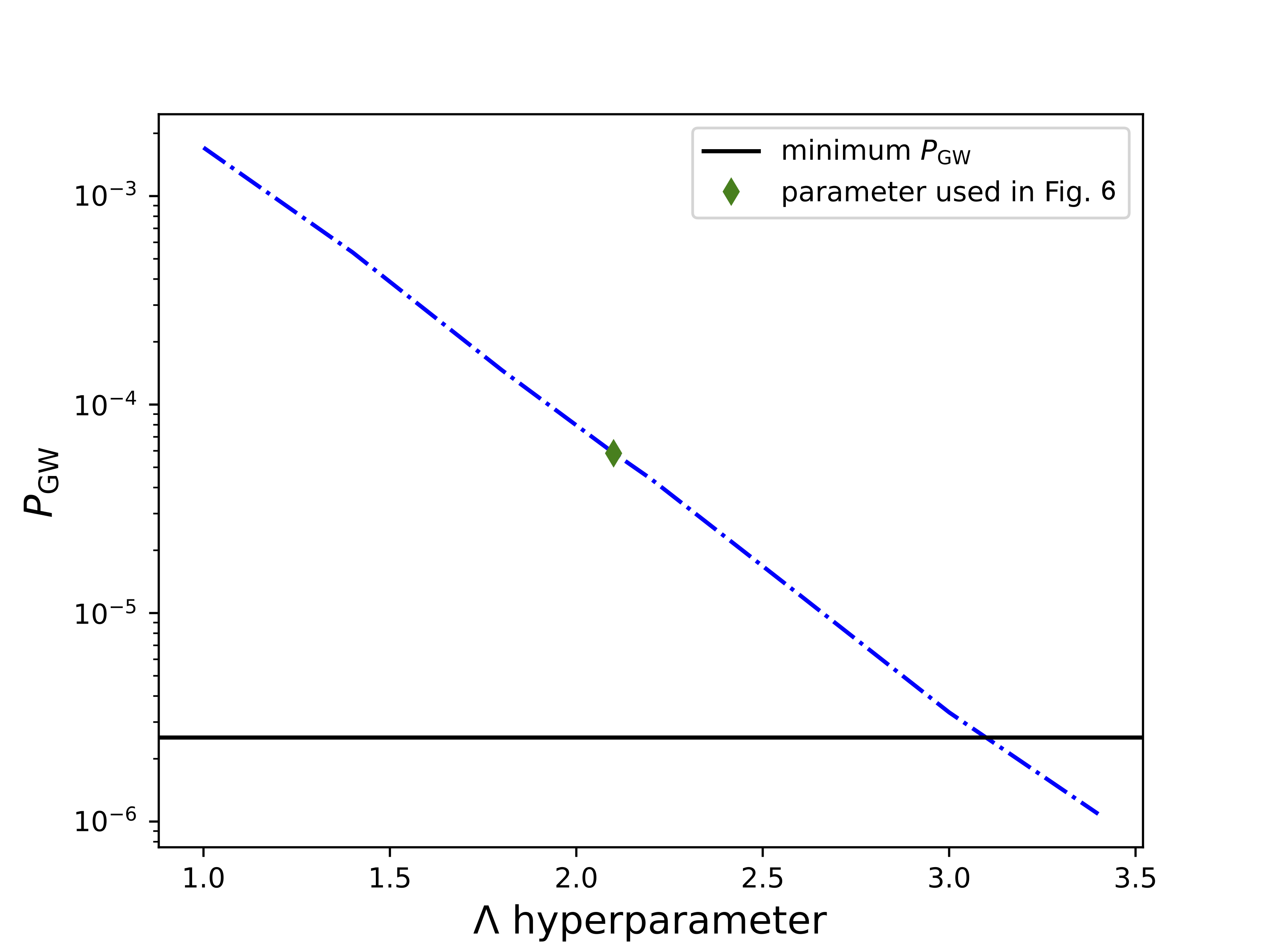}
    \fi
    \caption{Probability of detecting a \gwh in the \pf search as a function of the hyperparameter $\Lambda$. The horizontal line indicates the minimum $P_{\rm GW}$ at which a constraint would exist.
      }
    \label{fig:lambdapgw}
  \end{center}
\end{figure}

\subsection{Implications for light primordial black holes}
\label{sec:pbh_results}
 
Following the methodology outlined in section \ref{sec:pbh_methodology}, for four different choices of $m_1$, we plot the upper limit on $\ftilde(\mone,\mtwo)$ as a function of $m_2$ in Fig.~\ref{fig:pbh-ftilde}. Our results complement those that come from different searches of O3a \cite{Miller:2024fpo,Miller:2024jpo} and O4a data \cite{LIGOScientific:2025vwc} that focus on shorter-duration, planetary-mass \pbh inspirals, and sub-solar mass searches (($[0.1,1]M_\odot$) \cite{Phukon:2021cus,Nitz:2021vqh,Nitz:2021mzz,LIGOScientific:2021job,LIGOScientific:2022hai,Nitz:2022ltl,Wang:2023qgw}. Our limits are similar to those obtained in O3 \cite{bib:cwallskyO3FourPipelines}, primarily because LIGO’s low-frequency sensitivity has not greatly improved from O3 to O4a. Since our constraining power primarily comes from low frequencies -- where the frequency evolution of an inspiraling asteroid-mass system remains nearly monochromatic over the duration of O4a -- our limits have not improved much with respect to the O3 search~\cite{bib:cwallskyO3FourPipelines}.

\begin{figure}[htbp]
  \begin{center}
    \ifshowfigs
    \includegraphics[width=\linewidth]{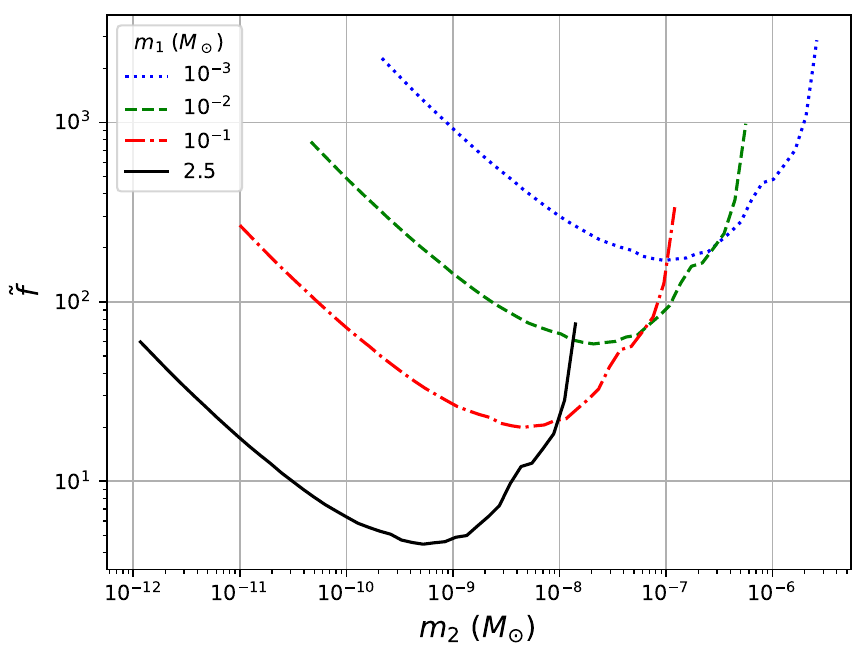}
    \fi
    \caption{Constraint on $\ftilde$ from non-detection of asteroid-mass PBH binaries for \pf. Codes to obtain these constraints are publicly available \cite{miller_2025_15559328}.}
    \label{fig:pbh-ftilde}
  \end{center}
\end{figure}

\section{\label{sec:conclusions}Summary and outlook}

In summary, we have performed three independent all-sky searches for continuous gravitational waves,
ranging from 20 Hz to as high as 2000 Hz, while probing spin-down magnitudes as high as $10^{-8}$ Hz/s.
No credible gravitational wave signals are observed, allowing upper limits to be placed on
possible signal amplitudes.
Over a large fraction of the parameter space covered, these limits are the most stringent to date,
with improvements in strain sensitivity due primarily to
the improved noise floors of the LIGO interferometers over previous LIGO data sets.
Fig.~\ref{fig:combined_upper_limits} shows the strain amplitude upper limits obtained from all three
search pipelines, along with results from previous searches in O3 data.

The most stringent population-averaged strain upper limits
reach \loweststrain\ near \loweststrainfreq\ Hz, matching
the best previous constraints up to $\sim$1700 Hz while extending coverage to a broader spin-down range. At higher frequencies,
the new limits improve upon previous results by factors of approximately $\sim$\ratiolow.
  At the highest frequencies ($\sim$2000 Hz) we are sensitive to neutron stars with an equatorial
ellipticity $\epsilon$ as small as $\sim$$\sci{4}{-7}$ at distances as far away as $\sim$4.7 kpc.
For ellipticities as high as $10^{-6}$, we are sensitive to 
neutron stars at the distance of the galactic center ($\sim$8.5 kpc).

The upper limits on strain amplitude from these searches
have been interpreted in three specific astrophysical scenarios to
constrain the populations of 1) galactic neutrons stars, 2) millisecond pulsars contributing
to the GeV excess and 3) asteroid-scale primordial black holes. 

Similar all-sky searches in the full O4 data ($>$2 year run) are expected to improve upon the sensitivities achieved for
the $\sim$eight months analyzed here. Beyond the O4 run, significant improvements in detector noise are expected
leading up to the fifth LIGO-Virgo-KAGRA run O5 planned for later in this decade. As search ranges extend to ever greater
distances in the galaxy over ever broadening frequency sub-bands, the prospects for discovery should
brighten~\cite{bib:RilesReview,bib:PagliaroEtalReview,bib:OwenReview}. 

\section{Acknowledgments}
 \newif\ifcoreonly\coreonlyfalse
 \newif\ifkagra\kagratrue
 \newif\ifheader\headerfalse
\ifheader
\begin{center}{\bf\Large
\ifkagra
Conference proceedings acknowledgements for \\ the LIGO Scientific Collaboration, the Virgo Collaboration and the KAGRA Collaboration
\else
Conference proceedings acknowledgements for \\ the LIGO Scientific Collaboration and the Virgo Collaboration
\fi
}\end{center}
\fi
This material is based upon work supported by NSF's LIGO Laboratory, which is a
major facility fully funded by the National Science Foundation.
The authors also gratefully acknowledge the support of
the Science and Technology Facilities Council (STFC) of the
United Kingdom, the Max-Planck-Society (MPS), and the State of
Niedersachsen/Germany for support of the construction of Advanced LIGO 
and construction and operation of the GEO\,600 detector. 
Additional support for Advanced LIGO was provided by the Australian Research Council.
The authors gratefully acknowledge the Italian Istituto Nazionale di Fisica Nucleare (INFN),  
the French Centre National de la Recherche Scientifique (CNRS) and
the Netherlands Organization for Scientific Research (NWO)
for the construction and operation of the Virgo detector
and the creation and support  of the EGO consortium. 
The authors also gratefully acknowledge research support from these agencies as well as by 
the Council of Scientific and Industrial Research of India, 
the Department of Science and Technology, India,
the Science \& Engineering Research Board (SERB), India,
the Ministry of Human Resource Development, India,
the Spanish Agencia Estatal de Investigaci\'on (AEI),
the Spanish Ministerio de Ciencia, Innovaci\'on y Universidades,
the European Union NextGenerationEU/PRTR (PRTR-C17.I1),
the ICSC - CentroNazionale di Ricerca in High Performance Computing, Big Data
and Quantum Computing, funded by the European Union NextGenerationEU,
the Comunitat Auton\`oma de les Illes Balears through the Conselleria d'Educaci\'o i Universitats,
the Conselleria d'Innovaci\'o, Universitats, Ci\`encia i Societat Digital de la Generalitat Valenciana and
the CERCA Programme Generalitat de Catalunya, Spain,
the Polish National Agency for Academic Exchange,
the National Science Centre of Poland and the European Union - European Regional
Development Fund;
the Foundation for Polish Science (FNP),
the Polish Ministry of Science and Higher Education,
the Swiss National Science Foundation (SNSF),
the Russian Science Foundation,
the European Commission,
the European Social Funds (ESF),
the European Regional Development Funds (ERDF),
the Royal Society, 
the Scottish Funding Council, 
the Scottish Universities Physics Alliance, 
the Hungarian Scientific Research Fund (OTKA),
the French Lyon Institute of Origins (LIO),
the Belgian Fonds de la Recherche Scientifique (FRS-FNRS), 
Actions de Recherche Concert\'ees (ARC) and
Fonds Wetenschappelijk Onderzoek - Vlaanderen (FWO), Belgium,
the Paris \^{I}le-de-France Region, 
the National Research, Development and Innovation Office of Hungary (NKFIH), 
the National Research Foundation of Korea,
the Natural Sciences and Engineering Research Council of Canada (NSERC),
the Canadian Foundation for Innovation (CFI),
the Brazilian Ministry of Science, Technology, and Innovations,
the International Center for Theoretical Physics South American Institute for Fundamental Research (ICTP-SAIFR), 
the Research Grants Council of Hong Kong,
the National Natural Science Foundation of China (NSFC),
the Israel Science Foundation (ISF),
the US-Israel Binational Science Fund (BSF),
the Leverhulme Trust, 
the Research Corporation,
the National Science and Technology Council (NSTC), Taiwan,
the United States Department of Energy,
and
the Kavli Foundation.
The authors gratefully acknowledge the support of the NSF, STFC, INFN and CNRS for provision of computational resources.

\ifkagra
This work was supported by MEXT,
the JSPS Leading-edge Research Infrastructure Program,
JSPS Grant-in-Aid for Specially Promoted Research 26000005,
JSPS Grant-in-Aid for Scientific Research on Innovative Areas 2402: 24103006,
24103005, and 2905: JP17H06358, JP17H06361 and JP17H06364,
JSPS Core-to-Core Program A.\ Advanced Research Networks,
JSPS Grants-in-Aid for Scientific Research (S) 17H06133 and 20H05639,
JSPS Grant-in-Aid for Transformative Research Areas (A) 20A203: JP20H05854,
the joint research program of the Institute for Cosmic Ray Research,
University of Tokyo,
the National Research Foundation (NRF),
the Computing Infrastructure Project of the Global Science experimental Data hub
Center (GSDC) at KISTI,
the Korea Astronomy and Space Science Institute (KASI),
the Ministry of Science and ICT (MSIT) in Korea,
Academia Sinica (AS),
the AS Grid Center (ASGC) and the National Science and Technology Council (NSTC)
in Taiwan under grants including the Science Vanguard Research Program,
the Advanced Technology Center (ATC) of NAOJ,
and the Mechanical Engineering Center of KEK.
\fi

Additional acknowledgements for support of individual authors may be found in the following document: \\
\texttt{https://dcc.ligo.org/LIGO-M2300033/public}.
For the purpose of open access, the authors have applied a Creative Commons Attribution (CC BY)
license to any Author Accepted Manuscript version arising.
We request that citations to this article use 'A. G. Abac {\it et al.} (LIGO-Virgo-KAGRA Collaboration), ...' or similar phrasing, depending on journal convention.

This document has been assigned LIGO Laboratory document number \texttt{LIGO-P2500416-v6}.

\appendix

\section{Details concerning the \pf\ search}
\label{sec:powerflux_appendix}

We present in the following some details concerning the \pf\ search.

\subsection{PowerFlux search configuration}
\label{sec:powerflux_config_app}
  
Table~\ref{tab:powerflux_tuning_info} gives the detailed configuration for the initial stage of the \pf\ search, showing parameters that depend explicitly on coarse frequency bands.

\begin{table*}[htb]
  \begin{center}
    {\setlength{\tabcolsep}{11pt}
  \begin{tabularx}{\textwidth}{lcccc}\hline\hline
    \T\B                                         &  20-60 Hz            &  60-475 Hz          & 475-1475 Hz        & 1475-2000 Hz \\
    \hline
    \T\B Frequency bin width                     &  0.139 mHz           &  0.139 mHz          & 0.278 mHz          & 0.556 mHz  \\
    \T\B Number of spin-down templates\strut     &   220                 &  110                &  33                & 110     \\
    \T\B Spin-down step size (Hz/s)\strut        &  $\sci{5.0}{-11}$  & $\sci{1.000}{-10}$  & $\sci{3.333}{-10}$ & $\sci{1.000}{-10}$   \strut        \\
     \T\B H1/L1 frequency mismatch tolerance (mHz) & 2.5                &   2.5               &  2.5               & 2.5         \\
     \T\B H1/L1 spin-down mismatch tolerance (Hz/s) &  $\sci{3.0}{-10}$ &  $\sci{3.0}{-10}$   &   $\sci{3.0}{-10}$ & $\sci{3.0}{-10}$    \\
    \hline\hline
  \end{tabularx}
  }
  \caption{Information on the band-specific parameters for the initial search (stage 0) and outlier follow-up stages.
    The number of spin-down steps and the spin-down step size refer to the templates used in the stage-0 search.
    The tolerances refer to the consistency requirements between H1 and L1 outliers used to define candidates selected
    for stage-1 follow-up. Other search parameters relevant to all search bands are described in the text.}
\label{tab:powerflux_tuning_info}
\end{center}
\end{table*}

\begin{table*}[htbp]
\begin{center}
\begin{tabular}{llccccc}\hline\hline
\T\B Stage & Instrument sum & {Phase coherence $\delta$} & \multicolumn{1}{c}{Spin-down step} & \multicolumn{1}{c}{Sky refinement} & \multicolumn{1}{c}{Frequency refinement} & \multicolumn{1}{c}{SNR increase} \\
 & & \multicolumn{1}{c}{rad} & \multicolumn{1}{c}{Hz/s} & $\alpha\times\delta_0$  &  & \multicolumn{1}{c}{\%}\\
\hline
\multicolumn{7}{c}{\T\B 20-60\,Hz frequency range, 7200\,s SFTs, 0.0625\,Hz frequency bands} \\
  0 & Initial/upper limit incoherent & NA      & $\sci{5.0}{-11}$ & $1$   & $1/2$ & -- \\
  1 & incoherent                        & $\pi/2$ & $\sci{2.5}{-11}$   & $1/4\times1/4$ & $1/8$ & 10 \\
  2 & coherent                          & $\pi/2$ & $\sci{1.0}{-11}$   & $1/4\times1/4$ & $1/8$ & 10  \\
 \hline
\multicolumn{7}{c}{\T\B 60-475\,Hz frequency range, 7200\,s SFTs, 0.0625\,Hz frequency bands} \\
  0 & Initial/upper limit incoherent & NA      & $\sci{1.000}{-10}$ & $1$   & $1/2$ & -- \\
  1 & incoherent                        & $\pi/2$ & $\sci{5.0}{-11}$ & $1/4\times1/4$ & $1/8$ & 20 \\
  2 & coherent                          & $\pi/2$ & $\sci{1.0}{-11}$   & $1/4\times1/4$ & $1/8$ & 10  \\
 \hline
  \multicolumn{7}{c}{\T\B 475-1475\,Hz frequency range, 3600\,s SFTs, 0.125\,Hz frequency bands} \\
  0 & Initial/upper limit incoherent & NA      & $\sci{3.333}{-10}$  & $1$   & $1/2$ & -- \\
  1 & incoherent                        & $\pi/2$ & $\sci{1.0}{-10}$  & $1/4\times1/4$ & $1/8$ & 20 \\
  2 & coherent                          & $\pi/2$ & $\sci{5.0}{-11}$   & $1/4\times1/4$ & $1/8$ & 15  \\
 \hline
 \multicolumn{7}{c}{\T\B 1475-2000\,Hz frequency range, 1800\,s SFTs, 0.25\,Hz frequency bands} \\
  0 & Initial/upper limit incoherent & NA      & $\sci{1.000}{-10}$  & $1$   & $1/2$ & -- \\
  1 & incoherent                        & $\pi/2$ & $\sci{3.0}{-11}$  & $1/4\times1/4$ & $1/8$ & 20 \\
  2 & coherent                          & $\pi/2$ & $\sci{1.5}{-11}$   & $1/4\times1/4$ & $1/8$ & 10  \\
 \hline\hline
\end{tabular}
\caption[Outlier follow-up parameters]{\pf\ outlier follow-up parameters. Stage 1 and higher use a loose-coherence algorithm for demodulation. The sky (both right ascension $\alpha$ and declination $\delta_0$) and frequency refinement parameters are relative to values used in the stage 0 \pf\ search}
\label{tab:powerflux_followup_parameters}
\end{center}
\end{table*}

The power calculation of the data can be expressed as a bilinear form of the input matrix $\left\{a_{t,f}\right\}$
constructed from the SFT coefficients with indices representing time and frequency:

\begin{equation}
P[f] = \sum_{t_1, t_2,D_i,D_j} a_{t_1, f+\Delta f(t_1)}^{(D_i)} a_{t_2, f+\Delta f(t_2)}^{(D_j)*} K_{t_1, t_2, f}^{D_iD_j}\ed
\end{equation}

\noindent In this expression $\Delta f(t)$ is the detector-frame frequency drift due to the effects from both Doppler shifts and the first
frequency derivative. The sum is taken over all times $t$ corresponding to the midpoints of the SFT time intervals.
The kernel $K_{t_1, t_2, f}^{(D_iD_j)}$ includes the contribution of time-dependent SFT noise weights, antenna response, signal
polarization parameters, and relative phase terms~\cite{bib:loosecoherence,bib:LooseCoherenceWellModeledSignals} for detectors $D_{i,j}$ (= H1, L1).
Separate power sums are computed for H1, L1 and combined H1-L1 data.

The fast first-stage (stage 0) \pf\ algorithm uses a kernel with diagonal terms only (including separate single-detector
contributions $D_i=D_j$). The second stage (stage 1)
increases effective coherence time while still allowing for controlled deviation in
phase~\cite{bib:loosecoherence} via kernels that increase effective coherence length by inclusion of
limited single-detector, off-diagonal terms. The third stage (stage 2) maintains the stage-1 effective coherence time, but
adds SFT coefficients from H1 and L1 data coherently ($D_i\ne D_j$) to improve SNR and parameter resolution.

The effective coherence length is captured in a parameter $\delta$~\cite{bib:loosecoherence}, which describes the degree of phase drift
allowed between SFTs. A value of $\delta=0$ corresponds to a fully coherent case,
and $\delta=\pi$ corresponds to incoherent power sums.

Depending on the terms used, the data from different interferometers can be combined incoherently (such as in stages 0
and 1, see Table \ref{tab:powerflux_followup_parameters}) or coherently (as used in stage 2). The coherent
combination is more computationally expensive but improves parameter estimation.

\subsection{Validation of the \pf\ upper limits}
\label{sec:powerflux_upperlimits_app}

Figure~\ref{fig:powerflux_ul_vs_strain} shows results of a high-statistics
``software injections'' simulation run performed as described in~\cite{bib:cwallskyS5}.
Correctly established upper limits lie above the dashed diagonal lines (defining equality between upper limit obtained and
true injection strain) in each panel, corresponding to
four selected sub-bands [SFT coherence times]: 20-60 Hz [7200s], 60-475 Hz [7200s], 475-1475 Hz [3600s] and 1475-2000 Hz [1800s].
Performance for the 7200s-SFT 20-60 Hz and 60-475 Hz bands are shown separately because of the proliferation
of spectral line artifacts below 60 Hz, primarily in the H1 data, and because of the large mismatch
in H1 and L1 noise floors below $\sim$40 Hz. The breakpoint frequencies of 475 Hz and 1475 Hz for
decreasing SFT coherence time are those used in the O1 PowerFlux
search~\cite{bib:cwallskyO1paper1,bib:cwallskyO1paper2}, marking the starts of bands disturbed by 1st and 3rd violin mode harmonics.
Additional band-specific parameters for the initial stage of the search are listed in Table~\ref{tab:powerflux_tuning_info}.

\begin{figure*}[htbp]
  \begin{center}
    \ifshowfigs
    \includegraphics[width=3.3in]{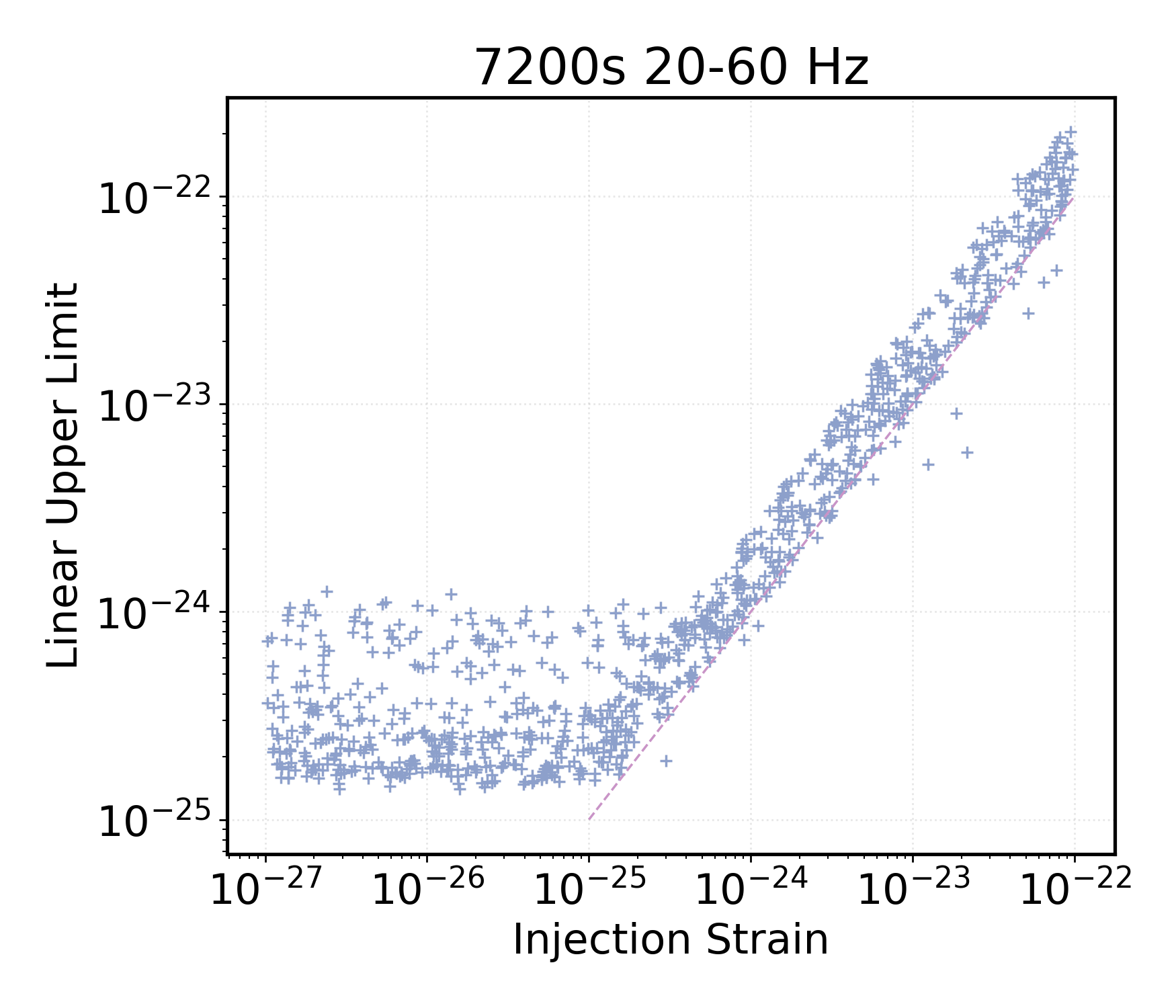}
    \includegraphics[width=3.3in]{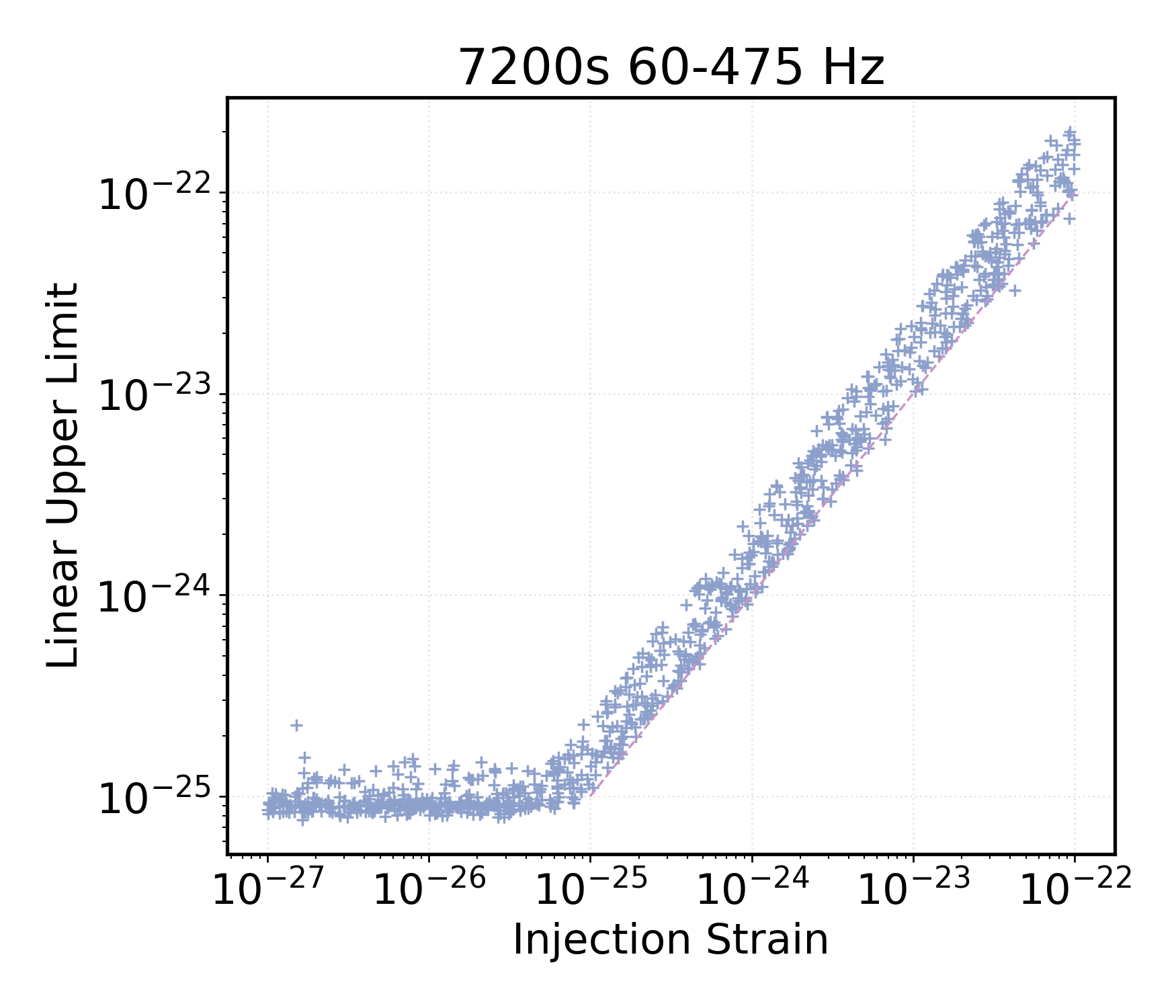}
    \includegraphics[width=3.3in]{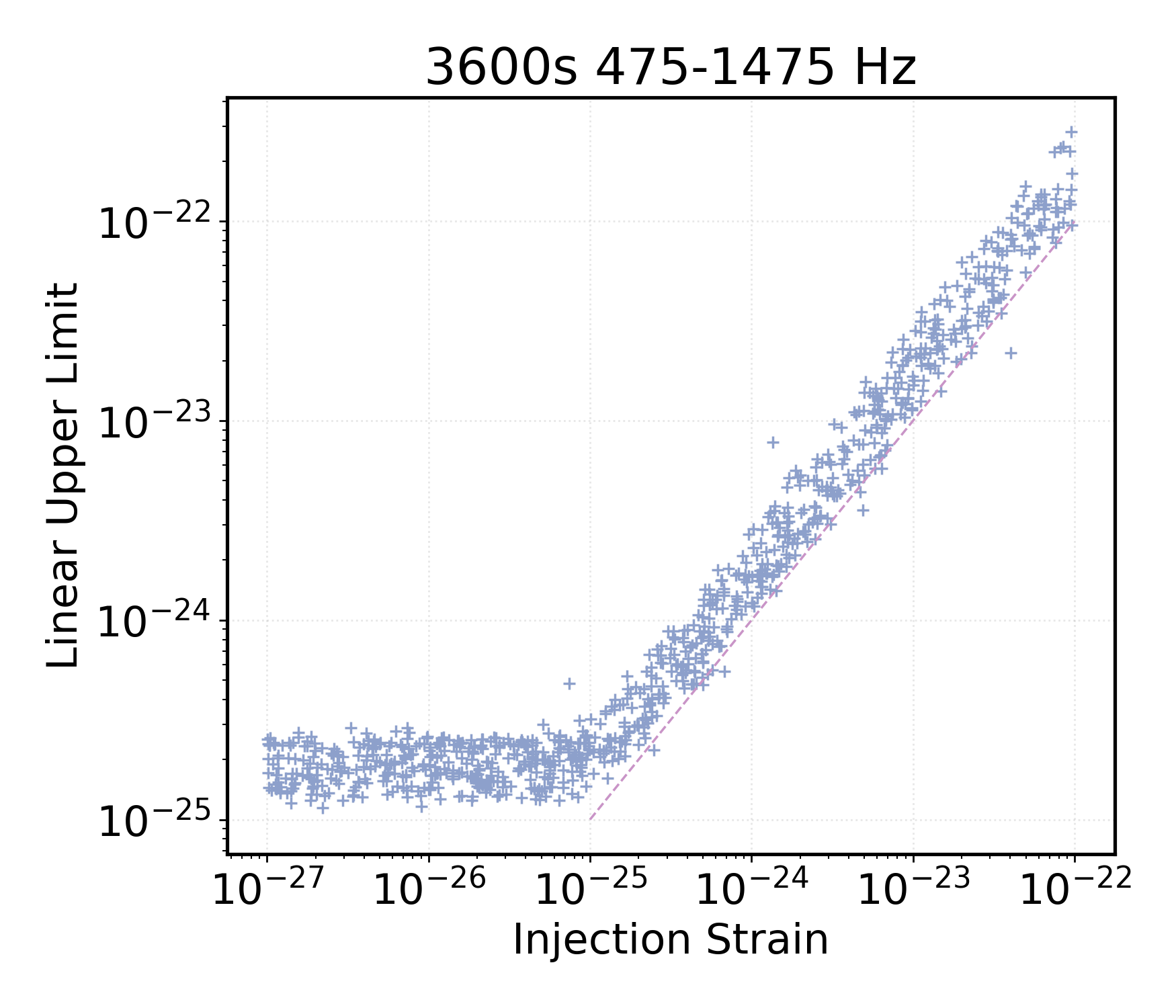}
    \includegraphics[width=3.3in]{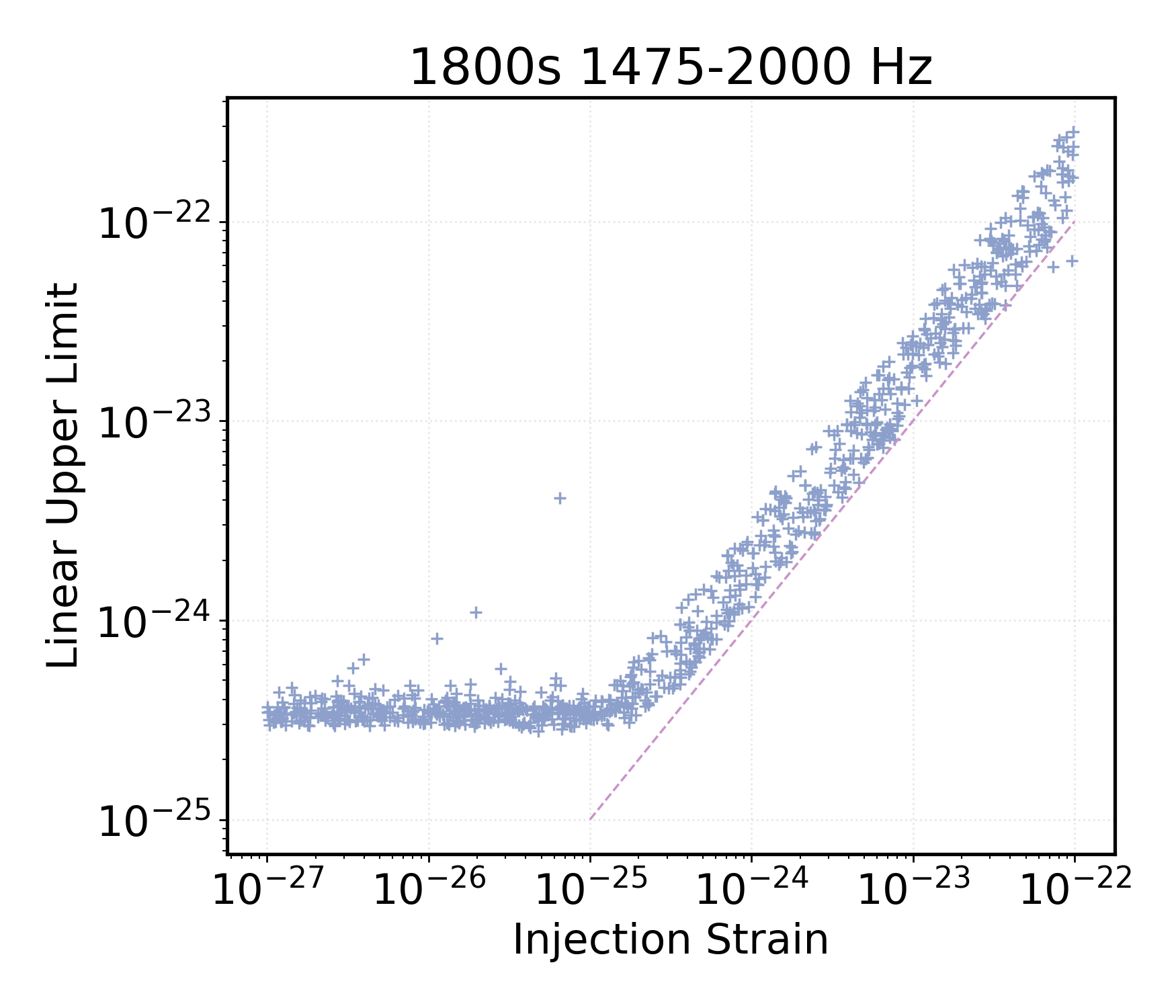}
    \fi
    \caption[Upper limit versus injected strain]{Illustration of \pf\ upper limit validation for very low, low, mid and high frequencies.
      Each point in the upper left, upper right, lower left and lower right panels represents a separate
      injection in the 20-60\,Hz, 60-475\,Hz, 475-1475\,Hz and 1475-2000\,Hz frequency ranges, respectively.
      Each established upper limit (vertical axis) is compared against the
      injected strain value defined by the horizontal axis (diagonal line defines equality of upper limit obtained and
      true injected strain).}
    \label{fig:powerflux_ul_vs_strain}
  \end{center}
\end{figure*}

\subsection{Validation of the \pf\ outlier follow-up}
\label{sec:powerflux_outlier_app}

Testing of the stages 0--2 pipeline is performed for frequency bands searched via software injections
using the same follow-up procedure. The recovery criteria also require that an outlier close to the true injection location (within 2.5~mHz in frequency
$f$, $\sci{3}{-10}$~Hz/s in spin-down and 28.5~rad$\cdot$Hz$/f$ in sky location) be found and successfully pass
through all stages of the detection pipeline.

Injection recovery efficiencies from simulations covering the major sub-bands (20-475 Hz, 475-1475 Hz, 1475-2000 Hz) are
shown in Fig.~\ref{fig:powerflux_injection_recovery} for stages 0, 1 and 2, which confirm that 95\%\ signal
recovery is comparable to the 95\%\ upper limit in all bands, as desired, except for the region
below 60 Hz for which spectral line artifacts heavily contaminate the H1 data set.
Injections in vetoed frequency bands (see section~\ref{sec:powerflux_outlier_followup} and Table~\ref{tab:powerflux_outlier_badbands}) are not included in these graphs.

\begin{figure*}[htbp]
 \begin{center}
    \ifshowfigs
    \includegraphics[width=3.3in]{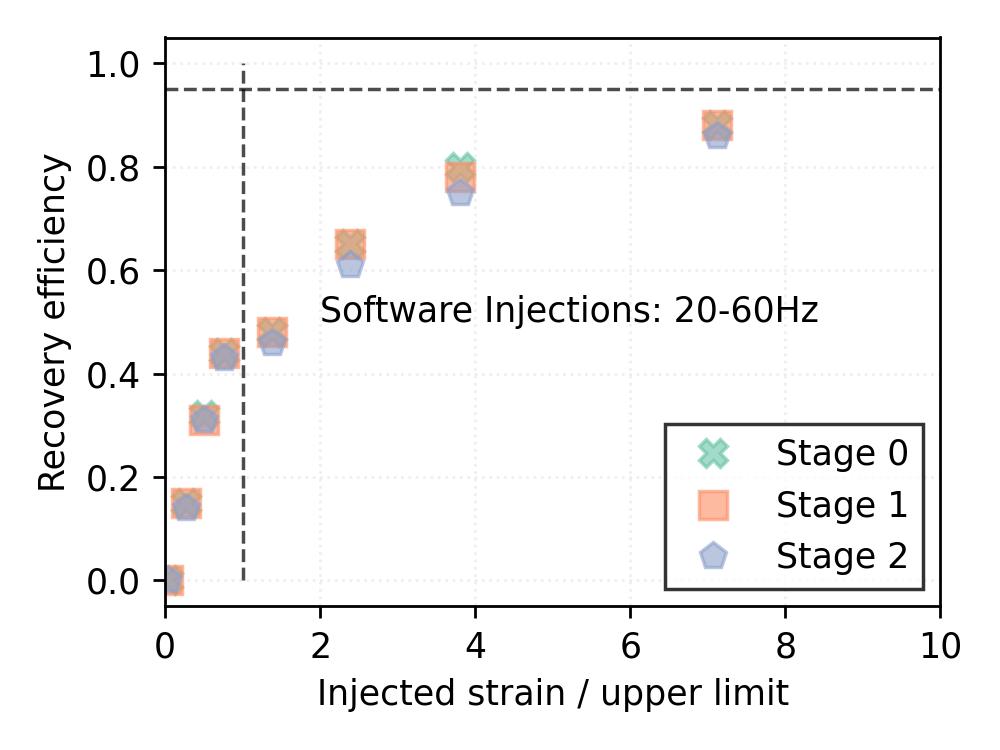}
    \includegraphics[width=3.3in]{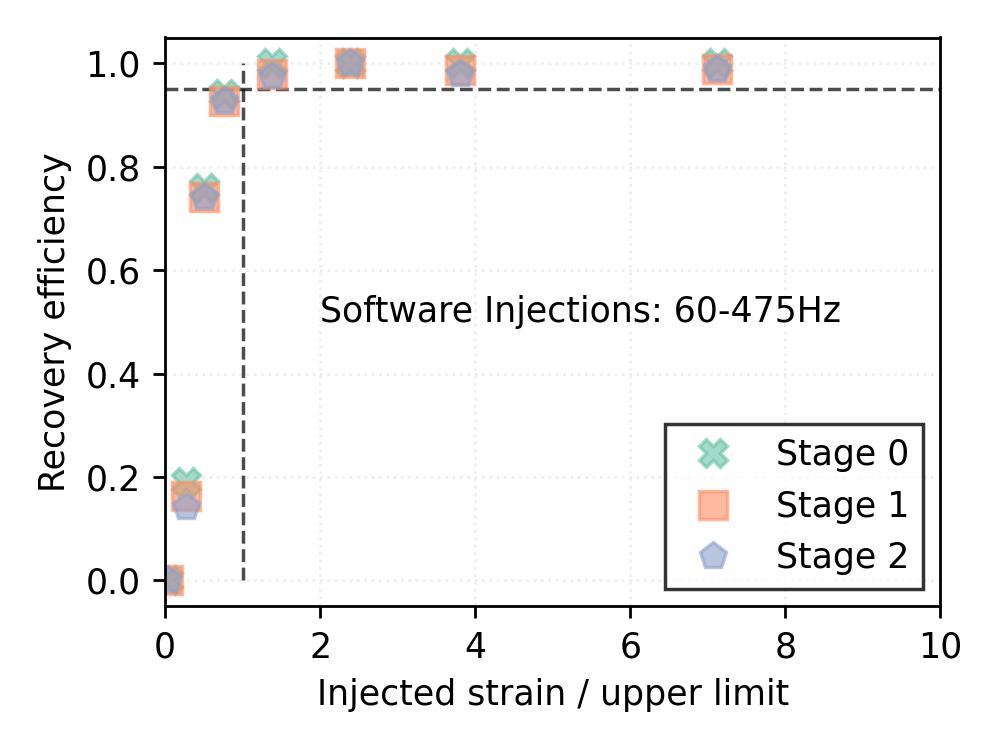}
    \includegraphics[width=3.3in]{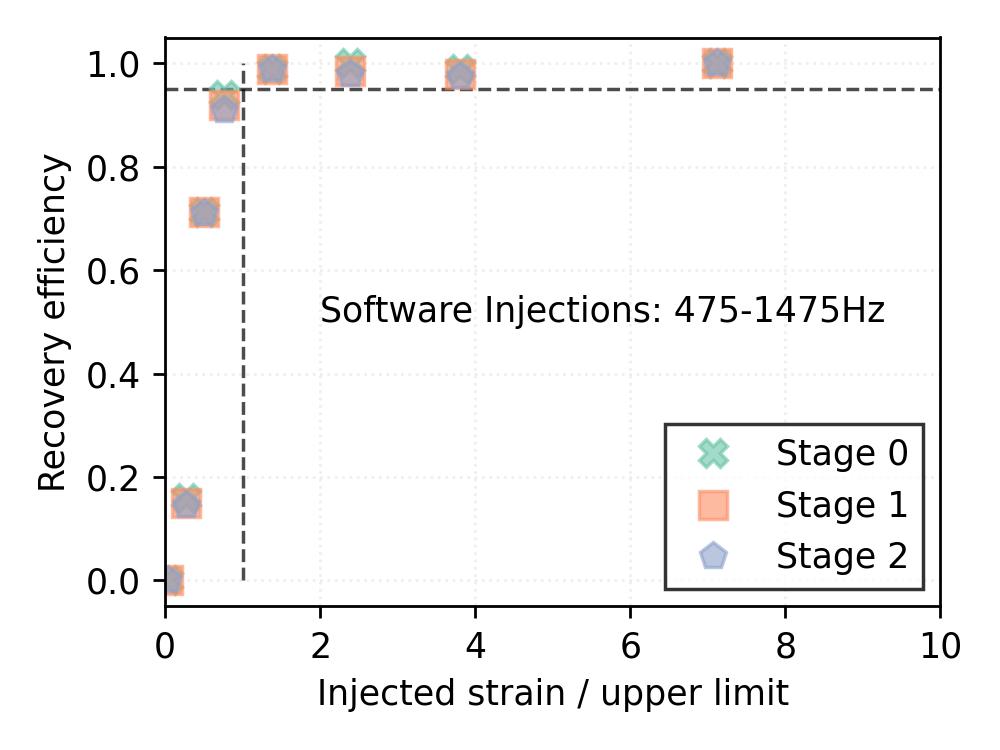}
    \includegraphics[width=3.3in]{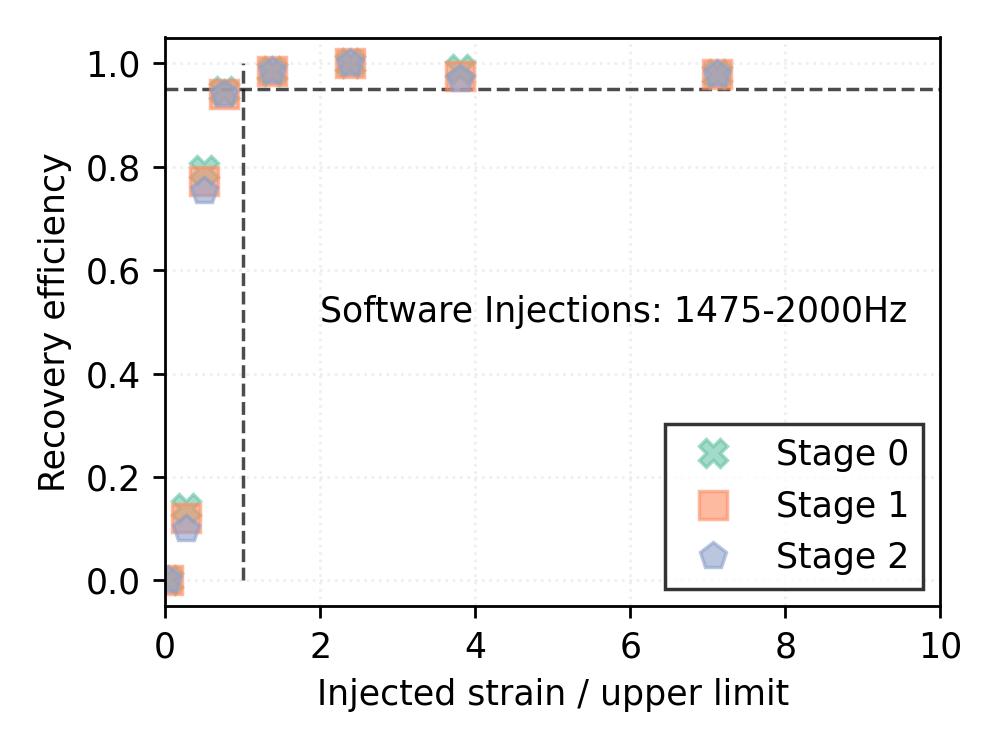}
    \fi
    \caption[Injection recovery]{Injection (software simulations) recovery efficiencies
      in the 20-60\,Hz, 60-475\,Hz, 475-1475\,Hz and 1475-2000\,Hz frequency bands are shown
      in the upper left, upper right, lower left and lower right panels, respectively, for stages 0, 1 and 2 of the search.
      The injected strain divided by the 95\%\ CL upper limit in its band (without injection)
      is shown on the horizontal axis.
      The percentage of surviving injections is shown on the vertical axis,
      with a horizontal dashed line drawn at the $95$\% level. The vertical dashed line
      marks a relative strain of unity. Ideally, the recovery efficiencies should lie
      to the left of the vertical line or above the horizonal line. One observes, however,
      that the ideal recovery efficiency is significantly degraded below 60 Hz, where large instrumental
      artifacts in the H1 data and only modest expected Doppler modulations make clean signal recovery more challenging.
      Most of the degradation stems from the band below 27 Hz for which H1 line contamination is severe and for
      which the H1 noise floor is substantially higher than the L1 noise floor.}
   \label{fig:powerflux_injection_recovery}
  \end{center}
\end{figure*}

\subsection{Vetoed bands in the \pf\ search}
\label{sec:powerflux_vetoedbands_app}

Table~\ref{tab:powerflux_outlier_badbands} shows particular frequency bands (about 13\%\ of the original search band) for which no
outlier follow-up is attempted for the initial or later stages because of instrumental artifacts
that lead to such an unmanageable flood of initial or later-stage outliers. 

Figure~\ref{fig:powerflux_vetobands} shows the regions vetoed because of excessive outlier counts
due to instrumental artifacts over the full search band, with a magnification of an example
100-Hz band (1900-2000 Hz) that includes a ``forest'' of 4th harmonics of violin modes
from both interferometers.

\begin{table*}[htb]
  \begin{center}
    \def\mc#1{\multicolumn{1}{c}{#1}}
\def\ml#1{\multicolumn{1}{l}{#1}}
{\setlength{\tabcolsep}{5pt}
\begin{tabularx}{\textwidth}{rrl@{\hskip0.3in}rrl@{\hskip0.3in}rrl@{\hskip0.3in}rrl}\hline\hline
\mc{$f_{\rm low}^{\rm veto}$} & \mc{$f_{\rm high}^{\rm veto}$} & \ml{$\Delta f^{\rm veto}$} & \mc{$f_{\rm low}^{\rm veto}$} & \mc{$f_{\rm high}^{\rm veto}$} & \ml{$\Delta f^{\rm veto}$} & \mc{$f_{\rm low}^{\rm veto}$} & \mc{$f_{\rm high}^{\rm veto}$} & \ml{$\Delta f^{\rm veto}$} & \mc{$f_{\rm low}^{\rm veto}$} & \mc{$f_{\rm high}^{\rm veto}$} & \ml{$\Delta f^{\rm veto}$} \\ 
\mc{(Hz)} & \mc{(Hz)} & \ml{(Hz)}  & \mc{(Hz)} & \mc{(Hz)} & \ml{(Hz)}  & \mc{(Hz)} & \mc{(Hz)} & \ml{(Hz)}  & \mc{(Hz)} & \mc{(Hz)} & \ml{(Hz)}  \\ 
\hline
27.097 & 27.903 & 0.805
 & 301.720 & 302.580 & 0.860
 & 898.260 & 900.940 & 2.680
 & 1799.640 & 1800.310 & 0.670
\\
27.247 & 27.613 & 0.365
 & 302.870 & 303.630 & 0.761
 & 906.059 & 907.141 & 1.081
 & 1899.628 & 1900.316 & 0.688
\\
27.332 & 27.673 & 0.341
 & 305.819 & 306.581 & 0.761
 & 909.459 & 910.541 & 1.082
 & 1919.628 & 1920.372 & 0.744
\\
27.547 & 27.883 & 0.336
 & 307.119 & 307.881 & 0.761
 & 918.458 & 919.142 & 0.684
 & 1920.628 & 1921.372 & 0.744
\\
27.707 & 28.073 & 0.366
 & 314.619 & 315.482 & 0.863
 & 922.308 & 923.022 & 0.715
 & 1921.628 & 1922.372 & 0.744
\\
28.047 & 28.393 & 0.346
 & 410.100 & 410.500 & 0.400
 & 958.724 & 959.296 & 0.572
 & 1929.657 & 1932.843 & 3.186
\\
29.807 & 30.133 & 0.326
 & 416.673 & 417.064 & 0.390
 & 959.724 & 960.276 & 0.552
 & 1935.256 & 1937.894 & 2.637
\\
32.017 & 32.653 & 0.636
 & 419.379 & 419.767 & 0.388
 & 960.724 & 961.276 & 0.552
 & 1939.156 & 1941.394 & 2.238
\\
33.178 & 33.488 & 0.310
 & 419.558 & 420.442 & 0.884
 & 986.751 & 1019.752 & 33.001
 & 1947.255 & 1948.065 & 0.810
\\
33.267 & 33.593 & 0.327
 & 432.807 & 436.994 & 4.187
 & 1022.248 & 1025.753 & 3.505
 & 1948.755 & 1949.645 & 0.890
\\
34.722 & 35.053 & 0.332
 & 501.100 & 505.601 & 4.501
 & 1224.797 & 1225.773 & 0.975
 & 1952.155 & 1955.346 & 3.191
\\
35.246 & 35.754 & 0.507
 & 506.899 & 509.201 & 2.302
 & 1229.997 & 1230.943 & 0.946
 & 1952.605 & 1957.346 & 4.741
\\
35.546 & 35.879 & 0.332
 & 509.799 & 513.801 & 4.002
 & 1259.424 & 1261.776 & 2.352
 & 1957.654 & 1960.346 & 2.692
\\
37.742 & 38.057 & 0.315
 & 516.298 & 517.202 & 0.903
 & 1463.704 & 1474.297 & 10.594
 & 1960.054 & 1962.346 & 2.292
\\
37.767 & 38.078 & 0.311
 & 598.740 & 600.220 & 1.480
 & 1480.702 & 1486.999 & 6.297
 & 1961.654 & 1963.846 & 2.193
\\
39.696 & 40.164 & 0.468
 & 603.890 & 604.710 & 0.821
 & 1489.201 & 1490.599 & 1.398
 & 1964.154 & 1968.347 & 4.193
\\
40.646 & 41.164 & 0.518
 & 606.189 & 607.011 & 0.821
 & 1491.201 & 1501.600 & 10.399
 & 1967.453 & 1970.847 & 3.394
\\
44.694 & 45.027 & 0.334
 & 611.789 & 612.861 & 1.072
 & 1509.919 & 1513.071 & 3.152
 & 1971.203 & 1972.097 & 0.894
\\
45.145 & 45.555 & 0.409
 & 614.469 & 615.412 & 0.943
 & 1511.999 & 1513.201 & 1.203
 & 1977.652 & 1979.298 & 1.646
\\
49.835 & 50.165 & 0.330
 & 629.137 & 629.578 & 0.441
 & 1599.660 & 1600.290 & 0.630
 & 1980.152 & 1981.248 & 1.096
\\
59.344 & 60.556 & 1.212
 & 629.587 & 630.683 & 1.096
 & 1699.650 & 1700.300 & 0.650
 & 1981.212 & 1981.923 & 0.711
\\
99.820 & 100.180 & 0.360
 & 629.742 & 630.178 & 0.436
 & 1729.617 & 1730.268 & 0.651
 & 1984.851 & 1985.699 & 0.847
\\
119.638 & 120.362 & 0.724
 & 633.127 & 633.583 & 0.457
 & 1737.901 & 1738.559 & 0.658
 & 1985.301 & 1986.029 & 0.727
\\
179.332 & 180.568 & 1.236
 & 828.752 & 829.248 & 0.496
 & 1745.375 & 1754.825 & 9.450
 & 1987.951 & 1988.949 & 0.998
\\
239.576 & 239.934 & 0.358
 & 828.847 & 829.353 & 0.506
 & 1797.835 & 1798.505 & 0.670
 & 1993.451 & 1994.349 & 0.899
\\
267.123 & 285.178 & 18.055
 & 829.297 & 829.803 & 0.506
 & 1798.870 & 1799.580 & 0.710
 & 1993.951 & 1994.799 & 0.849
\\
299.200 & 300.430 & 1.230
 & 829.717 & 830.203 & 0.486
 & 1799.450 & 1800.210 & 0.760
 & 1999.610 & 2000.330 & 0.720
\\
\hline\hline
\end{tabularx}
}

    \caption{Frequency bands vetoed from \pf\ outlier follow-up because of excessive
      outlier counts from instrumental artifacts. The stage at
      which outlier followup is aborted is also shown for each band.
      Altogether, about 204 Hz out of the 1980-Hz original search band was vetoed, or about 10\%.}
\label{tab:powerflux_outlier_badbands}
\end{center}
\end{table*}

\begin{figure*}[htbp]
  \ifshowfigs
  \includegraphics[width=6.5in]{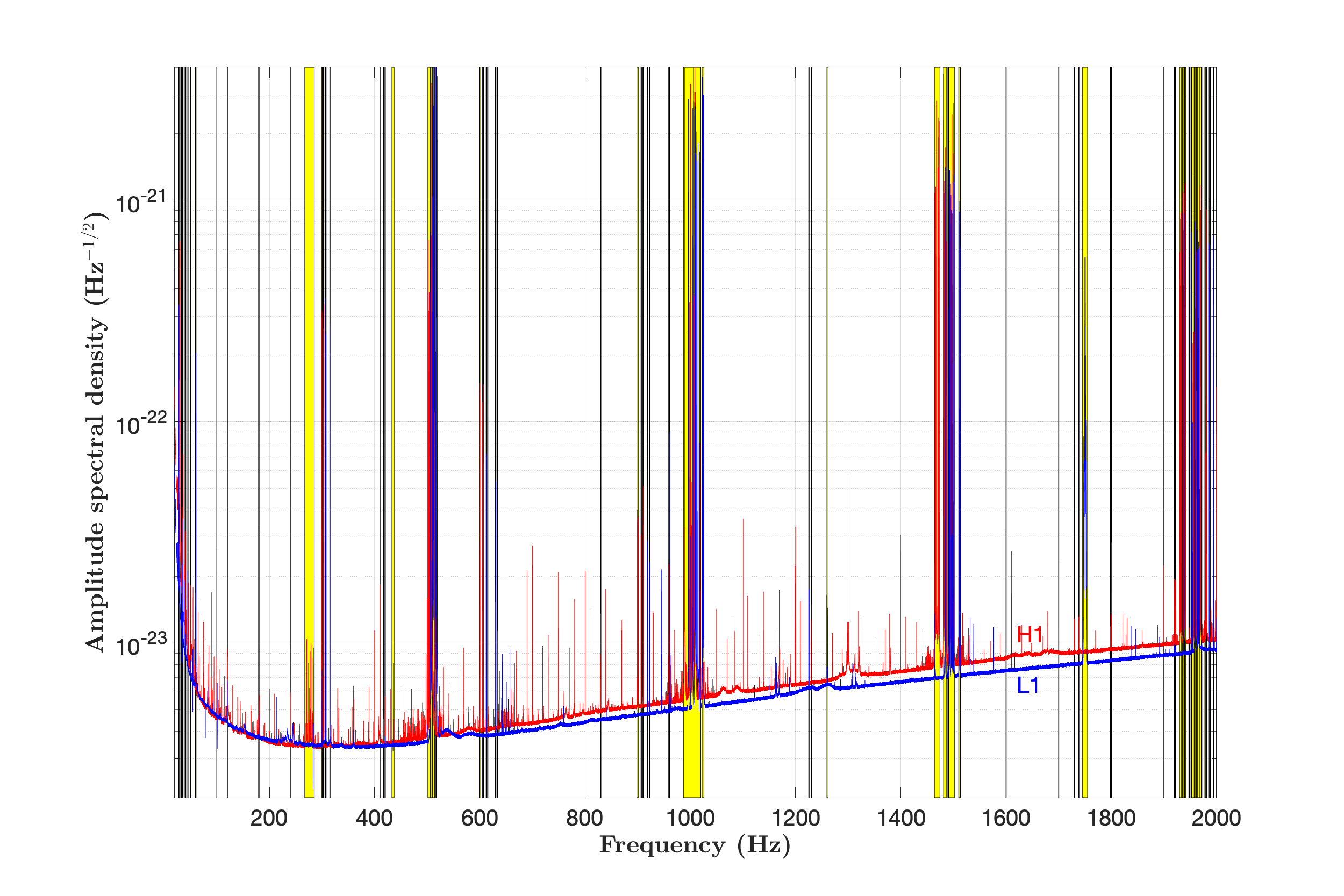}
  \includegraphics[width=6.5in]{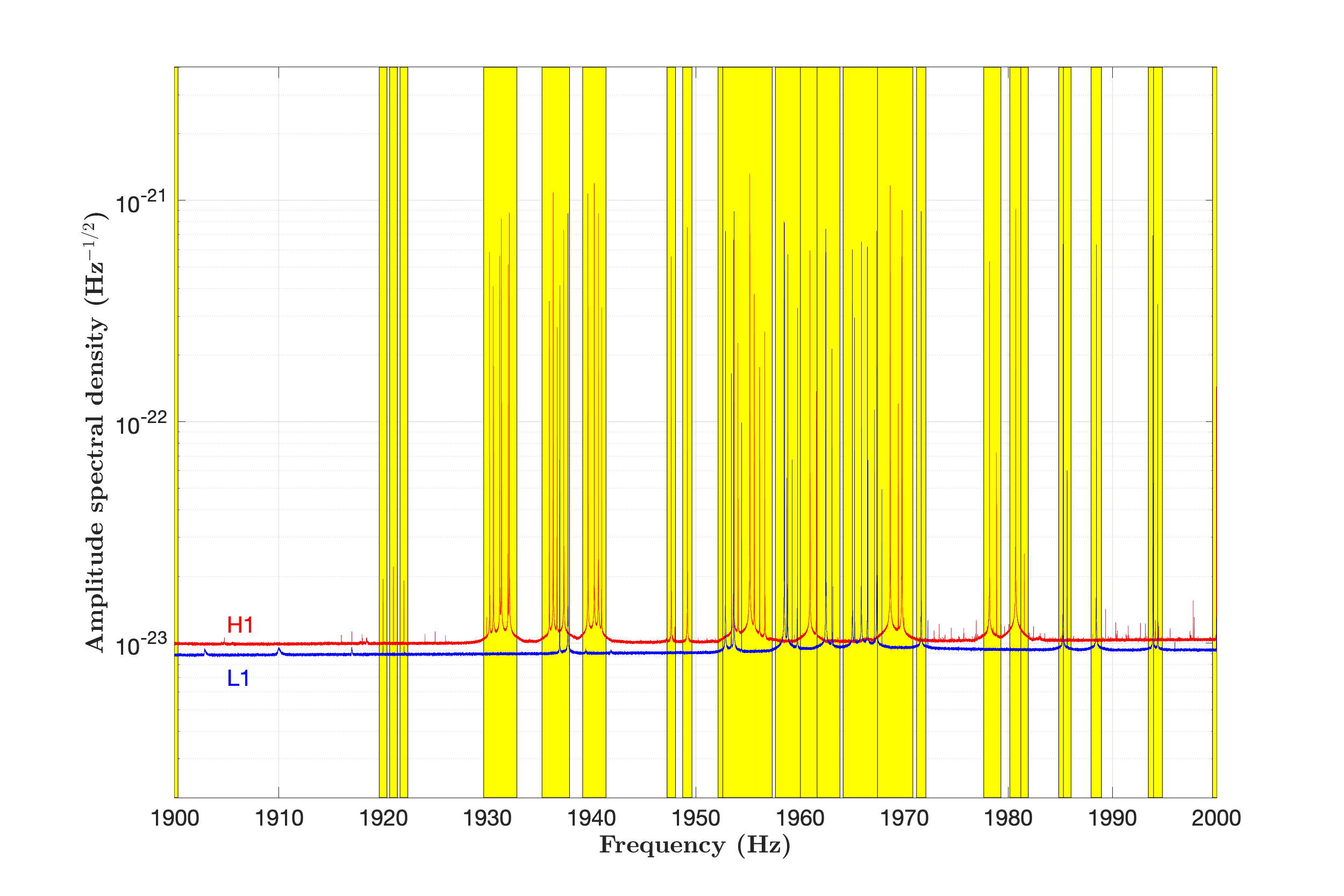}
  \fi
  \caption[Vetoed bands]{
    Shaded regions (yellow) overlaid on the run-averaged spectra (inverse-noise weighted) from the
    H1 (red) and L1 (blue) O3a data sets 
    indicate vetoed bands for which outlier follow-up is impeded by artifacts. The widest affected
    bands correspond to test-mass ``violin modes'' (ambient vibrations of the silica fibers from which LIGO mirrors are suspended)
    near multiples of 500 Hz.
    {\it Top panel:} Full 20-2000 Hz band. 
    {\it Bottom panel:} Magnification of example 1900-2000 Hz band, for which line artifacts
    are dominated by the 4th harmonics of violin modes. 
    (color online).}
  \label{fig:powerflux_vetobands}
\end{figure*}

\subsection{Upper limits results}
\label{sec:powerflux_upperlimits}

Figure~\ref{fig:powerflux_upper_limits} presents strict \pf\ upper limits
on signals, for different assumed polarizations, along with \pf\ upper limits from
searches in the O3 data set~\cite{bib:cwallskyO3a,bib:TripatheeRilesO3}.
The cyan curve shows the upper limits for a worst-case (linear) polarization. The black curve shows
upper limits for an optimally oriented source (circular polarization). Both curves represent strict frequentist upper
limits on these worst-case and best-case orientations with respect to location on the sky, spin-down and frequency
within each narrow search band. For sensitivity comparisons with other search pipeline results,
the purple curve represents approximate population-averaged upper limits
(over random sky locations and polarizations) and is
derived from the circular polarization curve with a simple scale factor (2.3), based on injection studies in test bands.

Each linear-polarization or circular-polarization point in Fig.~\ref{fig:powerflux_upper_limits} represents a maximum over the sky,
except for a small excluded portion of the sky near the ecliptic poles,
which is highly susceptible to detector artifacts due to stationary frequency evolution produced by the combination of
frequency derivative and Doppler shifts~\cite{bib:cwallskyS5}.

The O4a \pf\ results presented here improve upon the previous O3a \pf\ results in strain
sensitivity with median factors ranging from $\sim$1.6 to $\sim$1.8 at signal frequencies above 300 Hz,
where O4 detector noise improvement with respect to O3 performance is greatest.

\begin{figure*}[htbp]
  \begin{center}
    \ifshowfigs
    \includegraphics[width=6.8in]{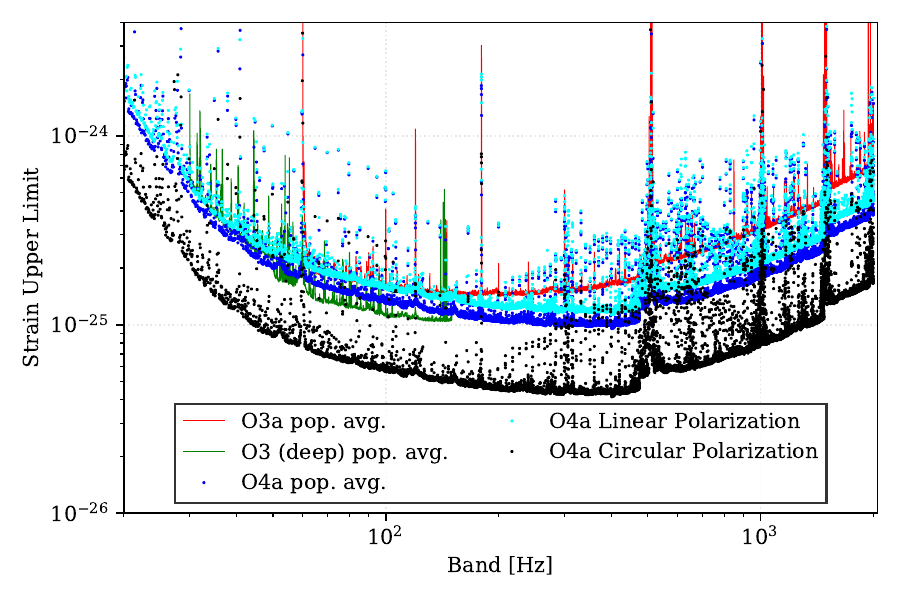}
    \fi
    \caption{\pf\ upper limits on gravitational strain amplitude for this O4a analysis and for previous O3 analyses.
      The cyan curve shows worst-case (linearly polarized) O4a
      $95$\%~CL upper limits in analyzed ($62.5$\,mHz, $125$\,mHz, $250$\,mHz) sub-bands for the
      broad (20-475 Hz, 475-1475 Hz, 1475-2000 Hz) search bands. The lowest black curve shows upper limits assuming a
      circularly polarized source. The dark blue curve shows approximate
      population-averaged all-sky upper limits inferred from the circularly polarized limits.
      For comparison, the red curve shows previous \pf\ population-averaged 95\%\ CL all-sky upper limits
      derived from the 6-month LIGO O3a data set~\cite{bib:cwallskyO3a}, while the green curve extending up to 200 Hz shows population-averaged
      limits from a \pf\ search of the full O3 data set~\cite{bib:TripatheeRilesO3}, using loose coherence in the initial search stage.
      }
    \label{fig:powerflux_upper_limits}
  \end{center}
\end{figure*}

\section{Details concerning the \fh\ search}

We present in the following some details concerning the \fh\ search.

\subsection{\fh\ method overview\label{app:fh_details}}
We describe here the main analysis steps,  with a particular emphasis on the new characteristics and to the specific choices used in this search. The first significant modification is that we have reorganized the \fh\ pipeline in order to make use of the BSD framework~\cite{bib:BSD}. In fact, the use of band-limited time series of the detectors calibrated data, already down-sampled and cleaned, gives the opportunity to divide the analysis in steps of 1 Hz and therefore to evaluate Fast Fourier Transforms (FFTs) with duration $\Tfft$ optimized for each 1 Hz band (the previous implementation used four different bands, each of fixed $\Tfft$ duration). 
The condition~\cite{bib:FH_Doppler_Antonucci2008} $\Tfft <= \sqrt{\frac{c}{2\, f_0 \,R_0}} /\Omega_{sid}$, where $f_0$ is the maximum frequency of the analysed band, $R_0$ the Earth radius at the detector location, $\Omega_{sid}$ the sidereal angular velocity, implies that the maximum Doppler effect (among all the possible sky positions) does not shift the frequency of the signal by more than one bin $\delta f=1/\Tfft$.
As a result, $\Tfft$, in the first step of the search, ranges from 16384 s\footnote{Actually, this value is smaller than the possible maximum, to avoid computational problems related to the dimensions of the resulting \fh maps.}, at 20 Hz, to 1024 s at 1024 Hz, which is the highest frequency of the search, and is re-evaluated for each 1 Hz band (see Fig.~\ref{fig:Tfft_O4a_fh}). 
The cleaning is done before computing FFTs and consists in the identification of short time-domain disturbances with an autoregressive procedure and their
subtraction from the data \cite{bib:SFTdatabase}.
Next, local maxima above a given threshold are selected using the square root of the normalized power of the data, as done in previous searches. This collection of peaks is the so-called ``peakmap''.
\begin{figure}[!ht]
    \centering
    \includegraphics[width=0.45\textwidth]{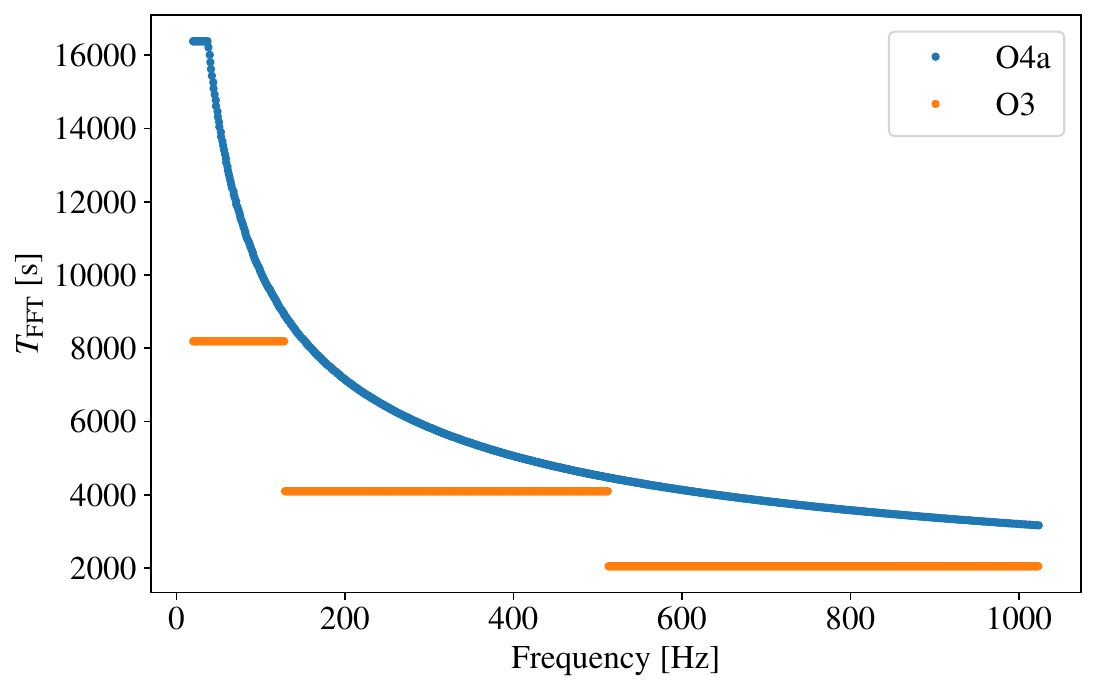}
    \caption{Comparison between the \fh $T_{\rm FFT}$ used in the O3 analysis~\cite{bib:cwallskyO3FourPipelines} compared with the O4a one. We are able to fine-tune the coherence time thanks to the BSD framework~\cite{bib:BSD}.}
    \label{fig:Tfft_O4a_fh}
\end{figure}
As already mentioned in Sec.~\ref{sec:frequencyhough_methodology}, for each sky position\footnote{Over a suitable grid, for which the bin size depends on the frequency and sky location.} the time-frequency peaks of the peakmap are properly shifted, to compensate for the Doppler effect due to the detector motion~\cite{bib:freqhough2}. The shifted peaks are then fed to the \fh\ algorithm~\cite{bib:freqhough2}, which transforms each peak to the frequency/spin-down plane of the source. The frequency and spin-down bins (which we will refer to as {\it coarse} bins in the following) are defined, respectively, as $\delta f ={1}/{\Tfft}$ and  $\delta \dot{f}={\delta f}/{T_{\textrm{obs}}}$, where $T_{\textrm{obs}}$ is the total run duration.
In practice, the nominal frequency resolution has been increased by a factor of 10 \cite{bib:freqhough2}, as {\fh} is not computationally bounded by the width of the frequency bin. The algorithm, moreover, adaptively weights any noise non-stationarity and the time-varying detector response~\cite{bib:adaptivefreqhough}. In addition, in this search, a new - more computationally effective - implementation of the {\fh} algorithm has been used for the first time, described in Appendix~\ref{app:newfh}.  
The whole analysis is split into $\mathcal{O}$(millions) of independent jobs, each of which covers a small portion of the parameter space, see Appendix~\ref{app:job}.
The output of a \fh\, transform is a 2-D histogram in the frequency/spin-down plane of the source.

Outliers, that are statistically significant points in this plane, are selected by dividing each 1 Hz band of the corresponding histogram into 20 intervals and taking, for each interval and for each sky location, the one or (in most cases) the two outliers with the highest histogram number count \cite{bib:freqhough2}, for each spin-down sub-range:
$[-10^{-8}, -7\times 10^{-9}]$ Hz/s, $ [-7\times  10^{-9},~ -4\times 10^{-9}]$ Hz/s, $ [-4\times 10^{-9},~  -1\times 10^{-9}]$ Hz/s, $ [-1\times 10^{-9},~ +2\times 10^{-9}]$ Hz/s. The second outlier is selected, as in previous searches, only if it is far enough from the highest one, suggesting the two have different origin.
All the steps described so far are applied separately to  the data of each detector involved in the analysis.
As in past analyses~\cite{bib:cwallskyO1paper2,bib:cwallskyO2}, coincidences among outliers of the two detectors are required, using a distance metric $d_{\rm FH}$ built in the four-dimensional
parameter space of sky position $(\lambda,~\beta)$ (in ecliptic coordinates), frequency $f$ and spin-down $\dot{f}$.\footnote{The distance is defined as
$$d_{\rm FH}=\sqrt{\left(\frac{\Delta f}{\delta f}\right)^2+\left(\frac{\Delta \dot{f}}{\delta \dot{f}}\right)^2+\left(\frac{\Delta \lambda}{\delta \lambda}\right)^2+\left(\frac{\Delta 
\beta}{\delta \beta}\right)^2},$$ $\Delta f$, $\Delta \dot{f}$, $\Delta \lambda$, and $\Delta \beta$ are the differences, for each parameter, among pairs of outliers of the two detectors, and $\delta f$, $\delta 
\dot{f}$, $\delta \lambda$, and $\delta \beta$ are the corresponding bin widths.}
Thanks to some optimizations done in the entire analysis chain, we have been able to eliminate the clustering before coincidences, that was used in the past searches to reduce the computational load of making coincidences.
Pairs of outliers with distance  $d_{\rm FH}\le 3$ are considered coincident. 

Coincident outliers are ranked according to the value 
a statistic built using a combination of the distance and the \fh\ histogram weighted number count of the coincident outliers \cite{bib:freqhough2}. After the ranking, the eight outliers in each 0.01 Hz band (we have enhanced by a factor of 10 the number of selected outliers at this stage, for the same reasons stated above) with the highest values of the statistic are selected and subject to the follow-up.

\subsection{New \fh\ implementation\label{app:newfh}}
The algorithm performing the \fh\ transform has been optimized with respect to the one used up through the O3 run. The \fh\ output map is a two-dimensional histogram over the frequency/spin-down parameter space in the source frame. In the previous implementation, each point in the input map was mapped as a straight line in the output space. Even if this mapping was implemented through a well-optimized loop, different points could write to the same memory locations, preventing this operation to be done in parallel. Following the inputs from~\cite{bib:LaRosaEtal} and under a flagship project inside \textsc{ICSC}~\cite{ICSC}, in the new implementation the loop has been inverted. Each point in the output map is computed by integrating along the corresponding line in the input map. This new implementation reaches the same performance as the previous one in terms of computing time when run on single-core CPUs. However, the new program can be executed in parallel exploiting modern parallel devices, in particular GPU devices, with a gain in speed of more than one order of magnitude in most cases.

\subsection{\fh\ follow-up details\label{app:fh_fu_details}}
After the first follow-up step described in Sec.~\ref{sec:fh_fu}, a sequence of vetoes is applied to refined outliers. 
First a new peakmap is created around the refined outlier frequency, using data corrected with updated parameters. This peakmap is projected onto the frequency axis and only those outliers whose Critical Ratio (CR) values have increased after the procedure are retained.
The CR is defined as $CR = \frac{x-\mu}{\sigma}$,
where $x$ is the value of the peakmap projection in a given frequency bin, $\mu$ is the average value and $\sigma$ the standard deviation of the projection.
The projection is then analyzed by splitting the frequency range into sub-bands, and the outlier is kept only if it ranks among the top two peaks in both detectors.
This significance veto is followed by a consistency check, which rejects coincident outliers whose CRs, properly weighted by the detector noise level, differ by more than a factor of five. To eliminate contamination from instrumental artifacts, outliers falling within known disturbed bands are also discarded. 
Additionally, coincident outlier pairs separated by a follow-up distance greater than a predefined threshold are excluded. Outliers associated with hardware injections are identified through similar criteria and analyzed separately for diagnostic purposes. Only those outliers that pass all these tests proceed to the subsequent follow-up stages, where the coherence time is further increased, and that are based on \pyfstat{}~\cite{bib:AshtonPrix,bib:pyFstat,Mirasola:2024lcq} in a similar way to what is described in Sec.~\ref{sec:powerflux_methodology}.
The second stage is run with $\Tcoh = 0.5$~day~\cite{bib:cwallskyO3FourPipelines,bib:cwallskybinaryO3aBinarySkyHough,bib:cwAXMPO3,
KAGRA:2022osp, Whelan:2023,bib:O3CrossCorr,bib:cwallskyO3FourPipelines}\footnote{Note that this choice results in a lower coherence time than the first stage for $f\lesssim 49$~Hz. Since \pyfstat{} coherently combines detector's data, we maintain the stage's effectiveness as supported by having detected a hardware injection (Inj10) in that freqency band (see Sec.~\ref{sec:fhufollowup}).}.
We set Gaussian priors over $f$, $\dot{f}$, $\alpha$, $\delta$ following~\cite{Mirasola:2024lcq} where their standard deviation is set to the parameter's resolutions.
Additionally, we consider the second-order spin-down contribution for a set of outliers (see Sec.~\ref{sec:fhulresults}) by including an additional Gaussian prior for this parameter.

Since \pyfstat{} coherently combines H1 and L1 data, we center the priors at the average of the single-detector results of the previous stage.
The MCMC configuration is set to the suggested one in~\cite{Mirasola:2024lcq}
\begin{equation*}
    \Nwalk=50,\,\Ntemp=1, \,\Ntot=300\,,
\end{equation*}
see~\cite{bib:AshtonPrix} for more details on such parameters.
The loudest template of the search is then taken as the follow-up stage's output and compared with the noise distribution to estimate its false-alarm probability ($\pfa$).
Following the discussion in~\cite{Mirasola:2024lcq}, we evaluated the $\pfa$ for each outlier by running $500$ off-sourced MCMCs.
In a similar manner to the suggested procedure of~\cite{bib:TenorioEtal,bib:IsiEtalSkyShift}, we performed the off-sourcing using the same priors of the search with only randomly shifting in each case the $\alpha$ value requiring that $\alpha-\alpha_{\rm cand}\geq \frac{\pi}{2}$ to avoid overlapping with the outlier's parameter space.
The loudest templates from all the runs are histogrammed and fitted with a Gumbel distribution similar to the procedure described in Appendix~A of~\cite{Mirasola:2024lcq}.
Only those outliers that fall below $\pfa<1\%$ are deemed as outliers and checked carefully before being further analysed in a third follow-up stage.
We examine the cumulative distribution of the detection statistic to discard outliers linked to detectors' time-dependent instabilities, similarly to~\cite{bib:cwallskyO3FourPipelines}.
Outliers with sudden variations of the cumulative $2\F$ are discarded, while the others are followed up in another stage.

As a further check, we also compare the results from the MCMC stages using Eq.~(11) of~\cite{bib:cwallskyO3FourPipelines}.

\subsection{Inclusion of the second order spin-down\label{app:fh_space}}

For the first time, we interpret the all-sky results in terms of an explicit spin-down power-law model for CW emission.
Assuming that the source is spinning down following the power law, $\dot{f} \propto -f^n$~\cite{LIGOScientific:2008hqb, bib:palomba1}, we can relate the expected second-order spin-down of the source to its frequency $f$ and $\dot{f}$ as
\begin{equation}
    \ddot{f} = n \, \frac{\dot{f}^2}{f}\,,
\label{eq:fddot_power_law}
\end{equation}
where $n$ is the braking index of the NS, the value of which depends on the driving spin-down mechanism. For instance, $n=3$ for a NS losing rotational energy only through electromagnetic dipole emission and $n=5$ for a NS (sometimes referred to as \textit{gravitar}) for which rotational energy is lost only through GWs.
The current implementation of the \fh~\cite{bib:freqhough2} and the first-stage follow-up procedure do not explicitly accommodate $\ddot{f}\neq0$.
We can identify the parameter-space region where the contribution of $\ddot{f}$ can be neglected by comparing Eq.~\eqref{eq:fddot_power_law} with the natural resolution $\delta\!\ddot{f} = 2\,\delta\!f/T_{\rm obs}^2$.
When $\delta\!\ddot{f} > \ddot{f}$, then the search is not affected by the outlier's second-order spin-down.
For a gravitar, this condition is equivalent to 
\begin{equation}
    \dot{f}^2 < \frac{2}{5}\,\frac{f_0}{T_{\rm FFT}\,T_{\rm obs}^2}\,.
\label{eq:selection_gravitar_cands}
\end{equation}
Outliers with parameters such that Eq.~\eqref{eq:selection_gravitar_cands} is verified at the first follow-up stage are analysed separately, including $\ddot{f}$ from the second follow-up stage on, where needed.
This corresponds to outliers falling within the overlap between the blue and orange regions in Fig.~\ref{fig:overlap_FH_par_space_gravitar_condition-label}. 
All the others have been analysed without accounting for any $\ddot{f}$ contribution.

\begin{figure}[!ht]
    \centering
    \includegraphics[width=0.45\textwidth]{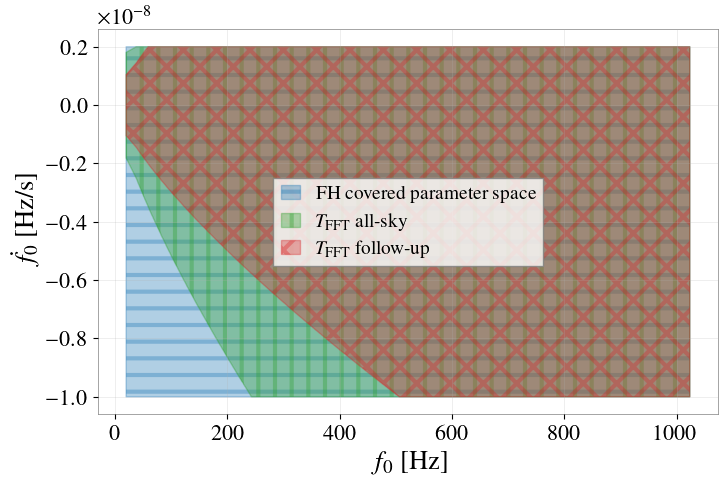}
    \caption{Parameter space covered by the \fh in the O4a search (blue with diagonal hatching). The orange region with star hatching corresponds to the part of parameter space for which Eq.~\eqref{eq:selection_gravitar_cands} is verified using the $\Tfft$ of the first follow-up stage (three times that of the initial search, see Fig.~\ref{fig:Tfft_O4a_fh}). Here, the second stage of the follow-up, which is based on \pyfstat{} and uses larger coherence times, has explicitly taken into account $\ddot{f}$ when needed. The portion of green region with vertical hatching not overlapped with the orange region denotes the part of parameter space for which Eq.~\eqref{eq:selection_gravitar_cands} is verified using the $\Tfft$ of the initial search, but not verified with the $\Tfft$ of the first follow-up stage. The blue-only part of the parameter space corresponds to the situation in which Eq.~\eqref{eq:selection_gravitar_cands} is never verified. All the follow-up stages in these two cases have been carried on without taking into account $\ddot{f}$.}
    \label{fig:overlap_FH_par_space_gravitar_condition-label}
\end{figure}

\subsection{\fh\ Sensitivity formula\label{app:sens}}
The analytical formula for the search sensitivity, given by Eq.~\eqref{eq:hul}, corrects an error in the original expression (equation 67 in~\cite{bib:freqhough2}). The error was due to a wrong integrand in the inner integral in equation 19 of~\cite{bib:freqhough2}, where the probability of selecting a peak in presence of a signal of given amplitude was used, instead of the probability of selecting a peak in absence of signals. Using the correct integrand yields the following expression for the upper limit, at given confidence level $\Gamma$, averaged over all the source parameters:
\begin{align}
h_{\mathrm{UL},\Gamma} = \frac{5}{2}\sqrt{\frac{\pi}{2.4308}}  \frac{S^{1/2}_n}{N^{1/4}\sqrt{T_\mathrm{FFT}}} \times\\  \left(\frac{p_0(1-p_0)}{p_1^2}\right)^{1/4}
     \sqrt{CR_\mathrm{max} -\sqrt{2}\mathrm{erfc}^{-1}(2\Gamma)},
     \label{eq:fh_sens_full}
\end{align}
where 
\begin{equation}
    p_0 = e^{-\theta} - e^{-2\theta} + \frac{1}{3}e^{-3\theta}
\end{equation}
and 
\begin{equation}
p1 = \theta \left(\frac{1}{2}e^{-\theta} - \frac{1}{2}e^{-2\theta} + \frac{1}{6}e^{-3\theta}\right) + \frac{1}{4}e^{-2\theta} - \frac{1}{9}e^{-3\theta},
\end{equation}
which differs from the expression originally found in~\cite{bib:freqhough2}. Note also, in Eq.~\eqref{eq:hul}, the absence of the term $\theta^{-1/2}$ present in equation 67 of~\cite{bib:freqhough2}. Taking $\theta=2.5$ and the confidence level $\Gamma=0.95$ we obtain Eq.~\eqref{eq:hul}. Numerically, the new expression gives estimated strain amplitude limits about 5$\%$ smaller than the original expression (for $\theta=2.5$). A more detailed discussion of the FrequencyHough analytical sensitivity estimation and of its various generalization is presented in \cite{ref:houghsens}.

\subsection{\fh\ job submission framework\label{app:job}}
The computing workflow for the \fh\ search has been traditionally performed using storage and computing resources mainly provisioned by the INFN--CNAF Tier--1~\cite{CNAF} computing centre and strictly tied with its specific infrastructure. Starting with O4a, the workflow has been adapted to be more general and to also comply with computing centers in the IGWN \cite{IGWN} community. We tried to achieve a uniform submission framework which enables us to have the campaign progressing with different computing resource providers, such as the INFN--CNAF, the IGWN federation of sites, and possibly others, such as ICSC~\cite{ICSC} or CINECA~\cite{CINECA}. 

To pursue this objective several improvements or additions with respect to the previous model as adopted for O3, which was entirely performed at CNAF, were considered :
\begin{enumerate}
  \item make the analysis code independent from site specific details, such as where the input files are located, how these should be accessed etc. The analysis code should start assuming that everything needed has been put in place already, and it should end assuming that the produced output will be properly handled and safely stored after it has done.
  \item make every job behave as predictably as possible: in particular, the expected runtime of each job should be lower than the maximum allowed runtime at the site
  \item whenever possible, recover and reuse partial results from evicted jobs
  \item have the analysis campaign proceed in parallel over several distinct sites (such as CNAF) or network of sites (such as IGWN or ICSC)
  \item have the computing campaign progress automatically, as unattended as possible  
  \item be resilient against paradigm shifts, such as protocol changes of some computing services. During the progress of the O4a analysis campaign, some auxiliary services muted their access interface and authentication method: CVMFS for data was deprecated and replaced by OSDF/Pelican, data access credentials shifted from X509 proxy to SCITOKENS etc. Impact on researcher's code of these changes must be minimized at most, and the submission framework should be easily adaptable to comply with such changes. 
\end{enumerate}

The above items were addressed in chronological order and the resulting submission tool has been successfully adopted to complete the analysis for O4a as also a few more use cases. We plan to thoroughly describe this work in a dedicated document.

\subsection{\fh\ upper limits from injections\label{app:ul_fh}}
In order to validate upper limits presented in Sec.~\ref{sec:fhulresults}, several injections for each 1~Hz frequency band were done. We inject 100 signals in real data, with the same amplitude $h_0$, with parameters randomly drawn from those used in the real search. We compute the analysis on these new datasets and we count the number of signals that we are able to recover, meaning those that produce an outlier in the \fh\ maps, with a parameter space distance $\le 3$ from one of the injections. This procedure is repeated for different amplitudes for each band, thus we obtain a curve of the fraction of signals recovered as a function of $h_0$. We hence perform a nonlinear fit using the three-parameters sigmoidal function:
\begin{equation}
    \label{eq:fh_gompertzmodel}
    f(x) = \frac{100\%}{\bigl(1+e^{-a(x-b)}\bigr)^c},
\end{equation}
which is the Gompertz model. Here $a$ determines the steepness of the curve in the central region, and $b$ is a characteristic strain in that region (50\%\ efficiency when $c=1$). The parameter $c$ takes into account potential asymmetries of the sigmoid, and it is 1 when the model Eq.~\eqref{eq:fh_gompertzmodel} reduces to the logistic function. After this procedure, we are able to determine the 95\% upper limit on signal strain. The whole procedure is repeated for 40 different frequency bands; in Fig.~\ref{fig:fh_fitUl} we show an example of the fit for a specific band (95 Hz). In Fig.~\ref{fig:fh_ul}, we show the results obtained for all the bands where we computed upper limits with software injections. One sees that the theoretical estimation is a conservative choice.
\begin{figure}[!ht]
    \centering
    \includegraphics[width=0.47\textwidth]{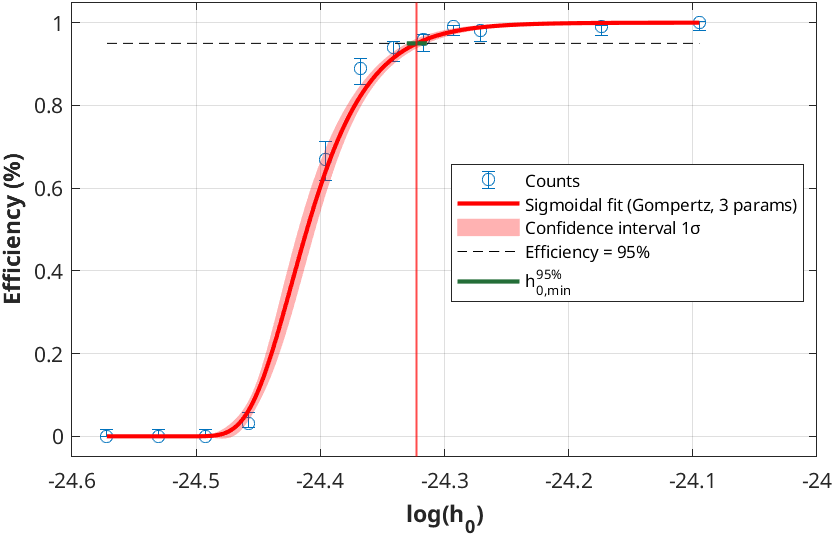}
    \caption{Computation of $h_{0,\rm{min}}^{95\%}$ from \fh\ software injections for a specific frequency band (31 Hz). Blue points represent the fraction of signals recovered out of each 100-injections block. The red line and shaded region are the fit and its confidence interval. The green segment indicates the $h_{0,\rm{min}}^{95\%}$ estimation and its uncertainties.
    }
    \label{fig:fh_fitUl}
\end{figure}

\begin{figure}[!h]
    \centering
    \includegraphics[width=1\linewidth]{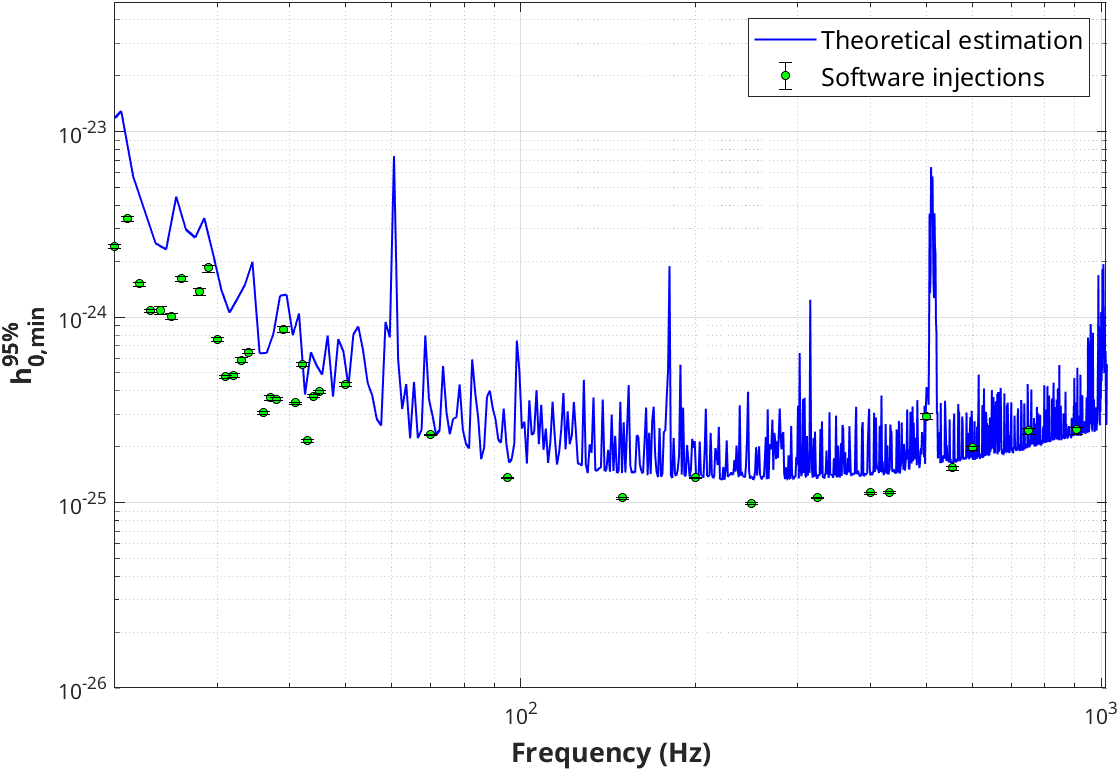}
    \caption{Comparison between theoretical estimation of \fh\ upper limits (see Eq.~\eqref{eq:hul}, blue curve) and its validation via software injections in some frequency bands (green points). The blue curve is obtained by taking the maximum outlier for each 1 Hz band. This estimation is conservative with respect to the limit obtained from injections.}
    \label{fig:fh_ul}
\end{figure}

\section{SOAP methodology \label{ssect:SOAPm}}

{\soap} \cite{bib:ViterbiGlasgow} is a fast, model-agnostic search for long duration signals based on the Viterbi algorithm \cite{Viterbi1967}.

\subsubsection{Search pipeline \label{ssect:SOAPm:SOAP}}

{\soap} searches through each of the summed and narrow-banded spectrograms described in Sec.~\ref{ssect:SOAPm:data} by rapidly identifying the track through the time frequency plane that gives the maximum sum of a detection statistic.
In this search the statistic used is known as the `line aware' statistic~\cite{bib:ViterbiGlasgow}, which uses multiple detectors data to compute the Bayesian statistic $p(\rm{signal})/[p(\rm{noise}) + p(\rm{line})]$, penalising instrumental line-like combinations of spectrogram powers. 
Since each of the four bands described in Sec.~\ref{ssect:SOAPm:data} takes the sum of a different number of SFT bins, the $\chi^2$ distributions that make up the Bayesian statistic are adjusted such that they have $2\times N\times M$ degrees of freedom, where $M$ is the number of summed frequency bins and $N$ is the number of summed time segments.

{\soap} then returns three main outputs for each sub-band: the Viterbi track, the Viterbi statistic and a Viterbi map. 
The Viterbi track is the time-frequency track which gives the maximum sum of statistics along the track, and is used for the parameter estimation stage in Sec.~\ref{ssect:SOAPm:parest}. 
The Viterbi statistic is the sum of the individual statistics along the track, and is one of the measures used to determine the candidates for followup in Sec.~\ref{ssect:SOAPm:cand}.
The Viterbi map is a time-frequency map of the statistics in every time-frequency bin which has been normalised along every time step. 
This is representative of the probability distribution of the signal frequency conditional on the time step and is used as input to the convolutional networks described in Sec.~\ref{ssect:SOAPm:CNN}.

\subsubsection{Convolutional neural network post processing\label{ssect:SOAPm:CNN}}

One post processing step in {\soap} consists of \acp{CNN} which take in combinations of three data types: the Viterbi map, the two detectors' spectrograms and the Viterbi statistic.
The aim of this technique is to improve the sensitivity to isolated neutron stars by reducing the impact of instrumental artifacts on the detection statistic.
This part of the analysis does add some model dependency, so is limited to search for signals that follow the standard CW frequency evolution.
The structure of the networks is described in~\cite{Bayley2020},  where the output is a detection statistic which lies between 0 and 1.
The networks are trained on $\sim 1 \times 10^5$ examples of continuous wave signals injected into real data, where the data is split in the same way as described in Sec.~\ref{ssect:SOAPm:data}.
Each of the sub-bands is duplicated, and a simulated continuous GW is injected into one of the two sub-bands such that the network has an example of noise and noise+signal cases. 
The sky positions, the frequency, frequency derivative, polarisation, cosine of the inclination angle and SNR of the injected signals are all uniformly drawn in the ranges described in~\cite{Bayley2020}.
These signals are then injected into real O4a data before the data processing steps described in Sec.~\ref{ssect:SOAPm:data}.
As the neural network should not be trained and tested on the same data, each of the training sub-bands is split into two categories (`odd' and `even'), where the sub-bands are placed in these categories alternately such that an 'odd' sub-band is adjacent to two `even' sub-bands. 
This allows a network to be trained on `odd' sub-bands and tested on `even' sub-bands and vice-versa.
The outputs from each of these networks can be combined and used as another detection statistic to be further analysed as described in Sec.~\ref{ssect:SOAPm:cand}.

\subsubsection{Parameter estimation\label{ssect:SOAPm:parest}}

The parameter estimation stage uses the Viterbi track to estimate the Doppler parameters of the potential source. 
Because of the complicated and correlated noise which appears in the Viterbi tracks, defining a likelihood is challenging. 
To avoid this difficulty, likelihood-free methods are used, in particular a machine learning method known as a conditional variational auto-encoder.
This technique was originally developed for parameter estimation of compact binary coalescence signals \cite{gabbard_bayesian_2022}, and can return Bayesian posteriors rapidly ($< 1$s).
In our implementation, the conditional variational auto-encoder is trained on isolated NS signals injected into many sub-bands, and returns an estimate of the Bayesian posterior in the frequency, frequency derivative and sky position \cite{bib:SOAP2}.
This acts both as a further check that the track is consistent with that of an isolated NS, and provides a smaller parameter space for a followup search.

\date{\today}

\bibliographystyle{unsrt}
\bibliography{references}

\iftoggle{endauthorlist}{
\iftoggle{fullauthorlist}{
  \let\author\myauthor
  \let\affiliation\myaffiliation
  \let\maketitle\mymaketitle
  \title{The LIGO Scientific Collaboration, Virgo Collaboration, and KAGRA Collaboration}
  
  \newpage
  \maketitle
}

}

\end{document}